\pdfoutput=1
\documentclass[11pt,letterpaper]{article}

\usepackage{amsmath,amssymb,amsthm}
\usepackage{booktabs}     %
\usepackage{deepthink}

\usepackage[nameinlink]{cleveref}

\usepackage[
  backend=biber,
  style=alphabetic,
  maxbibnames=100,
  minbibnames=100,
  maxcitenames=2,
  mincitenames=2,
  natbib=true      %
]{biblatex}
\expandafter\def\csname ver@paralist.sty\endcsname{0000/00/00 v0 stub}
\usepackage{amsmath,amsthm,amssymb,amsbsy,amscd,bm}
\usepackage{paralist}
\usepackage{color}
\usepackage{cleveref}
\usepackage{graphicx}
\usepackage{xspace}
\graphicspath{{./figs/}}
\usepackage{algorithm}
\usepackage{algorithmic}
\usepackage{comment}
\usepackage{multirow}
\usepackage{enumitem}
\usepackage{fancyhdr}
\usepackage{subcaption}
\usepackage{wrapfig}

\newtheorem{theorem}{Theorem}

\theoremstyle{remark}
\newtheorem{remark}{Remark}

\newcommand{\e}{\begin{equation}}
\newcommand{\ee}{\end{equation}}
\newcommand{\en}{\begin{equation*}}
\newcommand{\een}{\end{equation*}}
\newcommand{\eqn}{\begin{eqnarray}}
\newcommand{\eeqn}{\end{eqnarray}}
\newcommand{\bmat}{\begin{bmatrix}}
\newcommand{\emat}{\end{bmatrix}}

\DeclareMathAlphabet\mathbfcal{OMS}{cmsy}{b}{n}

\newcommand{\mb}{\bm}

\newcommand{\mrm}{\mathrm}

\newcommand{\eps}{\epsilon}

\newcommand{\calK}{\mathcal{K}}

\graphicspath{{./figs/}}

\newlength{\imgwidth}
\newboolean{twoColVersion}
\setboolean{twoColVersion}{false}
\newcommand{\twoCol}[2]{\ifthenelse{\boolean{twoColVersion}} {#1} {#2} }

\newcommand{\xmath}[1] {\ensuremath{#1}\xspace}

\newcommand{\blmath}[1] {\xmath{\bm{#1}}}

\newcommand{\x} {\blmath{x}}

\newcommand{\I} {\blmath{I}}

\newcommand{\cA} {\xmath{\mathcal{A}}} %

\newcommand{\cL} {\xmath{\mathcal{L}}} %
\newcommand{\cN} {\xmath{\mathcal{N}}} %

\newcommand{\xnext} {\xmath{\x_{k+1}}}
\newcommand{\xk} {\xmath{\x_k}}

\definecolor{customcolor}{RGB}{128, 0, 50}

\newcommand{\pmstd}[1]{{\scriptsize$\,\pm\,$#1}}

\newcommand{\figNSvis}{Fig.~\ref{fig:ns_sevir_combined}(a)}
\newcommand{\figSEVIRvis}{Fig.~\ref{fig:ns_sevir_combined}(b)}

\usepackage{etoc}      %

\title{ForcingDAS: Unified and Robust Data Assimilation via Diffusion Forcing}
\vspace{0.5em}

\authorblock{%
  \href{https://umjiayx.github.io/}{\textbf{Yixuan Jia}}\textsuperscript{1},\;
  \href{https://csy2077.github.io/siyichen.github.io/}{\textbf{Siyi Chen}}\textsuperscript{1},\;
  \href{https://bamb00zled82.github.io/}{\textbf{Yida Pan}}\textsuperscript{1},\;
  \href{https://heimine.github.io/}{\textbf{Xiao Li}}\textsuperscript{1},\;
  \href{https://shilianghe007.github.io/}{\textbf{Lianghe Shi}}\textsuperscript{1},\;
  \href{https://sites.google.com/view/jcy132/\%ED\%99\%88}{\textbf{Chanyong Jung}}\textsuperscript{1},\;
  \href{https://www.linkedin.com/in/haijie-yuan-763489290/}{\textbf{Haijie Yuan}}\textsuperscript{2},\\
  \href{https://sites.google.com/view/ismailalkhouri/about}{\textbf{Ismail Alkhouri}}\textsuperscript{3,1},\;
  \href{https://name.engin.umich.edu/people/wu-yue-cynthia/}{\textbf{Yue Cynthia Wu}}\textsuperscript{1},\;
  \href{https://sites.google.com/site/sairavishankar3/}{\textbf{Saiprasad Ravishankar}}\textsuperscript{2},\;
  \href{https://web.eecs.umich.edu/~fessler/}{\textbf{Jeffrey A.~Fessler}}\textsuperscript{1},\;
  \href{https://qingqu.engin.umich.edu/}{\textbf{Qing Qu}}\textsuperscript{1}
}

\affiliation{%
  \textsuperscript{1}\,University of Michigan, Ann Arbor
  \quad 
  \textsuperscript{2}\,Michigan State University
  \quad 
  \textsuperscript{3}\,Los Alamos National Laboratory
}

\abstracttext{%
Data assimilation (DA) estimates the state of an evolving dynamical system from noisy, partial observations, and is widely used in scientific simulation as well as weather and climate science. In practice, filtering methods rely on frame-to-frame transition models. However, these models accumulate errors over long horizons, and they are further fragile when observations are non-Markovian. At the same time, learned DA methods typically commit to a single regime, either filtering (nowcasting, real-time forecasting) or smoothing (retrospective reanalysis), which splits what should be a shared prior across application-specific pipelines. To address both issues, we introduce \textbf{ForcingDAS}, a \emph{unified} and \emph{robust} DA framework. Built on Diffusion Forcing with an independent noise level assigned to each frame, ForcingDAS learns a joint-trajectory prior instead of frame-to-frame transitions. This allows it to capture long-horizon temporal dependencies and reduce error accumulation. In addition, the same trained model spans the full filtering to smoothing spectrum at inference time. Specifically, nowcasting, fixed-lag smoothing, and batch reanalysis are selected through the inference schedule alone, without retraining. We evaluate ForcingDAS on 2D Navier–Stokes vorticity, precipitation nowcasting, and global atmospheric state estimation. Across all settings, a single model is competitive with or outperforms both learned and classical baselines that are specialized for individual regimes.%
}

\date{}  %
\correspondence{Yixuan Jia, \href{mailto:jiayx@umich.edu}{jiayx@umich.edu}}
\resources{\href{https://github.com/umjiayx/ForcingDAS}{Code} | \href{https://umjiayx.github.io/ForcingDAS/}{Project Page}}

\headerlogo{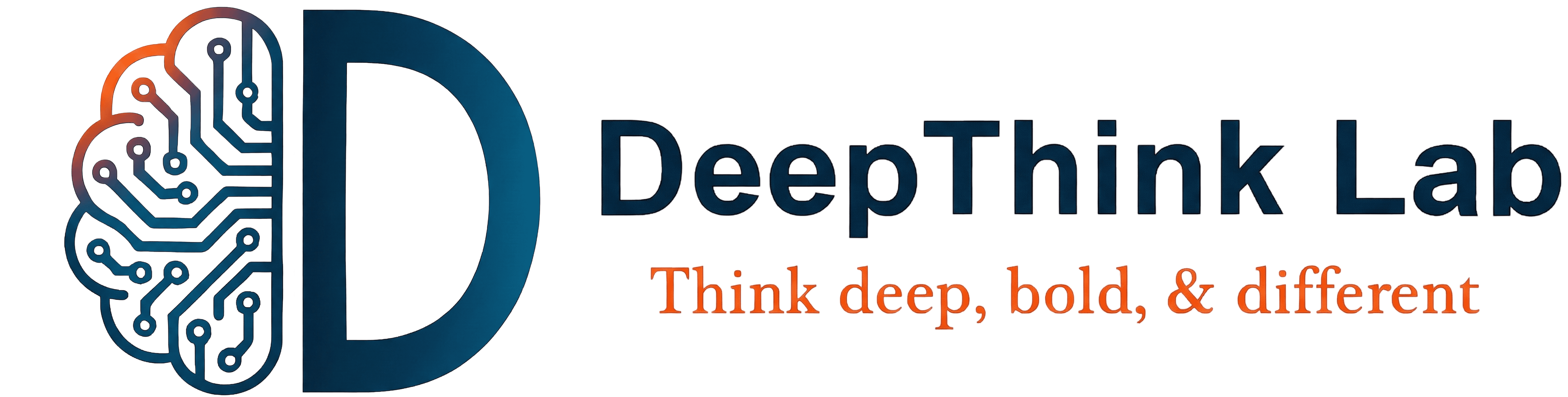}{https://deepthink-umich.github.io}

\begin{document}
\etocdepthtag.toc{mtmain}

\makeDeepthinkHeader

\begin{figure}[!h]
    \centering
    \captionsetup{font=footnotesize, labelfont=bf}
    \vspace{-1em}
    \includegraphics[width=0.84\linewidth]{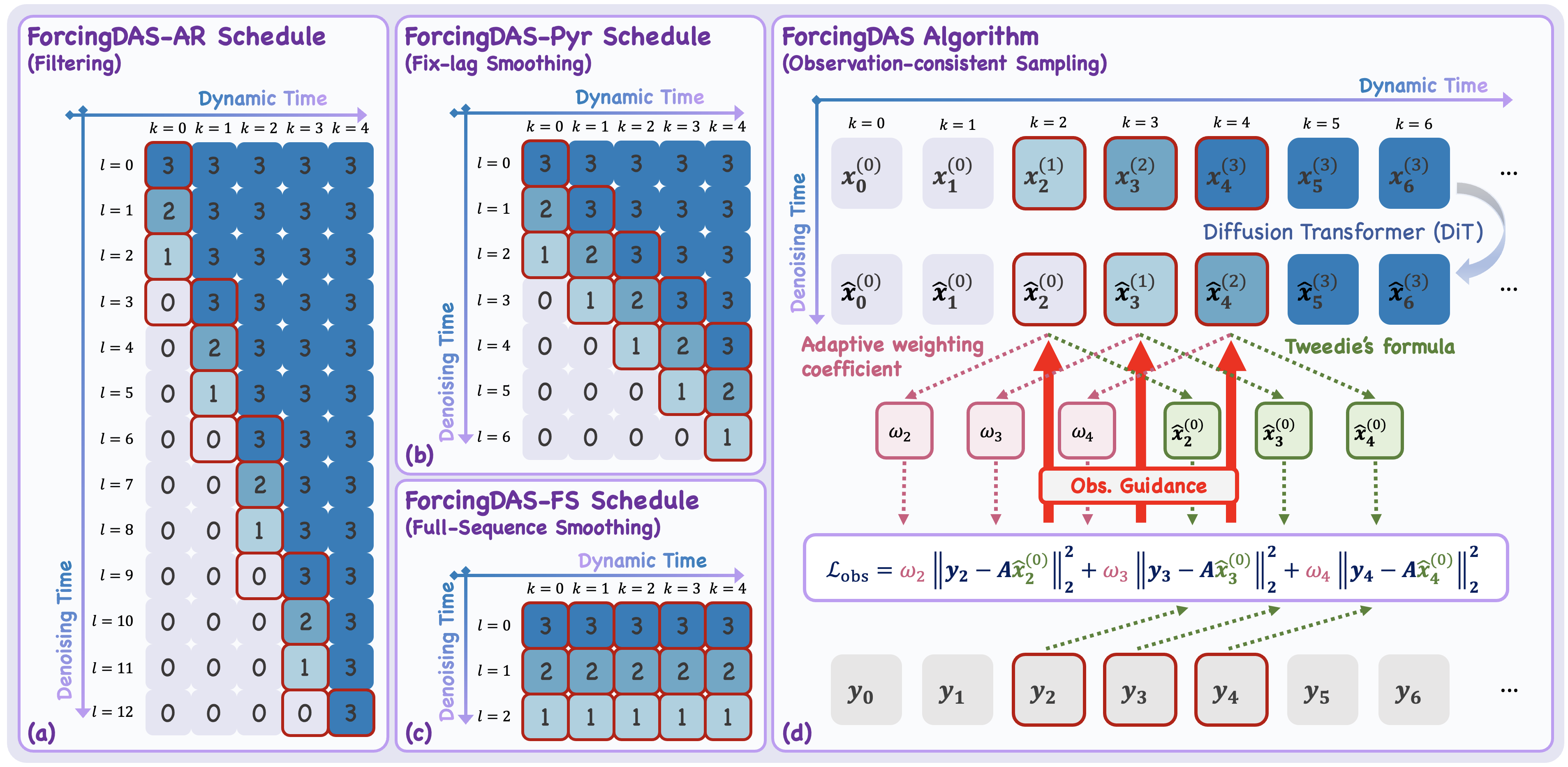}
    \caption{\textbf{ForcingDAS at a glance.} (a--c)~The scheduling matrix $\bm{S}^{(u)}$ (rows: denoising iteration $\ell$; columns: trajectory frame $k$; cells: diffusion step) selects the DA regime via the uncertainty scale $u$: autoregressive filtering ($u = T$), pyramid fixed-lag smoothing ($0 < u < T$), and full-sequence smoothing ($u = 0$). For illustration only, $T = 3$ (and $u = 1$ in (b)). Active frames at each iteration are outlined in red. (d)~Observation-consistent sampling: per-frame Tweedie estimates $\hat{\bm{x}}^{(0)}_k$ are scored against observations $\bm{y}_k$ through adaptive weights $w_k$ tied to each frame's noise level (Eq.~\ref{eq:reweight}); the aggregated loss $\mathcal{L}_{\text{obs}}$ is backpropagated through the shared DiT backbone, routing later-frame observation gradients to earlier active frames.
    }
    \label{fig:teaser}
\end{figure}

\setcounter{tocdepth}{2}
\newpage
{\small
\etocsettagdepth{mtmain}{subsection}
\etocsettagdepth{mtappendix}{none}
\tableofcontents
}
\section{Introduction}
\label{sec:intro}

Accurate state estimation from noisy, incomplete data is critical across many scientific domains, such as weather forecasting \citep{bocquet2015data,reichle2008data,lahoz2010data,law2015data,carrassi2018data}, oceanography \citep{cummings2005operational,cummings2013variational}, and seismology \citep{werner2009earthquake,banerjee2023parameter}.   Mathematically, a discrete-time stochastic dynamical system is described by:
\begin{align}
    \xnext &= \Psi(\xk) + \bm \xi_k,
    \label{eq:stochastic_dynamics_model} \\
     \bm y_{k+1} &= \cA(\xnext) + \bm \eta_{k+1},
    \label{eq:data_model}
\end{align}
where $\xk \in \mathbb{R}^D$ is the state with $1\leq k\leq K$, and $\Psi(\cdot)$ is the transition map typically governed by nonlinear partial differential equations (PDEs) \citep{brunton2015closed,duraisamy2019turbulence}, $\bm \xi_k \in \mathbb{R}^D$ is stochastic forcing, and $\bm y_k \in \mathbb{R}^M$ is partial observation through the sensing operator $\cA(\cdot)$ with noise $\bm \eta_k \sim \cN(\mathbf 0, \sigma_y^2 \I_M)$.

\textbf{Data assimilation (DA)} integrates physical models with observational data to yield physically consistent state estimates \citep{rabier2005overview, geer2018all, fletcher2017data, carrassi2018data}. Depending on the temporal availability of observations, DA is generally classified into several distinct regimes: (\emph{i}) \underline{\emph{Filtering}}, which recursively estimates the current state $\xk$ ``online'' using observations $\bm y_{1:k}$ available up to the present time, serving as the basis for real-time forecasting and nowcasting; (\emph{ii}) \underline{\emph{Smoothing}}, an ``offline'' batch process that refines past trajectories using both past and future observations $\bm y_{1:K}$, is used for retrospective reanalysis and climate science (e.g., the ERA5 reanalysis dataset is produced this way).

\begin{figure*}[t]
    \centering
    \includegraphics[width=\linewidth]{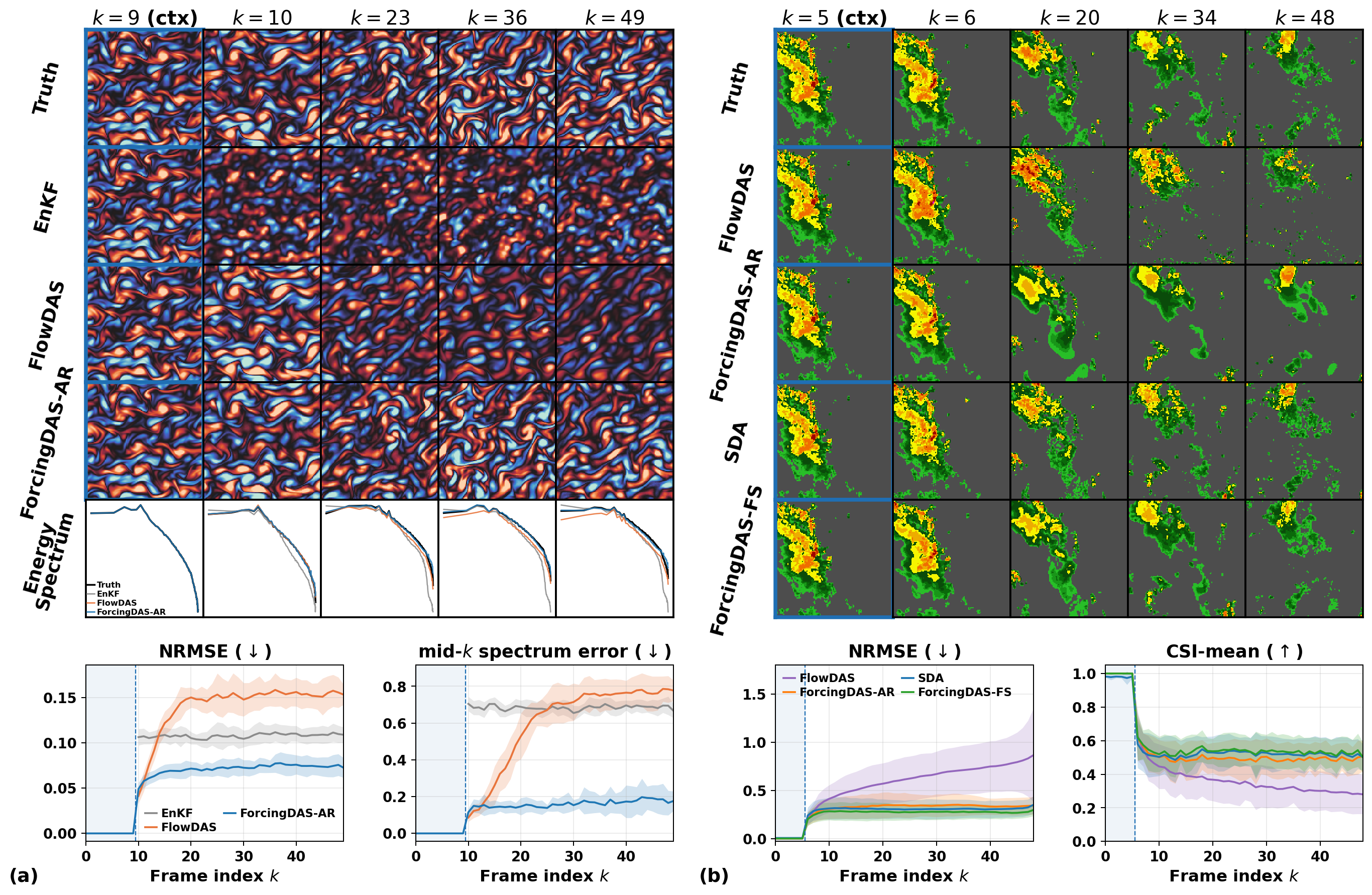}
    \vspace{-0.1in}
    \caption{\textbf{ForcingDAS is unified and robust.} \textbf{(a)}~Navier--Stokes vorticity (SO-5\%) and \textbf{(b)}~SEVIR VIL precipitation (SO-10\%), with clean context frames marked in blue. A single trained ForcingDAS model runs as a filter (AR) and as a full-sequence smoother (FS) without retraining, and its error stays flat over the whole window, whereas the per-step learned filter FlowDAS accumulates error. Curves show mean $\pm$ 1 s.d.\ over the 16 held-out trajectories (a) or events (b); numbers are in Tables~\ref{tab:ns_main} and~\ref{tab:sevir_main}.}
    \label{fig:ns_sevir_combined}
    \label{fig:ns_visualization_main}
    \label{fig:sevir_results}
\end{figure*}

\paragraph{Limitations of existing DA solvers.}
Despite their importance, classical and learned DA solvers are typically designed around one regime. Classical methods carry well-known bottlenecks: the (Ensemble) Kalman Filter \citep{kalman1960new, evensen2003ensemble} assumes linear-Gaussian dynamics, particle filters \citep{gordon1993novel} scale poorly with state dimension, and 4D-Var \citep{talagrand1987variational, courtier1994strategy} requires hand-crafted adjoint models that are expensive to build and run at an operational scale.
Recently, learned solvers alleviate the computational cost but still lack an adjustable regime: trajectory diffusion priors such as SDA \citep{rozet2023score} and its latent extension Appa \citep{andry2025appa} denoise a whole window at one shared noise level, so they are built for reanalysis and obtain a filtering estimate only by re-sampling a window posterior at every step; whereas per-step models such as FlowDAS \citep{flowdas} and Fourier Neural Processes \citep{chen2024fnp} commit to learned one-step transitions, which (a) accumulate errors over long horizons since each frame is committed before the next observation arrives, and (b) condition only on a short, fixed window of recent frames. This limitation matters in real-world applications such as weather forecasting. Each measured frame (e.g., a few coarse-grid surface variables observed at a fixed cadence) captures only a partial slice of a much larger atmospheric state, which includes unresolved subgrid dynamics, unobserved vertical and chemical variables, and processes evolving faster than the observation interval. As a result, past observations can carry information about hidden degrees of freedom that are not contained in the current frame alone; consequently, the future is not merely determined by the present observation. In summary, across all these methods, the regime is fixed at design or training time, or limited to a few discrete tasks, rather than being a continuous choice made at inference.

\paragraph{Why do we need a unified DA approach?}
Ideally, one aims to learn a single model that generalizes across all DA paradigms, given that the underlying dynamical system remains consistent regardless of the observation window. For example, a weather center runs nowcasting (filtering) and reanalysis (smoothing) on the same atmosphere, and the two should share a single learned prior over the dynamics. Training separate models for each DA paradigm not only fragments operational pipelines but also fails to fully exploit the shared dynamical structure inherent in the data. Therefore, in this work, we propose training a unified model across all paradigms (i.e., filtering, smoothing, and forecasting), and the specific paradigm is only chosen at inference, analogous to a foundation model where one set of weights serves multiple downstream tasks. 

\paragraph{Our approach: ForcingDAS.} Following this principle, we propose \textbf{ForcingDAS}, a unified DA framework. We train a single joint-trajectory diffusion prior over full trajectories in which each frame $\xk$ carries its own diffusion-time noise level $t_k$, in contrast to standard video diffusion where all frames share one noise level \citep{chen2024diffusion}.
At inference, this per-frame structure lets a scheduling matrix over the noise levels select the assimilation regime without retraining: a sequential schedule yields autoregressive filtering; a sliding-window pyramid schedule yields fixed-lag smoothing, with future observations refining past states via backward gradients; and a synchronous schedule yields full-sequence smoothing. Two further components close the gap to a usable solver: causality-aware training, which biases training noise toward the monotone patterns the scheduling matrix induces at inference, and noise-level-aware observation guidance, which scales each frame's correction by its current signal quality. See \Cref{fig:teaser} for a visual overview.

 In summary, our contributions can be highlighted as follows.
\begin{enumerate}[nosep, leftmargin=*]
    \item \textbf{A unified data-assimilation framework.} We propose ForcingDAS (\S\ref{sec:method}), a unified model in which filtering, fixed-lag smoothing, and full-sequence smoothing are points on one continuum, selected at inference time by a single scalar of the scheduling matrix. %
    \item \textbf{Robustness over long horizons.} Joint denoising across frames mitigates the error accumulation that has limited autoregressive learned DA, and modeling full trajectories rather than per-step transitions allows the prior to capture dependencies on hidden degrees of freedom that the current frame alone does not reveal (Fig.~\ref{fig:sevir_results}).
    \item \textbf{Strong empirical results on three benchmarks.} A single ForcingDAS model is competitive with or outperforms specialized learned and classical baselines on 2D Navier–Stokes vorticity, SEVIR-VIL precipitation nowcasting, and ERA5 global atmospheric state estimation. 
\end{enumerate}

\section{Background} \label{sec:background}

\subsection{Data Assimilation}
\label{sec:bg:da}

The goal of DA is to estimate the state trajectory $\bm{x}_{1:K} = (\bm{x}_1, \ldots, \bm{x}_K)$ of the dynamical system \eqref{eq:stochastic_dynamics_model} from partial observations $\bm{y}_{1:K} = (\bm{y}_1, \ldots, \bm{y}_K)$ described in \eqref{eq:data_model}. Two regimes are most relevant~\citep{asch2016data}. \textbf{Filtering} targets $p(\bm{x}_k \mid \bm{y}_{1:k})$, 
the current-state posterior given only past and present observations, and is typically run online: a forecast from a predictive model is corrected by the latest observation to produce a state estimate consistent with both the dynamics and the data. \textbf{Smoothing} targets $p(\bm{x}_k \mid \bm{y}_{1:k+\Delta K})$, refining $\bm{x}_k$ using observations up to $\Delta K$ steps ahead; \emph{fixed-lag smoothing} fixes $\Delta K$ to a rolling look-ahead window, while \emph{full-sequence smoothing} sets $\Delta K = K - k$ and uses the entire observed sequence $\bm{y}_{1:K}$, the standard setting for offline reanalysis. An extended discussion of related work covering classical and learned DA solvers is deferred to Appendix~\ref{app:related}.

\subsection{Diffusion Forcing (DF)}
\label{sec:bg:perframe}

A standard diffusion model~\citep{ho2020denoising, sohl2015deep} corrupts a clean data sample $\bm{x}^{(0)}$ at a diffusion step $t$ via $\bm{x}^{(t)} = \sqrt{\bar{\alpha}_t}\,\bm{x}^{(0)} + \sqrt{1-\bar{\alpha}_t}\,\bm{\epsilon}$, where $\bar{\alpha}_t$ is the cumulative signal-retention coefficient and $t \in \{1, \ldots, T\}$ indexes the diffusion step. A neural network $\bm{\epsilon}_{\bm{\theta}}(\bm{x}^{(t)}, t)$ is trained to predict the noise; samples are drawn by iterating the reverse process (e.g., DDIM~\citep{song2020denoising}) from pure noise back to clean data.

Diffusion Forcing (DF)~\citep{chen2024diffusion} generalizes this to temporal sequences by assigning an \emph{independent} diffusion step to each frame in a trajectory $\bm{x}_{1:K}$:
\begin{equation}
    \bm{x}_k^{(t_k)} = \sqrt{\bar{\alpha}_{t_k}}\, \bm{x}_k + \sqrt{1 - \bar{\alpha}_{t_k}}\, \bm{\epsilon}_k, \quad \bm{\epsilon}_k \sim \cN(\mathbf{0}, \I), \quad k = 1, \ldots, K.
    \label{eq:perframe_noise}
\end{equation}
A single denoising network $\bm{\epsilon}_{\bm{\theta}}(\bm{x}_{1:K}^{(\bm{t})}, \bm{t})$ takes the full noisy trajectory together with the per-frame diffusion-step vector $\bm{t} = (t_1, \ldots, t_K)$ and predicts the noise on each frame. %
The temporal backbone uses \emph{causal attention}: frame $k$ attends only to frames $k' \leq k$. At sampling time, a \textbf{scheduling matrix} $\bm{S} \in \{0, \ldots, T\}^{K \times L}$ specifies how each frame's diffusion step descends across the $L$ reverse-process iterations (i.e., $S_{k,\ell}$ is the diffusion step of frame $k$ at iteration $\ell$, with $S_{k,0} = T$ and $S_{k,L} = 0$); different scheduling matrices recover different sampling strategies from the same trained network. A more complete recap of DF and its relation to ForcingDAS is given in Appendix~\ref{app:df_recap}.

\section{Proposed Method}
\label{sec:method}

Building on \Cref{sec:background}, we introduce \textbf{ForcingDAS}, a unified DA framework that elevates DF from a generative modeling paradigm to a complete and versatile DA solver, seamlessly unifying filtering, fixed-lag smoothing, and full-sequence smoothing within a single trained model and selecting the regime entirely at inference through the scheduling matrix. We first identify three challenges that prevent DF from being used as a plug-and-play DA solver, give an overview of how ForcingDAS addresses them (\S\ref{sec:method:overview}), and then provide extra implementation details (\S\ref{sec:method:impl}).

\paragraph{Challenges of Applying Diffusion Forcing to DA}
Although the per-frame noise structure makes DF a natural candidate prior for DA, three challenges prevent it from being used as a plug-and-play DA solver. (\emph{i}) \underline{\emph{Causal mismatch between training and inference.}} Every DA scheduling matrix imposes a causally monotone pattern at inference due to the underlying physics, but DF training samples $\bm{t}$ i.i.d.\ uniformly, leaving the model effectively less trained on the configurations it sees most at inference; this gap is benign on natural-video data but matters for scientific dynamics, as we discuss in detail in \S\ref{sec:method:cat}. (\emph{ii}) \underline{\emph{No mechanism for observation guidance.}} DF models only the prior $p_{\bm{\theta}}(\bm{x}_{1:K})$, lacking a mechanism to integrate noisy partial measurements $\bm{y}_{1:K}$ during the reverse sampling process. However, enforcing these observation constraints is essential for state correction and accurate prediction. (\emph{iii}) \underline{\emph{Limited adaptivity to different DA regimes.}} Using a single trained model to unify filtering, fixed-lag smoothing, and full-sequence smoothing requires a principled family of scheduling matrices. However, as DF employs arbitrary scheduling matrices, it does not explicitly adapt to the corresponding DA regimes. %

\subsection{Overview of Our Method}
\label{sec:method:overview}

We address the three challenges above in turn, with one targeted innovation for each.

\paragraph{Causality-Aware Training (CAT).}
\label{sec:method:cat}
Standard DF draws the per-frame diffusion step vector $\bm{t}$ i.i.d. from $\mathrm{Uniform}\{0,\ldots,T\}^{K}$, whereas every DA scheduling matrix imposes a \emph{causally monotone} pattern at inference (earlier frames at no-higher diffusion steps than later ones). This discrepancy is generally acceptable for natural videos, as adjacent frames are often redundant. However, it significantly impacts the scientific systems considered in this work (NS, SEVIR, ERA5). Because these systems exhibit broadband spatial spectra and strong forward-in-time dependence~\citep{mojgani2024interpretable}, a clean historical state conveys fundamentally more information than a noisy current observation. Therefore, it is critical to train the model using the identical clean-past and noisy-future conditioning it experiences at inference time. We therefore replace the i.i.d.\ sampling with a mixture that biases training toward causally monotone patterns while keeping i.i.d.\ samples as a regularizer:
\begin{equation}
    \mathcal{L}_{\text{CAT}} = \mathbb{E}_{\bm{x}_{1:K}}\; \mathbb{E}_{\bm{t} \sim p_\rho}\; \sum_{k=1}^{K} \lambda(t_k) \left\| \bm{\epsilon}_k - \bm{\epsilon}_{\bm{\theta}}\!\left(\bm{x}_{1:K}^{(\bm{t})},\, \bm{t}\right)_k \right\|^2,
    \qquad p_\rho = \rho\, p_{\text{sorted}} + (1-\rho)\, p_{\text{iid}},
    \label{eq:cat_loss}
\end{equation}
where $p_{\text{iid}}$ is the standard $\mathrm{Uniform}\{1,\ldots,T\}^K$, $p_{\text{sorted}}$ is its pushforward under sorting in non-decreasing order, $\rho \in [0,1]$ is the \emph{causal ratio}, and $\lambda(t_k)$ is an optional per-step weight; $\rho = 0$ recovers standard DF. Detailed motivation and ablations on $\rho$ are in Appendix~\ref{app:forecasting}.

\paragraph{Observation-Consistent Sampling.}
\label{sec:method:guidance}
DF's reverse process samples from the trajectory prior $p_{\bm{\theta}}(\bm{x}_{1:K})$ but provides no mechanism to condition on partial measurements $\{\bm{y}_k\}$. We close this gap with a measurement-likelihood gradient added to each DDIM step, in the spirit of test-time gradient guidance for inverse problems~\citep{chung2022diffusion}. A constant guidance scale, however, is suboptimal in the per-frame setting where different frames are simultaneously at different diffusion steps: strong corrections distort heavily noised frames, while weak corrections miss fine detail on clean ones. We derive an adaptive weighting from the Tweedie estimate's error variance. Recall $\hat{\bm{x}}_k^{(0)} = (\bm{x}_k^{(t_k)} - \sqrt{1-\bar{\alpha}_{t_k}}\,\bm{\epsilon}_{\bm{\theta}}) / \sqrt{\bar{\alpha}_{t_k}}$; the estimation error $\hat{\bm{x}}_k^{(0)} - \bm{x}_k$ has variance scaling as $(1-\bar{\alpha}_{t_k})/\bar{\alpha}_{t_k}$, so the total variance of the observation residual is $\sigma_y^2 + \gamma\,(1-\bar{\alpha}_{t_k})/\bar{\alpha}_{t_k}$ with $\gamma > 0$ controlling the contribution of the prediction uncertainty. Weighting each frame's squared residual by the inverse of this variance, and taking the square root as in the unsquared residual norm of \citet{chung2022diffusion}, yields
\begin{equation}
    \mathcal{L}_{\text{obs}} = \Big( \sum_{k=1}^{K} w(t_k) \cdot \left\| \bm{y}_k - \cA(\hat{\bm{x}}_k^{(0)}) \right\|_2^2 \Big)^{1/2}, \quad w(t_k) = \frac{1}{\sigma_y^2 + \gamma\,\frac{1 - \bar{\alpha}_{t_k}}{\bar{\alpha}_{t_k}}},
    \label{eq:reweight}
\end{equation}
so that frames with reliable Tweedie estimates receive stronger guidance and heavily noised frames are down-weighted. The gradient correction at each DDIM step is
\begin{equation}
    \bm{x}_{1:K}^{(\bm{t}-1)} \leftarrow \text{DDIM-step}(\bm{x}_{1:K}^{(\bm{t})}) - \zeta \nabla_{\bm{x}_{1:K}^{(\bm{t})}} \mathcal{L}_{\text{obs}}.
    \label{eq:guidance_step}
\end{equation}
For schedules with multiple simultaneously active frames at different diffusion steps that we discuss in the following, the residuals in Eq.~\eqref{eq:reweight} are summed over the active window before differentiation; causal attention then routes each later frame's gradient backward to earlier active frames, producing the smoothing pathway analyzed in Appendix~\ref{app:asymmetry}.

\paragraph{Unified DA View of the Scheduling Matrix.}
\label{sec:method:framework}
Building on these innovations for training and sampling in the DA setting, we now establish a unifying framework. We demonstrate that the scheduling matrix inherently acts as a DA-algorithm selector, defining a single-parameter family that seamlessly spans the entire filtering-smoothing spectrum. While the original DF formulation permits arbitrary scheduling matrices, it lacks this algorithmic interpretation. By making this mapping explicit, we enable a single scalar $u \geq 0$ for \emph{uncertainty scale} to recover any standard DA regime without requiring any model retraining. We describe the elements of the scheduling matrix $\mb S$ as 
\begin{equation}
    S_{k,\ell}^{(u)} = \mathrm{clip}\!\Big(T - \ell + u \cdot (k-1),\; 0,\; T\Big), \quad k = 1,\ldots,K,\;\; \ell = 0,\ldots,L-1,
    \label{eq:scheduling}
\end{equation}
where $L = T + u(K-1)$ and $\mathrm{clip}(\cdot, 0, T)$ saturate each entry to the valid diffusion-step range $[0, T]$.
As shown in \Cref{fig:teaser}, the schedule continuously interpolates between three DA regimes by varying $u$: (i) \emph{filtering} ($u = T$), where each frame fully denoises before the next begins, with posterior $p(\bm{x}_k \mid \bm{y}_{1:k})$; (ii) \emph{fixed-lag smoothing} ($0 < u < T$, pyramid), where $W \approx \lceil T/u \rceil$ frames are concurrently active at staggered diffusion steps, and observations on later active frames refine earlier ones through windowed observation aggregation; and (iii) \emph{full-sequence smoothing} ($u = 0$), where all frames descend in lockstep, every observation jointly refines every frame, and the inference reduces to a batch smoother under the constraint of causal attention (see Appendix~\ref{app:asymmetry} for the past--future asymmetry that this entails).

\paragraph{Advantages of ForcingDAS.}
In contrast to specialized solvers, ForcingDAS offers two distinct advantages: (\emph{i}) \underline{\emph{Unification across the DA spectrum}}, our approach enables a single trained model to perform filtering, fixed-lag, and full-sequence smoothing. By adjusting the uncertainty scale $u$ at inference, the model adapts to different observation windows, such as real-time forecasting or retrospective reanalysis, similar to how foundation models support multiple downstream tasks. (\emph{ii}) \underline{\emph{Robustness over long horizons}}. By modeling the joint trajectory prior rather than a per-step transition, ForcingDAS employs joint denoising to suppress the autoregressive error accumulation that limits traditional learned solvers.

\vspace{-0.1in}
\subsection{Implementation Details}
\label{sec:method:impl}

\paragraph{Training.}
For each minibatch trajectory $\bm{x}_{1:K}$, we sample the per-frame diffusion-step vector $\bm{t} = (t_1, \ldots, t_K)$ from the CAT mixture in Eq.~\eqref{eq:cat_loss}: with probability $\rho$ we sort an i.i.d.\ $\mathrm{Uniform}\{1,\ldots,T\}$ draw into a non-decreasing staircase, otherwise we keep the i.i.d.\ draw; with probability $\rho_c$ we additionally clamp a random number of leading frames to diffusion step zero, so the model is exposed to inference-style clean-context conditioning. We then corrupt the trajectory according to Eq.~\eqref{eq:perframe_noise}, evaluate the score-matching loss in Eq.~\eqref{eq:cat_loss}, and update the network parameters. Pseudocode (Algorithm~\ref{alg:training}), the DiT-3D backbone with causal temporal attention, frame stacking, noise clipping, and full optimizer settings are in Appendix~\ref{app:impl}.

\paragraph{Inference.}
Given a trained model and observations $\{\bm{y}_k\}$, we initialize $\bm{x}_k^{(T)} \sim \cN(\mathbf{0}, \I)$ for $k = 1, \ldots, K$ and step through the $L$ scheduled iterations of $\bm{S}^{(u)}$ in Eq.~\eqref{eq:scheduling}. At iteration $\ell$, the active set $\calK_{\text{active}}$ collects the frames whose diffusion step decreases between $\ell$ and $\ell+1$. For these frames, we (\emph{i}) compute the Tweedie's estimate $\hat{\bm{x}}_k^{(0)}$ via the denoising network, (\emph{ii}) form the noise-level-aware reweighted observation residual in Eq.~\eqref{eq:reweight} aggregated over $\calK_{\text{active}}$, (\emph{iii}) backpropagate to obtain a single gradient through the shared denoising network, and (\emph{iv}) apply one DDIM step combined with the gradient correction in Eq.~\eqref{eq:guidance_step}. Iterating over $\ell = 0, \ldots, L-1$ produces a clean trajectory $\bm{x}_{1:K}^{(0)}$ that is consistent with both the trained prior and the observations. Pseudocode (Algorithm~\ref{alg:sampling}) and per-experiment hyperparameters $(\zeta, \gamma)$ are in Appendix~\ref{app:impl}.

\vspace{-0.1in}
\section{Experiments}
\label{sec:exp}

We evaluate ForcingDAS on three benchmarks. 2D Navier--Stokes vorticity (\S\ref{sec:exp:ns}) is a PDE whose vorticity field is the full state, and for which a classical ensemble filter can run the exact dynamics; SEVIR-VIL precipitation nowcasting (\S\ref{sec:exp:sevir}) and ERA5 global atmospheric state estimation (\S\ref{sec:exp:era5}) are real-world benchmarks, where the observed sequence is a low-dimensional projection of a higher-dimensional latent state. Existing learned DA methods on ERA5 (Appendix~\ref{app:related:da}, Table~\ref{tab:era5_landscape}) each commit to a single regime, either filtering or window-based 4D-Var. We evaluate a single trained model in three regimes (filtering, fixed-lag smoothing, and full-sequence smoothing) under our \emph{with-context} protocol; the harder \emph{cold-start} setting is in Appendix~\ref{app:cold_start}. The fixed-lag window is $\min(K, \lceil T/u\rceil)$ frames for an assimilation horizon of $K$ frames and $T=100$ DDIM steps: we use a 10-frame lag ($u=10$) on NS and ERA5 and a 20-frame lag ($u=5$) on SEVIR. For all learned methods, sampling hyperparameters are selected on held-out tuning data disjoint from the evaluation set, or taken from the original paper. %
\subsection{Navier--Stokes Vorticity}
\label{sec:exp:ns}

We use NS as a controlled benchmark on a PDE-governed dynamical system: the vorticity field is the full state of 2D incompressible flow with known dynamics. Even in this setting, sparse-observation filtering is challenging because per-step error accumulates across long horizons. 

\begin{figure}[t]
    \centering
    \includegraphics[width=\linewidth]{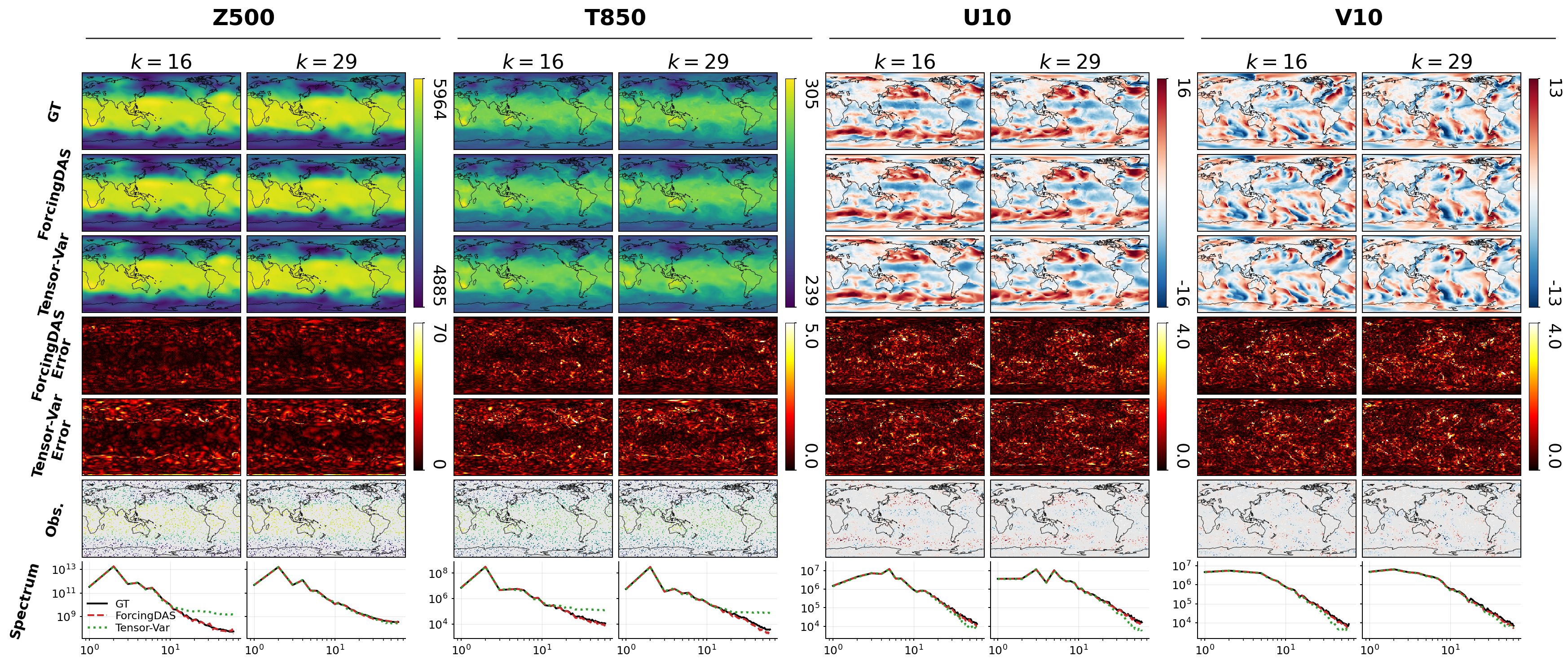}
    \caption{ERA5 SO-10\% with-context assimilation on the first of the 16 evaluation trajectories, for all four variables (Z500 in m, T850 in K, U10 and V10 in m\,s$^{-1}$), at a mid-window frame ($k=16$) and the last frame ($k=29$). ForcingDAS runs as a full-sequence smoother (FS), the regime matched to Tensor-Var. The last row is the per-frame power spectrum along latitude (log--log). Numbers are in Table~\ref{tab:era5_main}; the cold-start counterpart is Fig.~\ref{fig:era5_cold_start}.}
    \label{fig:era5_z500_comparison}
\end{figure}

\paragraph{Setup, baselines, and metrics.}
We consider 2D incompressible Navier--Stokes vorticity at $128 \times 128$ resolution, evaluated on 16 held-out trajectories. In the main protocol each trajectory has $K = 50$ frames, the first 10 clean as context and the remaining 40 assimilated and scored. The operator is sparse pixel masking at a $5\%$ observed ratio (\textbf{SO-5\%}, one fixed mask) with additive Gaussian noise of $\sigma_y = 0.05$ in the $[0,1]$ data space ($1.83$ in raw vorticity units). Observations are generated once per trajectory and every method assimilates the identical values, converted to its own units. The cold-start protocol ($K=30$, no clean context) is in Appendix~\ref{app:cold_start:ns}. We compare against the learned filter FlowDAS~\citep{flowdas}, which is built on a per-step transition model, and the classical Ensemble Kalman Filter (EnKF)~\citep{evensen2003ensemble}, which runs the exact PDE (Appendix~\ref{app:enkf_ns}). The appendix tables additionally report the Ensemble Kalman Smoother (EnKS, fixed-lag and full-sequence) and the learned smoother SDA~\citep{rozet2023score}. ForcingDAS-Pyr uses a 10-frame lag. We report NRMSE and the relative error of the radially averaged kinetic-energy spectrum: for each frame we compute the radial spectrum $E(k)$ of the predicted and ground-truth vorticity and the per-bin relative error $e(k) = |E_{\mrm{pred}}(k) - E_{\mrm{gt}}(k)| / \max(E_{\mrm{gt}}(k), \eps)$, averaged over wavenumber bands $\mathcal{K}$; we report mid-$k$ ($\mathcal{K} = [8, 32)$, the inertial range) and the integrated all-$k$ error ($\mathcal{K} = [0.5, 64)$). All numbers are mean $\pm$ std over the 16 trajectories. Dataset details and full hyperparameters are in Appendix~\ref{app:exp_details:ns}.

\paragraph{Results.}
Table~\ref{tab:ns_main} reports the with-context numbers. In the filtering regime ForcingDAS-AR halves FlowDAS's NRMSE ($0.071$ vs.\ $0.141$) and cuts the all-$k$ spectrum error from $0.617$ to $0.215$; it also improves on the EnKF ($0.108$), which runs the exact PDE. Because every method assimilates identical observations on the same trajectories, the comparison is paired: ForcingDAS-AR is better than both FlowDAS and EnKF on all 16 trajectories and on all three metrics (Wilcoxon signed-rank $p = 3.1\times10^{-5}$, the minimum attainable for $n=16$). Within each smoothing regime, ForcingDAS also improves on the matched smoothers on all 16 trajectories (Table~\ref{tab:ns_main}): ForcingDAS-Pyr on EnKS-FL over the same 10-frame lag ($0.057$ vs.\ $0.113$), and ForcingDAS-FS on EnKS-FS ($0.067$ vs.\ $0.116$) and on SDA given the same clean context ($0.067$ vs.\ $0.075$). \figNSvis{} shows why FlowDAS falls behind: its error grows over the first $\sim$10 assimilated frames, after which its fields lose the small-scale eddies, while ForcingDAS-AR stays close to the truth across the full window. EnKF's error is flat but high, and its spectrum decays too fast at high wavenumbers.

\vspace{-0.1in}
\subsection{Precipitation Nowcasting (SEVIR-VIL)}
\label{sec:exp:sevir}

\paragraph{Setup.}
We use the Vertically Integrated Liquid (VIL) product of the Storm Event Imagery (SEVIR) dataset~\citep{veillette2020sevir}, a vertically integrated radar-derived measure of precipitation intensity. We use a SEVIR-LR build that keeps all 49 frames of each event at 5-minute cadence (single channel, $128 \times 128$), with a random 90/10 split into training and held-out events. We evaluate on 16 held-out events with enough convective activity (more than 5\% of pixels above VIL 16; Appendix~\ref{app:exp_details:sevir:dataset}). In the main protocol, 6 clean context frames seed a 49-frame trajectory and the remaining 43 frames are assimilated and scored. We evaluate four observation operators with additive Gaussian noise $\sigma_y = 0.05$ in the $[0,1]$ data space: sparse pixel masks observing $10\%$ and $20\%$ of pixels (\textbf{SO-10\%}, \textbf{SO-20\%}) and bilinear down-sampling by $2\times$ and $4\times$ (\textbf{SR2x}, \textbf{SR4x}). The observations are generated once per event and every method assimilates the identical values. ForcingDAS-Pyr uses a 20-frame lag. The cold-start protocol (30 frames, no clean context) is in Appendix~\ref{app:cold_start:sevir}.

\paragraph{Baselines and metrics.}
We compare against the learned filter FlowDAS~\citep{flowdas}, retrained on our SEVIR build, and the learned smoother SDA~\citep{rozet2023score}. SEVIR has no operationally meaningful classical-DA baseline, since VIL dynamics are not given by a known PDE. SDA has no mechanism for clean context frames, so we give it the same 6 clean frames by replacement conditioning: at every sampling step the context frames are overwritten with the ground truth diffused to the current noise level (Appendix~\ref{app:exp_details:sevir:baselines}). We report NRMSE (mean $\pm$ std over events) and the Critical Success Index (CSI) averaged over the VIL thresholds $\{16, 74, 133, 160, 181, 219\}$ (CSI-mean). Following the SEVIR convention, CSI is pooled: hits, misses, and false alarms are summed over all assimilated frames of all 16 events before taking the ratio. Its $\pm$ is the std over 2{,}000 bootstrap resamples of the 16 events. CSI at each threshold, including the heavy-precipitation threshold 160, is in Table~\ref{tab:sevir_csi}.

\paragraph{Results.}
Table~\ref{tab:sevir_main} reports the with-context results. \emph{Filtering:} ForcingDAS-AR improves on FlowDAS on every operator and metric, roughly halving the NRMSE (e.g., SO-10\% $0.330$ vs.\ $0.616$), and it is better on every one of the 16 events for every operator (Wilcoxon $p = 3.1\times10^{-5}$). FlowDAS's error grows steadily over the 43 assimilated frames, whereas ForcingDAS stays flat (\figSEVIRvis{}). \emph{Full-sequence smoothing:} Given the same clean context, ForcingDAS-FS improves on the learned smoother SDA on every operator, in NRMSE (e.g., SR2x $0.169$ vs.\ $0.215$) and in CSI-mean, and has the lower NRMSE on 15, 14, 16, and 16 of the 16 events for SO-10\%, SO-20\%, SR2x, and SR4x. The same ordering holds in the cold-start protocol (Appendix~\ref{app:cold_start:sevir}).

\subsection{Global Atmospheric State Estimation (ERA5)}
\label{sec:exp:era5}

\begin{table}[t]
  \centering
  \caption{SEVIR VIL nowcasting with 6 clean context frames and 43 assimilated frames: NRMSE ($\downarrow$; mean {\scriptsize $\pm$ std} over 16 events) and CSI-mean ($\uparrow$) over thresholds $\{16, 74, 133, 160, 181, 219\}$ (pooled over all events and frames; {\scriptsize $\pm$} std over bootstrap resamples of the events). ForcingDAS-Pyr uses a 20-frame lag; SDA receives the same 6 clean frames by replacement conditioning. Best in \emph{Filtering} and best in \emph{Full-sequence Smoothing} per column in \textbf{bold}. Cold-start results are in Table~\ref{tab:sevir_cold_start}. CSI-160 and the other per-threshold CSIs are in Table~\ref{tab:sevir_csi}.}
  \label{tab:sevir_main}
  \footnotesize
  \setlength{\tabcolsep}{1.5pt}
  \resizebox{\linewidth}{!}{%
\begin{tabular}{l cc cc cc cc}
\toprule
 & \multicolumn{2}{c}{SO-10\%} & \multicolumn{2}{c}{SO-20\%} & \multicolumn{2}{c}{SR2x} & \multicolumn{2}{c}{SR4x} \\
\cmidrule(lr){2-3} \cmidrule(lr){4-5} \cmidrule(lr){6-7} \cmidrule(lr){8-9}
Method & NRMSE & CSI-m & NRMSE & CSI-m & NRMSE & CSI-m & NRMSE & CSI-m \\
\midrule
\multicolumn{9}{l}{\textit{Filtering}} \\
\quad FlowDAS & 0.616\pmstd{0.213} & 0.403\pmstd{0.020} & 0.479\pmstd{0.183} & 0.481\pmstd{0.023} & 0.363\pmstd{0.199} & 0.637\pmstd{0.017} & 0.504\pmstd{0.189} & 0.471\pmstd{0.024} \\
\quad ForcingDAS-AR & \textbf{0.330}\pmstd{0.107} & \textbf{0.557}\pmstd{0.024} & \textbf{0.258}\pmstd{0.080} & \textbf{0.625}\pmstd{0.020} & \textbf{0.199}\pmstd{0.054} & \textbf{0.684}\pmstd{0.017} & \textbf{0.313}\pmstd{0.100} & \textbf{0.558}\pmstd{0.026} \\
\multicolumn{9}{l}{\textit{Fix-lag Smoothing}} \\
\quad ForcingDAS-Pyr & 0.293\pmstd{0.093} & 0.584\pmstd{0.021} & 0.214\pmstd{0.063} & 0.671\pmstd{0.017} & 0.164\pmstd{0.046} & 0.720\pmstd{0.014} & 0.279\pmstd{0.087} & 0.559\pmstd{0.026} \\
\multicolumn{9}{l}{\textit{Full-sequence Smoothing}} \\
\quad SDA & 0.307\pmstd{0.099} & 0.573\pmstd{0.021} & 0.243\pmstd{0.079} & 0.639\pmstd{0.017} & 0.215\pmstd{0.077} & 0.667\pmstd{0.015} & 0.327\pmstd{0.109} & 0.551\pmstd{0.021} \\
\quad ForcingDAS-FS & \textbf{0.274}\pmstd{0.080} & \textbf{0.592}\pmstd{0.021} & \textbf{0.219}\pmstd{0.062} & \textbf{0.662}\pmstd{0.017} & \textbf{0.169}\pmstd{0.046} & \textbf{0.712}\pmstd{0.014} & \textbf{0.269}\pmstd{0.076} & \textbf{0.585}\pmstd{0.022} \\
\bottomrule
\end{tabular}}
\end{table}

\begin{table}[t]
\centering
\caption{ERA5 SO-10\% assimilation with 6 clean context frames (24 assimilated frames): latitude-weighted NRMSE ($\downarrow$, z-score units) and anomaly correlation coefficient (ACC, $\uparrow$) per variable. Mean {\scriptsize $\pm$ std} over 16 held-out trajectories from the 2016 evaluation year; every method assimilates identical observations. ForcingDAS-Pyr uses a 10-frame lag. Best in \emph{Filtering} and best in \emph{Full-sequence Smoothing} per column in \textbf{bold}. Cold-start results are in Appendix~\ref{app:cold_start:era5}.}
\label{tab:era5_main}
\footnotesize
\setlength{\tabcolsep}{2pt}
\resizebox{\linewidth}{!}{%
\begin{tabular}{l cc cc cc cc}
\toprule
 & \multicolumn{2}{c}{Z500} & \multicolumn{2}{c}{T850} & \multicolumn{2}{c}{U10} & \multicolumn{2}{c}{V10} \\
\cmidrule(lr){2-3} \cmidrule(lr){4-5} \cmidrule(lr){6-7} \cmidrule(lr){8-9}
Method & NRMSE & ACC & NRMSE & ACC & NRMSE & ACC & NRMSE & ACC \\
\midrule
\multicolumn{9}{l}{\textit{Filtering}} \\
\quad 3D-Var           & 0.067\pmstd{0.003} & 0.964\pmstd{0.004} & 0.114\pmstd{0.003} & 0.868\pmstd{0.020} & 0.478\pmstd{0.015} & 0.792\pmstd{0.012} & 0.536\pmstd{0.014} & 0.810\pmstd{0.012} \\
\quad FlowDAS          & 0.035\pmstd{0.001} & 0.990\pmstd{0.001} & 0.070\pmstd{0.001} & 0.948\pmstd{0.007} & 0.206\pmstd{0.005} & 0.958\pmstd{0.002} & 0.249\pmstd{0.005} & 0.956\pmstd{0.004} \\
\quad ForcingDAS-AR    & \textbf{0.026}\pmstd{0.000} & \textbf{0.994}\pmstd{0.001} & \textbf{0.054}\pmstd{0.001} & \textbf{0.970}\pmstd{0.004} & \textbf{0.159}\pmstd{0.003} & \textbf{0.975}\pmstd{0.001} & \textbf{0.193}\pmstd{0.004} & \textbf{0.974}\pmstd{0.002} \\
\multicolumn{9}{l}{\textit{Fix-lag Smoothing}} \\
\quad ForcingDAS-Pyr   & 0.023\pmstd{0.001} & 0.996\pmstd{0.000} & 0.048\pmstd{0.001} & 0.976\pmstd{0.003} & 0.148\pmstd{0.003} & 0.979\pmstd{0.001} & 0.181\pmstd{0.004} & 0.977\pmstd{0.002} \\
\multicolumn{9}{l}{\textit{Full-sequence Smoothing}} \\
\quad Tensor-Var       & 0.034\pmstd{0.001} & 0.990\pmstd{0.001} & 0.063\pmstd{0.002} & 0.957\pmstd{0.005} & 0.155\pmstd{0.003} & 0.976\pmstd{0.001} & 0.184\pmstd{0.003} & 0.976\pmstd{0.002} \\
\quad ForcingDAS-FS    & \textbf{0.022}\pmstd{0.001} & \textbf{0.996}\pmstd{0.000} & \textbf{0.049}\pmstd{0.001} & \textbf{0.975}\pmstd{0.003} & \textbf{0.149}\pmstd{0.003} & \textbf{0.978}\pmstd{0.001} & \textbf{0.182}\pmstd{0.004} & \textbf{0.977}\pmstd{0.002} \\
\bottomrule
\end{tabular}}
\end{table}

\paragraph{Dataset.}
ERA5~\citep{hersbach2020era5}, ECMWF's fifth-generation global atmospheric reanalysis, is the de facto benchmark for learned DA~\citep{pmlr-v235-huang24h, andry2025appa}. We use it as the ground-truth atmospheric state and run the same trained model as a filter (real-time forecasting) and as a smoother (retrospective reanalysis), which directly tests the unification claim. Data come from \href{https://weatherbench2.readthedocs.io/en/latest/data-guide.html}{WeatherBench2}~\citep{rasp2020weatherbench} at $1.5^{\circ}$ resolution with a 6-hourly cadence, on a $240 \times 120$ longitude--latitude grid. We train on 1979--2015 and evaluate on 16 trajectories of $K=30$ frames from the held-out 2016 evaluation year. The 16 windows are spread evenly across the year, one every 23 days, so that the reported std reflects seasonal variability. Four variables span upper-air dynamics, thermodynamics, and surface wind: \textbf{Z500} (geopotential at 500\,hPa), \textbf{T850} (temperature at 850\,hPa), and \textbf{U10} and \textbf{V10} (10\,m zonal and meridional wind), stacked as a $4 \times 240 \times 120$ input with per-variable z-score normalization. This is a controlled testbed rather than an operational DA system: operational systems assimilate real instrument observations into $0.25^{\circ}$ states with dozens of variables. See Appendix~\ref{app:exp_details:era5} for details and Appendix~\ref{app:related:da} for the broader landscape (Table~\ref{tab:era5_landscape}).

\paragraph{Observation setup.}
We simulate sparse observations with one random pixel mask covering $10\%$ of grid points, shared across the 4 channels and all frames, plus additive Gaussian noise of $\sigma_y = 0.05$ in z-score units (\textbf{SO-10\%}). The observations are generated once and every method assimilates the identical values. In the main protocol the first 6 frames are clean context and the remaining 24 are assimilated and scored. The cold-start protocol (no clean context) is in Appendix~\ref{app:cold_start:era5}. The equivalent noise in physical units is in Appendix~\ref{app:exp_details:era5:obs}.

\paragraph{Baselines and metrics.}
We compare against the classical filter \textbf{3D-Var}, which runs an independent variational analysis at each step with a spatially correlated background covariance and persistence cycling; the learned filter \textbf{FlowDAS}~\citep{flowdas}, which performs per-step posterior sampling with a learned stochastic-interpolant forecast model; and the learned smoother \textbf{Tensor-Var}~\citep{yang2025tensorvar}, which runs 4D-Var in a learned feature space with linear feature dynamics and a transformer inverse-observation network. We retrained Tensor-Var on our variables and grid. Its inverse-observation network is trained on the same fixed mask that is used at evaluation, whereas ForcingDAS never sees the mask during training. Implementation details are in Appendix~\ref{app:classical_da}. Per variable we report the latitude-weighted NRMSE (RMSE in z-score space, following WeatherBench2~\citep{rasp2020weatherbench}) and the latitude-weighted anomaly correlation coefficient (ACC) against the WeatherBench2 1990--2019 ERA5 climatology. Both are computed per trajectory over the assimilated frames and reported as mean $\pm$ std over the 16 trajectories; full definitions are in Appendix~\ref{app:exp_details:era5:metrics}.

\paragraph{Results.}
Table~\ref{tab:era5_main} reports the with-context results. \emph{Filtering:} ForcingDAS-AR has the lowest NRMSE and highest ACC among the filters on all four variables. Paired over the 16 trajectories, it is better than both FlowDAS and 3D-Var on every trajectory, every variable, and both metrics (Wilcoxon $p = 3.1\times10^{-5}$). The largest margins over FlowDAS are on the surface winds (U10 $0.159$ vs.\ $0.206$; V10 $0.193$ vs.\ $0.249$). 3D-Var's Gaussian background covariance interpolates the smooth synoptic-scale Z500 field reasonably well (ACC $0.964$), but it cannot recover finer structure: its NRMSE is 2.5--3$\times$ that of ForcingDAS-AR on every variable. \emph{Full-sequence smoothing:} ForcingDAS-FS improves on Tensor-Var on all four variables. The margins are large on Z500 ($0.022$ vs.\ $0.034$) and T850 ($0.049$ vs.\ $0.063$) and small but consistent on the surface winds: FS is better on 16/16 trajectories for U10 ($0.149$ vs.\ $0.155$) and on 14/16 for V10 ($0.182$ vs.\ $0.184$). The same trained model thus serves all three regimes without retraining. Fig.~\ref{fig:era5_z500_comparison} shows ForcingDAS and Tensor-Var on a representative trajectory for all four variables. Its spectrum row shows a difference that NRMSE does not capture: ForcingDAS-FS follows the ground-truth spectrum up to the highest wavenumbers, whereas Tensor-Var carries excess small-scale power on Z500 and T850 and too little on the surface winds.

\vspace{-0.1in}
\section{Conclusion and Limitations}
\label{sec:conclusion}

\vspace{-0.1in}
We introduced \textbf{ForcingDAS}: a single joint-trajectory diffusion prior with per-frame noise levels, trained once, that supports filtering, fixed-lag smoothing, and full-sequence smoothing at inference through a single noise-scheduling matrix and no retraining. Causality-aware training matches the noise patterns the model sees at inference, and noise-level-aware observation guidance scales each frame's correction by its signal quality, together mitigating the long-horizon error accumulation that limits autoregressive learned DA. On 2D Navier--Stokes vorticity, SEVIR radar nowcasting, and ERA5 global atmospheric reanalysis, a single ForcingDAS model is competitive with or outperforms specialized learned baselines (FlowDAS, SDA, Tensor-Var) and the classical 3D-Var and EnKF pipelines, suggesting that the filtering--smoothing choice need not be baked in but can be exposed as a control over inference. 

\paragraph{Limitations.} The main constraints are causal-only smoothing (future observations reach past states only through backward gradients), compute that scales with the active window for joint-denoising regimes, and a coarse ERA5 setup relative to operational systems. See Appendix~\ref{app:limitations} for the full discussion.

\section*{Acknowledgment}
YJ, SC, YP, LS, CJ, and QQ acknowledge funding support from NSF CAREER CCF-2143904, NSF CCF-2212066, NSF CCF-2212326, NSF IIS 2402950, ONR N000142512339, DARPA HR0011578254, and Google TPU Research Awards. SR acknowledges funding support in part from the National Science Foundation (NSF) grants CCF-2212065 and NSF CAREER CCF-2442240.

\printbibliography

\clearpage
\addcontentsline{toc}{section}{Appendix}
\newpage

\appendix

\setcounter{section}{0}
\setcounter{subsection}{0}
\setcounter{subsubsection}{0}
\setcounter{figure}{0}
\setcounter{table}{0}
\setcounter{equation}{0}
\setcounter{algorithm}{0}
\renewcommand{\thesection}{S\arabic{section}}
\renewcommand{\thesubsection}{\thesection.\arabic{subsection}}
\renewcommand{\thesubsubsection}{\thesubsection.\arabic{subsubsection}}
\renewcommand{\thefigure}{S\arabic{figure}}
\renewcommand{\thetable}{S\arabic{table}}
\renewcommand{\theequation}{S\arabic{equation}}
\renewcommand{\thealgorithm}{S\arabic{algorithm}}

\begin{center}
{\LARGE \bf Appendices}
\end{center}\vspace{-0.15in}
\par\noindent\rule{\textwidth}{1pt}

\etocdepthtag.toc{mtappendix}
\etocsettagdepth{mtmain}{none}
\etocsettagdepth{mtappendix}{subsection}
\etocsettocstyle{\vspace{0.5em}}{}
\tableofcontents
\newpage

\section{Related Work}
\label{app:related}

This appendix positions ForcingDAS along three axes: (i) classical and learned data assimilation, the problem ForcingDAS solves (\S\ref{app:related:da}); (ii) Diffusion Forcing and other per-frame-noise diffusion models, the family its trajectory prior belongs to (\S\ref{app:related:per_frame}); and (iii) diffusion-based inverse-problem solvers, which share its test-time guidance mechanism (\S\ref{app:related:diff_ip}).

\subsection{Classical and Learned Data Assimilation}
\label{app:related:da}

We organize related DA methods into classical (variational and ensemble) and learned (per-frame transitions and trajectory priors) approaches.

\paragraph{Classical variational methods (3D-Var, 4D-Var).}
3D-Var~\citep{courtier1998ecmwf} computes a single-time analysis by minimizing
$J(\bm{x}) = \tfrac{1}{2}(\bm{x}-\bm{x}_b)^\top \bm{B}^{-1}(\bm{x}-\bm{x}_b) + \tfrac{1}{2}(\bm{y}-\cA \bm{x})^\top \bm{R}^{-1}(\bm{y}-\cA \bm{x})$
with a fixed background $\bm{x}_b$ and background error covariance $\bm{B}$. 4D-Var~\citep{talagrand1987variational, courtier1994strategy} extends this to a window of length $W$, optimizing an initial state $\bm{x}_0$ that is propagated by a forward model $\Psi$ and scoring observations against the resulting trajectory. Both formulations require a hand-crafted $\bm{B}$ and (for 4D-Var) an adjoint of $\Psi$. We use 3D-Var as the classical filtering baseline on ERA5; implementation details are in Appendix~\ref{app:classical_da}.

\paragraph{Classical ensemble methods (EnKF, EnKS).}
The Ensemble Kalman Filter~\citep{evensen2003ensemble, burgers1998analysis} replaces the explicit $\bm{B}$ with a sample covariance from a forecast ensemble, which sidesteps the linear-Gaussian assumption of the original Kalman filter~\citep{kalman1960new} and works well when the forecast model is cheap to ensemble. The Ensemble Kalman Smoother~\citep{evensen2000ensemble} is the smoothing analog, updating an entire window jointly given all observations in that window. Both methods require an explicit dynamical model at inference time; particle filters~\citep{gordon1993novel, van2009particle} avoid the Gaussian assumption but are known to degenerate at high state dimension~\citep{bickel2008sharp, snyder2015performance}. We use the EnKF and two EnKS variants as physics-based baselines on Navier--Stokes (Appendix~\ref{app:enkf_ns}).

\paragraph{Learned filters: per-frame transition priors (FlowDAS).}
FlowDAS~\citep{flowdas} learns a single-step stochastic-interpolant transition and runs DPS-style guidance frame by frame, giving an online filter. Because its prior is a per-step transition, it commits to each frame before the next observation arrives, so per-step errors accumulate over long horizons (\S\ref{sec:exp:ns}); and when each observed frame is only a partial slice of a higher-dimensional latent state, a short, fixed window of recent frames does not determine the next one (\S\ref{sec:exp:sevir}, \S\ref{sec:exp:era5}).

\paragraph{Learned smoothers: variational and trajectory priors (Tensor-Var, SDA, Appa).}
Tensor-Var~\citep{yang2025tensorvar} learns a linear latent dynamics model coupled with a transformer-based inverse-observation network, and solves a 4D-Var problem in the learned feature space in closed form over a short window; it is a windowed smoother and cannot run as a real-time filter. Score-Based Data Assimilation (SDA)~\citep{rozet2023score} trains a frame-level diffusion score on full trajectories and combines it with observation guidance for offline reanalysis. The score is symmetric in time and shares one noise level across the window, which fits batch smoothing; a filtering estimate can only be read off as the last frame of a window posterior, re-sampled at each new assimilation time. ForcingDAS keeps the trajectory prior of SDA but uses the per-frame noise structure of Diffusion Forcing to expose filtering and fixed-lag smoothing as well as full-sequence smoothing from a single trained model.

\textbf{Appa.} Appa~\citep{andry2025appa} scales the SDA recipe to global weather: a 565M-parameter latent diffusion model over windows of 24 hourly latent ERA5 states (a 340M-parameter autoencoder compresses each state about $530\times$), with moment-matching posterior sampling for arbitrary observation operators. It shows that one trained trajectory prior can serve reanalysis, filtering, and forecasting, and is the closest prior work to ours in spirit. The two designs differ in how the DA regime is realized. (\emph{i}) \emph{Noise structure.} Appa shares one diffusion noise level across all states of a window: reanalysis composes window scores, filtering takes the last state of a window posterior, and forecasting slides windows autoregressively. ForcingDAS assigns each frame its own noise level, so the scheduling matrix moves continuously from filtering ($u = T$) through fixed-lag smoothing ($0 < u < T$) to full-sequence smoothing ($u = 0$); Appa does not study the intermediate fixed-lag regime. (\emph{ii}) \emph{Streaming cost.} In ForcingDAS-AR each new frame is denoised once, conditioned on the already-assimilated past, whereas a filtering estimate obtained as the marginal of a window posterior requires a new window sample at each assimilation time. (\emph{iii}) \emph{DA-specific training and guidance.} Because frames carry different noise levels, ForcingDAS adds causality-aware training, which matches the training noise patterns to those of the inference schedules, and noise-level-aware guidance, which scales each frame's observation correction by its current signal quality (\S\ref{sec:method}). (\emph{iv}) \emph{Complementary focus.} Appa targets operational scale ($0.25^{\circ}$, 71 variables, compared against IFS and GraphDOP), whereas our ERA5 study compares filtering and smoothing regimes against matched baselines in a controlled $1.5^{\circ}$, four-variable setting, so a direct benchmark is out of scope; our SDA baseline shares Appa's window prior and score composition in pixel space.

\paragraph{Learned latent and operator dynamics (Koopman priors, NODA).}
A related line of learned DA first learns the dynamics, or an operator for them, and then assimilates against that learned model. The neural Koopman prior~\citep{frion2024neural} trains an autoencoder whose latent dynamics are linear and uses it as the dynamical prior in variational DA, either as a weak constraint on the full trajectory or as a strong constraint that optimizes only the initial latent state. AIKAE~\citep{frion2025augmented} augments an invertible Koopman autoencoder with a second, non-invertible encoder and uses the pre-trained model in the same strong-constraint variational scheme. NODA~\citep{singh2024learning} learns a state-space solution operator for semilinear PDEs and, building on nonlinear observer theory, alternates a learned prediction step with an observation-driven correction step, so that one recursive model serves both forecasting and assimilation. These methods learn latent or operator-based dynamics and assimilate either by variational optimization over a window or by recursive prediction and correction, so the DA regime is fixed by the assimilation scheme. ForcingDAS instead learns a diffusion prior over whole trajectories with per-frame noise levels, assimilates by guided sampling, and selects filtering, fixed-lag smoothing or full-sequence smoothing at inference through the scheduling matrix.

\paragraph{Learned DA on ERA5.}
ERA5 has become the de facto evaluation benchmark for learned global atmospheric DA, and several recent systems run at full $0.25^{\circ}$ resolution with 69--82 input channels~\citep{pmlr-v235-huang24h, andry2025appa}. Table~\ref{tab:era5_landscape} summarizes the main entries, and we briefly characterize each below.

\textbf{DiffDA}~\citep{pmlr-v235-huang24h} couples a GraphCast forecast backbone with a diffusion-based analysis head and assimilates simulated column observations through a single-step posterior. \textbf{FengWu-Adas}~\citep{chen2023towards} integrates the FengWu forecast model with a learned analysis network that consumes a mix of simulated and real GDAS observations, and runs cyclically across many analysis steps. \textbf{FuXi-DA}~\citep{xu2024fuxida} is a generalized deep-learning DA framework targeted at real satellite observations: it assimilates FY-4B radiances on top of the FuXi forecast model, demonstrating that learned operators can handle real instrument geometry rather than only synthetic grid masks. \textbf{Appa}~\citep{andry2025appa} is a latent diffusion model for the global atmospheric state that supports zero-shot inverse problems with arbitrary observation operators at test time, and uses SDA-style score composition for posterior sampling; one trained model performs reanalysis, filtering, and forecasting as separate tasks (see the comparison above). \textbf{L4DVar}~\citep{huang2025l4dvar} runs variational DA inside a learned latent space with deterministic encoder / decoder maps, retaining the variational structure of 4D-Var while operating on a much lower-dimensional state. \textbf{ADAF}~\citep{xiang2024adaf} is a regional, high-resolution AI DA framework that combines surface station and GOES-16 satellite observations on a CONUS grid. \textbf{FengWu-4DVar}~\citep{pmlr-v235-xiao24a} couples the FengWu forecast model with a 4D-Var solver where the adjoint is obtained by automatic differentiation through the learned forward model, replacing the hand-coded operational adjoint.

The listed systems are built around one DA regime, either (cyclic) filtering or window-based 4D-Var, as is Tensor-Var, our ERA5 smoothing baseline; Appa extends this to reanalysis, filtering, and forecasting from one latent prior, handled as separate tasks. ForcingDAS instead makes the regime a continuous inference-time choice: a single schedule parameter moves the same trained model from filtering through fixed-lag smoothing to full-sequence smoothing. The systems in Table~\ref{tab:era5_landscape} operate at far larger scale than our study, which is a controlled comparison at $1.5^{\circ}$ with four variables (\S\ref{sec:exp:era5}).

\begin{table}[h]
\centering
\caption{Landscape of learned DA methods on ERA5. The last column lists the DA regime(s) each system supports; ForcingDAS spans the continuum from filtering to smoothing with a single schedule parameter. ``Vars'' counts total input channels (upper-air variables $\times$ pressure levels $+$ surface variables). ``Obs'' indicates the observation type used at test time. ``Params'' gives the approximate number of trainable parameters; ``N/R'' = not reported.}
\label{tab:era5_landscape}
\small
\setlength{\tabcolsep}{3pt}
\begin{tabular}{@{}lcccccc@{}}
\toprule
\textbf{Method} & \textbf{Res.} & \textbf{Grid} & \textbf{Vars} & \textbf{Params} & \textbf{Obs} & \textbf{DA regime} \\
\midrule
DiffDA       & $0.25^{\circ}$ & $721 \!\times\! 1440$ & 82 & ${\geq}$37M$^\dagger$ & Simul.\ columns & Filtering \\
FengWu-Adas  & $0.25^{\circ}$ & $721 \!\times\! 1440$ & 69 & N/R & Simul.\ + GDAS & Cyclic filtering \\
FuXi-DA      & $0.25^{\circ}$ & $721 \!\times\! 1440$ & 70 & N/R & FY-4B sat. & Filtering \\
Appa         & $0.25^{\circ}$ & $721 \!\times\! 1440$ & 71 & 565M & Arbitrary (0-shot) & Reanal./filter/forecast$^{*}$ \\
L4DVar       & $1.4^{\circ}$  & $128 \!\times\! 256$  & 69 & N/R & GDAS real & Latent 4D-Var \\
ADAF         & $0.05^{\circ}$ & $512 \!\times\! 1280$ & 4  & N/R & Sfc + GOES-16 & Filtering (CONUS) \\
FengWu-4DVar & $0.25^{\circ}$ & $721 \!\times\! 1440$ & 69 & N/R & Simulated & 4D-Var (AI adj.) \\
\midrule
\textbf{ForcingDAS (ours)}              & $1.5^{\circ}$  & $240 \!\times\! 120$  & 4  & ${\sim}$268M & Simul.\ sparse (10\%) & \textbf{Filter $\leftrightarrow$ Smoother} \\
\bottomrule
\multicolumn{7}{@{}l}{\footnotesize $^{*}$Separate tasks of one latent window prior with a shared noise level per window; fixed-lag smoothing is not studied.}\\
\multicolumn{7}{@{}l}{\footnotesize $^{\dagger}$GraphCast backbone is 36.7M; the adapted diffusion variant is likely larger (not reported).}
\end{tabular}
\end{table}

\subsection{Diffusion Forcing and Per-Frame-Noise Diffusion}
\label{app:related:per_frame}
\label{app:df_recap}

ForcingDAS uses Diffusion Forcing (DF)~\citep{chen2024diffusion} as its trajectory prior. This subsection recaps DF, marks which components ForcingDAS inherits and which it introduces, and positions the per-frame-noise diffusion methods that have appeared since. We follow the notation of \S\ref{sec:bg:perframe}: $k = 1, \ldots, K$ is the frame index, $t = 1, \ldots, T$ is the diffusion-step index, $t_k$ is the per-frame diffusion step, and $\bm{t} = (t_1, \ldots, t_K)$. The original DF paper instead uses $T$ for sequence length (our $K$), $K$ for the number of diffusion steps (our $T$), and $k_t$ for $t_k$.

\subsubsection{Problem formulation}
\label{app:df:formulation}
DF generalizes the standard diffusion forward process~\citep{ho2020denoising, sohl2015deep} from a single noise variable to a per-frame collection of independent ones. Given a clean trajectory $\bm{x}_{1:K} = (\bm{x}_1, \ldots, \bm{x}_K) \sim q$, draw independent noise levels $t_k \in \{0, \ldots, T\}$ and Gaussian noise $\bm{\epsilon}_k \sim \cN(\mathbf{0}, \I)$, and form
\begin{equation}
    \bm{x}_k^{(t_k)} = \sqrt{\bar{\alpha}_{t_k}}\, \bm{x}_k + \sqrt{1 - \bar{\alpha}_{t_k}}\, \bm{\epsilon}_k, \quad k = 1, \ldots, K,
    \label{eq:df_forward}
\end{equation}
with $\{\bar{\alpha}_t\}_{t=0}^{T}$ the cumulative signal-retention coefficient of a chosen variance schedule. The two limits $t_k = 0$ (clean) and $t_k = T$ (white noise) recover an unmasked and a fully masked frame, so $\bm{t}$ acts as a continuous frame-level mask, the ``masking along the noise axis'' view of \citet{chen2024diffusion}. Teacher-forced next-token prediction ($t_k \in \{0, T\}$) and full-sequence diffusion ($t_k = t$ shared across $k$) appear as special cases.

DF parameterizes the reverse process by a single joint denoiser $\bm{\epsilon}_{\bm{\theta}}(\bm{x}_{1:K}^{(\bm{t})}, \bm{t})$ that predicts the noise on every frame from the full noisy trajectory and the per-frame noise vector. The original DF instantiates this as a causal RNN with a hidden state $\bm{z}_k$ updated via $\bm{z}_k \sim p_{\bm{\theta}}(\bm{z}_k \mid \bm{z}_{k-1}, \bm{x}_k^{(t_k)}, t_k)$ (a Bayes-filter-style posterior that doubles as a conditional diffusion unit): per-frame features from a 2D U-Net (video) or a residual MLP (low-dimensional data) are passed through a GRU along the temporal axis. ForcingDAS uses a causal-attention transformer instead (Appendix~\ref{app:impl:arch}). Both choices are length-invariant in $K$.

\subsubsection{Training procedure}
\label{app:df:training}
Standard DF training samples $t_k \stackrel{\text{i.i.d.}}{\sim} \mathrm{Uniform}\{0, \ldots, T\}$ across frames, corrupts $\bm{x}_{1:K}$ via Eq.~\eqref{eq:df_forward}, and minimizes the per-frame score-matching loss
\begin{equation}
    \cL_{\mathrm{DF}}(\bm{\theta}) \;=\; \mathbb{E}_{\bm{x}_{1:K},\, \bm{t},\, \bm{\epsilon}_{1:K}} \sum_{k=1}^{K} \big\| \bm{\epsilon}_k - \bm{\epsilon}_{\bm{\theta}}\big( \bm{x}_{1:K}^{(\bm{t})}, \bm{t} \big)_k \big\|^2.
    \label{eq:df_train_loss}
\end{equation}
The defining feature is the per-frame independence of $\bm{t}$: it exposes the network to all jointly-noised configurations rather than the shared step of full-sequence diffusion or the clean-past / fully-masked-future configuration of teacher forcing, in principle yielding robustness to any test-time scheduling matrix. The validity of this objective is captured by the central theoretical claim of \citet{chen2024diffusion}:

\begin{theorem}[Informal, restated from~\citealt{chen2024diffusion}]
\label{thm:df_elbo}
Optimizing $\cL_{\mathrm{DF}}$ with $\bm{t}$ drawn i.i.d.\ uniformly maximizes a reweighted ELBO on $\ln p_{\bm{\theta}}(\bm{x}_{1:K}^{(\bm{t})})$ averaged over $\bm{t}$, and under suitable conditions simultaneously lower-bounds the joint log-likelihood for \emph{every fixed} $\bm{t}$.
\end{theorem}

Theorem~\ref{thm:df_elbo} licenses the use of a single trained DF network under arbitrary per-frame schedules and underpins the ForcingDAS claim that the same prior can serve as filter, fixed-lag smoother, or full-sequence smoother (\S\ref{sec:method:framework}). The training pseudocode in our notation is Algorithm~\ref{alg:df_training}; the ForcingDAS counterpart, which replaces the i.i.d.\ draw by the CAT mixture, is Algorithm~\ref{alg:training}.

\begin{algorithm}[h]
\caption{Diffusion Forcing training (recap of Alg.~1 of~\citealt{chen2024diffusion}).}
\label{alg:df_training}
\begin{algorithmic}[1]
\REQUIRE Dataset $\{\bm{x}_{1:K}\}$, total diffusion steps $T$, schedule $\{\bar{\alpha}_t\}$
\REPEAT
    \STATE Sample $\bm{x}_{1:K}$ from the dataset
    \STATE $t_k \sim \mathrm{Uniform}\{0, \ldots, T\}$, $\bm{\epsilon}_k \sim \cN(\mathbf{0}, \I)$, $k = 1, \ldots, K$ \hfill $\triangleright$ \textit{i.i.d.\ across frames}
    \STATE $\bm{x}_k^{(t_k)} \leftarrow \sqrt{\bar{\alpha}_{t_k}}\,\bm{x}_k + \sqrt{1 - \bar{\alpha}_{t_k}}\,\bm{\epsilon}_k$
    \STATE Update $\bm{\theta}$ on $\sum_{k=1}^{K} \|\bm{\epsilon}_k - \bm{\epsilon}_{\bm{\theta}}(\bm{x}_{1:K}^{(\bm{t})}, \bm{t})_k\|^2$
\UNTIL{converged}
\end{algorithmic}
\end{algorithm}

\subsubsection{Sampling procedure}
\label{app:df:inference}
DF sampling is controlled by a \emph{scheduling matrix} $\bm{S} \in \{0, \ldots, T\}^{K \times L}$ (the matrix $\mathcal{K} \in [K]^{M \times T}$ of \citet{chen2024diffusion}, with rows and columns transposed): $S_{k, \ell}$ is the diffusion step of frame $k$ at the $\ell$-th of $L$ reverse iterations, with $S_{k, 0} = T$ and $S_{k, L} = 0$ for every $k$. Generation initializes $\bm{x}_k^{(T)} \sim \cN(\mathbf{0}, \I)$ and iterates $\ell = 0, \ldots, L-1$; at each $\ell$ the active set $\mathcal{K}_{\mathrm{active}} = \{k : S_{k, \ell+1} < S_{k, \ell}\}$ collects the frames that descend, and a single DDIM~\citep{song2020denoising} step takes each $k \in \mathcal{K}_{\mathrm{active}}$ from $t_k = S_{k, \ell}$ to $t_k' = S_{k, \ell+1}$:
\begin{equation}
    \bm{x}_k^{(t_k')} = \sqrt{\bar{\alpha}_{t_k'}}\,\hat{\bm{x}}_k^{(0)} + \sqrt{1 - \bar{\alpha}_{t_k'} - \sigma_{t_k}^2}\,\bm{\epsilon}_{\bm{\theta}}(\bm{x}_{1:K}^{(\bm{t})}, \bm{t})_k + \sigma_{t_k}\,\bm{z}_k, \quad \bm{z}_k \sim \cN(\mathbf{0}, \I),
    \label{eq:df_ddim}
\end{equation}
with Tweedie estimate $\hat{\bm{x}}_k^{(0)} = (\bm{x}_k^{(t_k)} - \sqrt{1 - \bar{\alpha}_{t_k}}\,\bm{\epsilon}_{\bm{\theta}}(\cdot)_k) / \sqrt{\bar{\alpha}_{t_k}}$; non-active frames are left unchanged. \citet{chen2024diffusion} highlights three canonical schedules (synchronous full-sequence, autoregressive next-token, and \emph{pyramid}, with the far future kept noisier than the near future); ForcingDAS recasts all three as DA regimes (full-sequence smoothing, filtering, and fixed-lag smoothing) and embeds them into the one-parameter family $\bm{S}^{(u)}$ of Eq.~\eqref{eq:scheduling}.

DF further supports test-time guidance: any differentiable cost $c(\bm{x}_{1:K}^{(\bm{t})})$ contributes a gradient added to Eq.~\eqref{eq:df_ddim}. Because the causal denoiser makes a future frame's $\hat{\bm{x}}_{k'}^{(0)}$ a differentiable function of all $\bm{x}_k^{(t_k)}$ with $k \leq k'$, the gradient flows backwards in time and refines past frames. ForcingDAS reuses this ``long-horizon guidance'' mechanism for observation guidance (\S\ref{sec:method:guidance}, Appendix~\ref{app:asymmetry}). The Monte Carlo Guidance variant of \citet{chen2024diffusion}, which averages gradients over multiple future roll-outs, is not needed for DA and not used. The sampling pseudocode is Algorithm~\ref{alg:df_sampling}; the ForcingDAS counterpart with observation guidance is Algorithm~\ref{alg:sampling}.

\begin{algorithm}[h]
\caption{Diffusion Forcing sampling (recap of Alg.~2 of~\citealt{chen2024diffusion}).}
\label{alg:df_sampling}
\begin{algorithmic}[1]
\REQUIRE $\bm{\epsilon}_{\bm{\theta}}$, scheduling matrix $\bm{S} \in \{0, \ldots, T\}^{K \times L}$, optional cost $c(\cdot)$, scale $\zeta \geq 0$
\STATE $\bm{x}_k^{(T)} \sim \cN(\mathbf{0}, \I)$, $k = 1, \ldots, K$
\FOR{$\ell = 0, \ldots, L - 1$}
    \STATE $t_k \leftarrow S_{k, \ell}$, $t_k' \leftarrow S_{k, \ell+1}$; $\mathcal{K}_{\mathrm{active}} \leftarrow \{k : t_k' < t_k\}$
    \STATE Apply Eq.~\eqref{eq:df_ddim} to each $k \in \mathcal{K}_{\mathrm{active}}$
\ENDFOR
\STATE \textbf{Return} $\bm{x}_{1:K}^{(0)}$
\end{algorithmic}
\end{algorithm}

\subsubsection{What ForcingDAS inherits and what it introduces}
\label{app:df:demarcation}
ForcingDAS adopts from DF, without claim of novelty, the per-frame Gaussian forward process (Eq.~\eqref{eq:df_forward}), the joint denoiser conditioned on the noise-level vector $\bm{t}$, causal temporal dependency across frames, the scheduling-matrix abstraction, and the backward flow of guidance gradients through the causal denoiser. It introduces: (i) the reformulation of per-frame diffusion sampling as a DA solver through observation guidance with noise-level-aware reweighting (Eq.~\eqref{eq:reweight}) and windowed aggregation over frames active at staggered diffusion steps; (ii) the one-parameter family $\bm{S}^{(u)}$ that maps the uncertainty scale $u$ onto filtering, fixed-lag smoothing, and full-sequence smoothing, with DF's canonical schedules as special cases; (iii) causality-aware training (Eq.~\eqref{eq:cat_loss}), which biases the training noise distribution toward the causally monotone configurations used at DA inference; and (iv) the empirical validation of a single trained per-frame-diffusion prior across the DA spectrum against classical and learned DA baselines.

\subsubsection{Other per-frame-noise diffusion methods}

\paragraph{Rolling Diffusion~\citep{ruhe2024rolling}.}
Rolling Diffusion sweeps a window across a sequence with frames at monotonically increasing noise levels, so generation proceeds frame by frame in a streaming fashion. The active-window structure resembles ForcingDAS-Pyr ($0 < u < T$), but the design target is unconditional video generation rather than data assimilation: there is no observation operator and no notion of filtering or smoothing.

\paragraph{Diffusion Forcing Transformer (DFoT)~\citep{song2025dfot}.}
DFoT replaces the original DF U-Net with a transformer backbone and uses the per-frame noise levels to expose a ``history guidance'' knob at sampling time, which trades off how strongly clean past frames condition future ones. We share the per-frame noise structure but adapt it for DA rather than long-video generation. ForcingDAS could in principle use the DFoT backbone in place of our DiT-3D; the filtering--smoothing spectrum and the observation guidance are orthogonal to the choice of backbone.

\paragraph{Self-Forcing~\citep{huang2025selfforcing}.}
Self-Forcing trains an autoregressive video diffusion model on its own roll-outs to close the train--test exposure gap. It targets a single inference pattern (autoregressive, frame by frame) and does not expose a continuous schedule. CAT (\S\ref{sec:method:cat}) addresses a related but different gap: standard DF samples per-frame noise levels i.i.d., while every DA scheduling matrix imposes a causally monotone pattern at inference; CAT biases training toward the monotone configurations rather than toward roll-out trajectories.

None of these methods has been formulated as a data assimilation solver: there is no observation operator $\cA$, no observation guidance, and no filtering / fixed-lag / smoothing distinction.

\subsection{Diffusion-Based Inverse-Problem Solvers}
\label{app:related:diff_ip}

A broader line of work uses pretrained diffusion priors with test-time gradient guidance to solve inverse problems~\citep{chung2022diffusion,song2024solving,li2024decoupled,alkhouri2024sitcom}. These methods target single-image restoration (deblurring, inpainting, super-resolution) and do not handle a sequence of partial observations or distinguish filtering from smoothing. ForcingDAS uses the same gradient-guidance idea but extends it to trajectory-level data assimilation; the noise-level-aware reweighting of Eq.~\eqref{eq:reweight} arises from the per-frame diffusion-step staircase that single-image solvers do not encounter.

\newpage
\section{Analysis and Implementation of ForcingDAS}
\label{app:impl}

This appendix collects the analysis of how observations couple frames under causal attention (\S\ref{app:asymmetry}) and the implementation details summarized in \S\ref{sec:method:impl}: the network backbone (\S\ref{app:impl:arch}), further design choices (\S\ref{app:impl:design}), and pseudocode for training and inference (\S\ref{app:impl:algorithms}).

\subsection{Asymmetric Past--Future Coupling}
\label{app:asymmetry}

Causal temporal attention introduces a directional asymmetry in how observations couple the trajectory. A frame $k$ directly attends to every past frame $k' < k$: once past frames have been corrected by their observations, those corrections propagate into frame $k$ through the \emph{forward pass}, a strong, direct pathway. Future observations, by contrast, reach past frames only through the \emph{backward gradient} of the measurement loss: since the causal Jacobian $\partial \hat{\bm{x}}_{0,k'} / \partial \bm{x}_k^{(t_k)}$ is non-zero for $k \leq k'$ and zero otherwise (Eq.~\eqref{eq:causal_jacobian} below), guidance at a future frame $k'$ does produce a gradient on frame $k$, but an \emph{indirect} one that passes through the attention and nonlinear layers between the two frames. Consequently, even in its most smoothing-like mode ($u=0$), ForcingDAS lets future observations influence past states only through this backward gradient, whereas a model with bidirectional attention would also carry them through the forward pass. We give the formal chain-rule decomposition below (Remark~\ref{rem:crossframe}).

\paragraph{Forward direction: past $\to$ future.}
When frame $k$ is being denoised, causal attention lets it directly read information from all past frames $k' < k$. If past frames have already been corrected by their observations, those corrections propagate to frame $k$ through the forward pass, a direct information pathway. This is the \emph{filtering} channel.

\paragraph{Backward direction: future $\to$ past.}
When observation guidance is applied, the measurement loss at a future frame $k'$ ($k' > k$) generates gradients $\partial \mathcal{L}(\bm{y}_{k'}) / \partial \bm{x}_k^{(t_k)}$ that flow backward through the network. Because causal attention makes $\hat{\bm{x}}_{0,k'}$ a differentiable function of all inputs $\bm{x}_{k}$ with $k \leq k'$, these backward gradients are non-zero but \emph{indirect}: they pass through softmax Jacobians and nonlinearities. This is the \emph{smoothing} channel.

\begin{remark}[Cross-frame gradient structure]
\label{rem:crossframe}
Let $g_k(\bm{x}_{1:K}^{(\bm{t})}, \bm{t}) \triangleq \hat{\bm{x}}_{0,k}$ denote the Tweedie denoised estimate for frame $k$, viewed as a function of the full noisy input. Under causal temporal attention, the Jacobian satisfies
\begin{equation}
    \frac{\partial\, g_{k'}}{\partial\, \bm{x}_k^{(t_k)}}
    \begin{cases}
        \neq \mathbf{0} & \text{if } k \leq k', \\
        = \mathbf{0}    & \text{if } k > k'.
    \end{cases}
    \label{eq:causal_jacobian}
\end{equation}
Consequently, the observation-guidance gradient of frame $k'$'s measurement loss on frame~$k$ decomposes via the chain rule as
\begin{equation}
    \nabla_{\bm{x}_k^{(t_k)}} \!\left\|\bm{y}_{k'} - \cA(g_{k'})\right\|^2
    = \underbrace{-2\,\bm{J}_{\cA}^\top (\bm{y}_{k'} - \cA(g_{k'}))}_{\text{observation residual signal}} \cdot \underbrace{\frac{\partial\, g_{k'}}{\partial\, \bm{x}_k^{(t_k)}}}_{\text{cross-frame Jacobian}},
    \label{eq:cross_grad}
\end{equation}
where $\bm{J}_{\cA} = \partial \cA / \partial \hat{\bm{x}}_{0,k'}$. For $k' > k$, this gradient is the \emph{backward smoothing signal}: non-zero by Eq.~\eqref{eq:causal_jacobian}, and routed through the causal path from frame $k$ to frame $k'$ in the network.
\end{remark}

\subsection{Network Architecture}
\label{app:impl:arch}
\label{app:df:arch}
The denoising network $\bm{\epsilon}_{\bm{\theta}}$ is a 3D Diffusion Transformer (DiT-3D)~\citep{peebles2023scalable, chen2024diffusion} adapted to per-frame noise levels. Each frame is patchified into a sequence of spatial tokens; the per-frame diffusion step $t_k$ is mapped to a sinusoidal time embedding and injected into the corresponding frame's tokens via adaptive layernorm. Spatial self-attention runs over the full image within each frame, while temporal attention is restricted to the causal mask: frame $k$ attends only to frames $k' \leq k$. The same backbone is used across all three benchmarks. We use a transformer rather than the convolutional / recurrent backbone of the original DF because the scientific fields targeted here (2D Navier--Stokes vorticity, SEVIR radar, ERA5 weather) have broadband spatial spectra and rapidly evolving structure, for which global attention is empirically a stronger inductive bias.

The training setup is otherwise inherited from DF: the per-frame Gaussian forward process of Eq.~\eqref{eq:df_forward}, $\bm{\epsilon}$-parameterization, a cosine variance schedule, $T = 1000$ diffusion steps, AdamW optimization with mixed-precision training, and 100-step DDIM sampling. Per-dataset training data, compute, and sampling hyperparameters are listed in Appendix~\ref{app:exp_details}.

\subsection{Other Design Choices}
\label{app:impl:design}
\paragraph{Frame stacking.}
Following DF, the implementation supports stacking $F$ consecutive frames into a single super-frame with $C \cdot F$ channels, which reduces the temporal dimension fed to the network by a factor $F$ and assigns one noise level per super-frame. All experiments in this paper use $F = 1$, i.e., one noise level per frame.

\paragraph{Noise clipping.}
For $z$-score-normalized weather data with a wide dynamic range (ERA5), we clip the standard Gaussian noise to $[-15, 15]$ rather than the default $[-6, 6]$, which would otherwise truncate the heavy tails of the multi-variable normalized distribution.

\paragraph{Sliding window at inference.}
Frames are generated in chunks: one frame at a time for the autoregressive schedule, and the whole remaining horizon at once for the pyramid and full-sequence schedules. When the trajectory is longer than the number of frames $K_{\text{train}}$ the network was trained on, only the most recent $K_{\text{train}}$ frames up to the end of the current chunk are fed to the network; earlier frames influence the chunk only through the already-generated frames inside this window.

\subsection{Pseudocode}
\label{app:impl:algorithms}
Algorithm~\ref{alg:training} gives the training procedure (sample a CAT-mixture diffusion-step vector, corrupt, score-match), and Algorithm~\ref{alg:sampling} gives the inference procedure (read the scheduling matrix, identify active frames, compute the reweighted observation gradient, apply the DDIM step with the guidance correction). Compared with the DF recaps in Algorithms~\ref{alg:df_training}--\ref{alg:df_sampling}, the new steps are the sorting and context clamping in Algorithm~\ref{alg:training} (lines 4--5) and the observation guidance in Algorithm~\ref{alg:sampling} (lines 6a--6c).

\begin{algorithm}[h]
\caption{ForcingDAS Training}
\label{alg:training}
\begin{algorithmic}[1]
\REQUIRE Dataset of trajectories $\{\bm{x}_{1:K}\}$, causal ratio $\rho$, context clean ratio $\rho_c$
\REPEAT
    \STATE Sample trajectory $\bm{x}_{1:K}$ from dataset
    \STATE Sample diffusion steps $t_k \sim \mathrm{Uniform}\{1,\ldots,T\}$ for $k = 1,\ldots,K$
    \STATE With probability $\rho$: sort $\bm{t} \leftarrow \mathrm{sort}(t_1,\ldots,t_K)$ \hfill $\triangleright$ \textit{Causally-aligned noise}
    \STATE With probability $\rho_c$: set $t_k \leftarrow 0$ for $k = 1,\ldots,C$ where $C \sim \mathrm{Uniform}\{1,\ldots,C_{\max}\}$
    \STATE Sample $\bm{\epsilon}_k \sim \cN(\mathbf{0}, \I)$ for $k=1,\ldots,K$
    \STATE Corrupt: $\bm{x}_k^{(t_k)} \leftarrow \sqrt{\bar{\alpha}_{t_k}}\,\bm{x}_k + \sqrt{1-\bar{\alpha}_{t_k}}\,\bm{\epsilon}_k$
    \STATE Update $\bm{\theta}$ to minimize $\sum_{k=1}^K \lambda(t_k)\,\|\bm{\epsilon}_k - \bm{\epsilon}_{\bm{\theta}}(\bm{x}_{1:K}^{(\bm{t})}, \bm{t})_k\|^2$
\UNTIL{converged}
\end{algorithmic}
\end{algorithm}

\begin{algorithm}[h]
\caption{ForcingDAS Sampling (Inference)}
\label{alg:sampling}
\begin{algorithmic}[1]
\REQUIRE Observations $\{\bm{y}_k\}$, forward operator $\cA$, scheduling matrix $\bm{S}^{(u)}$, trained model $\bm{\epsilon}_{\bm{\theta}}$, guidance scale $\zeta$, noise parameter $\gamma$
\STATE Initialize $\bm{x}_k^{(T)} \sim \cN(\mathbf{0}, \I)$ for $k = 1,\ldots,K$
\FOR{$\ell = 0, 1, \ldots, L-1$}
    \STATE Read diffusion steps: $t_k \leftarrow S_{k,\ell}^{(u)}$, $t_k' \leftarrow S_{k,\ell+1}^{(u)}$ for $k = 1,\ldots,K$
    \STATE Identify active frames: $\calK_{\text{active}} \leftarrow \{k : t_k > t_k'\}$
    \STATE Compute $\hat{\bm{x}}_{0,k} \leftarrow g_k(\bm{x}_{1:K}^{(\bm{t})}, \bm{t})$ via the Tweedie estimate \hfill $\triangleright$ \textit{Prior step}
    \makeatletter
    \let\savedALClno\ALC@lno
    \renewcommand{\ALC@lno}{\footnotesize 6a:}%
    \STATE Compute per-frame weights: $w(t_k) \leftarrow \left(\sigma_y^2 + \gamma\,\frac{1-\bar{\alpha}_{t_k}}{\bar{\alpha}_{t_k}}\right)^{-1}$ \hfill $\triangleright$ \textit{Obs.\ guidance}
    \renewcommand{\ALC@lno}{\footnotesize 6b:}%
    \STATE Aggregate loss: $\mathcal{L}_{\text{obs}} \leftarrow \big(\sum_{k \in \calK_{\text{active}}} w(t_k)\,\|\bm{y}_k - \cA(\hat{\bm{x}}_{0,k})\|^2\big)^{1/2}$ \hfill $\triangleright$ \textit{Obs.\ guidance}
    \vspace{2pt}
    \renewcommand{\ALC@lno}{\footnotesize 6c:}%
    \STATE Compute gradient: $\bm{g} \leftarrow \nabla_{\bm{x}_{1:K}^{(\bm{t})}} \mathcal{L}_{\text{obs}}$ \hfill $\triangleright$ \textit{Obs.\ guidance}
    \setcounter{ALC@line}{6}%
    \let\ALC@lno\savedALClno
    \makeatother
    \STATE For each $k \in \calK_{\text{active}}$: \hfill $\triangleright$ \textit{DDIM update}
    \STATE \quad $\bm{x}_k^{(t_k')} \leftarrow \sqrt{\bar{\alpha}_{t_k'}}\,\hat{\bm{x}}_{0,k} + \sqrt{1-\bar{\alpha}_{t_k'}-\sigma_{t_k}^2}\,\bm{\epsilon}_{\bm{\theta}}(\cdot)_k + \sigma_{t_k}\bm{z}_k - \zeta\, \bm{g}_k$
\ENDFOR
\RETURN Estimated trajectory $\bm{x}_{1:K}^{(0)}$
\end{algorithmic}
\end{algorithm}

\newpage
\section{Experimental Details and Additional Results}
\label{app:exp_details}

This appendix collects, for each of the three benchmarks of \S\ref{sec:exp}: (i) the dataset and its splits; (ii) the observation model and the conversion between the data-space noise level and its physical-units counterpart; (iii) training compute and the sampling hyperparameters behind the reported numbers; (iv) baseline configurations; and (v) additional results and visualizations. It closes with an ablation of causality-aware training on probabilistic forecasting (\S\ref{app:forecasting}). Cold-start results are in Appendix~\ref{app:cold_start}.

\paragraph{Convention: data space vs.\ raw data space.}
Throughout the main text, we report the additive observation noise as a Gaussian of standard deviation $\sigma_y = 0.05$ in the dataset's \emph{data space}, i.e.\ the space in which both the diffusion model and the observation operator $\cA$ act at training and inference.
This is the space produced by the dataset loader: $[0,1]$-rescaled vorticity for NS (\S\ref{app:exp_details:ns}), $[0,1]$-scaled VIL for SEVIR (\S\ref{app:exp_details:sevir}), and per-channel $z$-score-normalized fields for ERA5 (\S\ref{app:exp_details:era5}).
For each system, the dataset normalization is an invertible affine map $x^{\text{data}} = (x^{\text{raw}} - a)/b$, where $b$ is a positive scalar (NS, SEVIR) or a per-channel scale ($\sigma_c^{\text{clim}}$ for ERA5).
Because every observation operator $\cA$ we evaluate (sparse pixel masking; bilinear down-sampling) is \emph{linear}, applying $\cA$ in data space and adding noise of std $\sigma_y$ is exactly equivalent to applying the same $\cA$ (same mask / same kernel) directly to the raw field and adding Gaussian noise of standard deviation
\begin{equation}
    \sigma_y^{\text{raw}} \;=\; b \cdot \sigma_y^{\text{data}},
    \qquad b \;=\;
    \begin{cases}
        x_{\max} - x_{\min} & \text{(min--max scaling, NS \& SEVIR)}, \\
        \sigma_c^{\text{clim}} & \text{(per-channel $z$-score, ERA5)}.
    \end{cases}
    \label{eq:noise_scale_conversion}
\end{equation}
That is, at every observed grid point $i$ (and channel $c$ for ERA5),
$y_i^{\text{data}} = x_i^{\text{data}} + \sigma_y^{\text{data}}\,\varepsilon_i = (x_i^{\text{raw}} - a)/b + \sigma_y^{\text{data}}\,\varepsilon_i$, so that $b\, y_i^{\text{data}} + a = x_i^{\text{raw}} + (b\,\sigma_y^{\text{data}})\,\varepsilon_i$, with the \emph{same} Gaussian sample $\varepsilon_i$.
Eq.~\eqref{eq:noise_scale_conversion} is therefore an exact restatement, not an approximation, and the per-benchmark subsections below tabulate the resulting $\sigma_y^{\text{raw}}$ in physical units.

\subsection{Navier--Stokes Vorticity}
\label{app:exp_details:ns}
\subsubsection{Dataset and physical setting}
\label{app:exp_details:ns:dataset}
We use 2D incompressible Navier--Stokes vorticity trajectories generated from the stochastic vorticity equation on the torus $\mathbb{T}^2 = [0, 2\pi]^2$~\citep{pmlr-v235-chen24n,rozet2023score}:
\begin{equation}
    \mathrm{d}\omega + (\bm{v} \cdot \nabla\omega)\,\mathrm{d}t = \nu\,\Delta\omega\,\mathrm{d}t - \alpha\,\omega\,\mathrm{d}t + \varepsilon\,\mathrm{d}\eta,
    \label{eq:ns_vorticity}
\end{equation}
where $\bm{v} = \nabla^{\perp}\psi = (-\partial_y\psi,\,\partial_x\psi)$ is the velocity field recovered from the stream function $\psi$ satisfying $-\Delta\psi = \omega$.
The parameters are viscosity $\nu = 10^{-3}$, linear drag $\alpha = 0.1$, and noise amplitude $\varepsilon = 1.0$.
The stochastic forcing $\mathrm{d}\eta$ is white-in-time and acts on eight fixed sinusoidal modes driven by independent Wiener processes $\{W_i(t)\}_{i=1}^8$:
\begin{equation}
\label{eq:ns_forcing}
\begin{aligned}
    \eta(t,x,y) = \;&W_1(t)\sin(6x) + W_2(t)\cos(7x) + W_3(t)\sin\!\big(5(x{+}y)\big) + W_4(t)\cos\!\big(8(x{+}y)\big) \\
    +\;& W_5(t)\cos(6x) + W_6(t)\sin(7x) + W_7(t)\cos\!\big(5(x{+}y)\big) + W_8(t)\sin\!\big(8(x{+}y)\big).
\end{aligned}
\end{equation}

The data are generated using a pseudo-spectral solver with Euler--Maruyama time stepping on a $256 \times 256$ grid at $\Delta t = 10^{-4}$, with snapshots stored every 0.5~time units.
The fields are then bilinearly downsampled to $128 \times 128$ for training and evaluation, and every second snapshot is kept, so consecutive frames are $1.0$ time units apart.
The trajectories are split 950/50 into training and held-out evaluation trajectories. We evaluate on 16 of the 50 held-out trajectories, which are not used for training or model selection and are disjoint from the 950 training trajectories of the ForcingDAS, FlowDAS, and SDA checkpoints alike. Four further held-out trajectories, disjoint from these 16, form a tuning set, used to select the ForcingDAS sampling hyperparameters (Appendix~\ref{app:exp_details:ns:hp}). We use $K = 50$ frames per trajectory for the with-context protocol (10 clean context frames, 40 assimilated) and $K = 30$ for the cold-start protocol (Appendix~\ref{app:cold_start:ns}).
Because the underlying PDE is fully specified, this experiment also admits strong physics-based DA references (the Ensemble Kalman Filter and Smoother), whose implementation is described in Appendix~\ref{app:enkf_ns}.

The training data have empirical extrema $[\,x_{\min},\, x_{\max}\,] = [\,-19.16,\, +17.42\,]$ and standard deviation $\sigma^{\text{raw}} = 3.07$ (rad/s), giving a min--max width of $b = x_{\max} - x_{\min} = 36.58$.
The dataset loader maps raw vorticity into the data space by min--max rescaling to $[0, 1]$,
$x^{\text{data}} = (x^{\text{raw}} - x_{\min}) / b;$
this is the input the diffusion model receives, and the space in which $\cA$ is evaluated at test time.

\subsubsection{Observation model and noise calibration}
\label{app:exp_details:ns:obs}

\paragraph{Operator.}
For NS we evaluate the 5\% sparse observation operator (cf.\ Setup in \S\ref{sec:exp:ns}): a random binary pixel mask retaining $5\%$ of grid points. The mask is a fixed (seeded) sample, shared across frames and trajectories, that observes 795 of the $16{,}384$ pixels.

\paragraph{Noise.}
The clean operator output is corrupted by additive Gaussian noise of std $\sigma_y^{\text{data}} = 0.05$ (the value we report in the main text), i.e.\ $\bm{y}_k = \cA(\bm{x}_k^{\text{data}}) + \bm{\varepsilon}_k$, $\bm{\varepsilon}_k \sim \cN(\mathbf{0}, 0.05^2 \I)$, in the $[0,1]$-rescaled vorticity space.

\paragraph{Equivalent noise in raw vorticity units.}
By Eq.~\eqref{eq:noise_scale_conversion} with $b = 36.58$, this is exactly equivalent to applying the same operator to the \emph{raw} vorticity field and adding Gaussian noise of standard deviation
\begin{equation}
    \sigma_y^{\text{raw}} \;=\; 0.05 \times 36.58 \;=\; 1.829 \;\;\text{(rad/s)}
    \;\;\approx\; 0.60 \times \sigma^{\text{raw}}.
\end{equation}
That is, the observation noise is roughly $60\%$ of the climatological standard deviation of the field, a moderate noise level relative to the dynamic range of the vorticity.

\paragraph{Shared observations.}
The observations are drawn once per trajectory from a seeded generator and stored, and every method (ForcingDAS, FlowDAS, SDA, EnKF, EnKS) assimilates these identical values; the evaluation script checks this and refuses to score a run otherwise. The baselines operate in different units (FlowDAS in raw vorticity divided by $3.07$; SDA, EnKF, and EnKS in raw vorticity), so the stored observations are converted into each method's units by the exact affine map of Eq.~\eqref{eq:noise_scale_conversion}. Every method therefore sees the same noise level, $\sigma_y^{\text{raw}} = 1.829$. Since all methods see the same observations on the same trajectories, per-trajectory differences between two methods are exactly paired, and we test them with the Wilcoxon signed-rank test.

\paragraph{Metrics.}
NRMSE is computed per frame as $\|\hat{\bm{x}}_k - \bm{x}_k\|_F / \|\bm{x}_k\|_F$ in the $[0,1]$ data space, averaged over the scored frames of a trajectory, and then reported as mean $\pm$ std over the 16 trajectories. The spectrum errors are the mean per-bin relative errors of the radially averaged kinetic-energy spectrum (computed from the vorticity through the stream function) over mid-$k$ $[8,32)$ and over all 64 bins $[0.5,64)$. Only assimilated frames are scored, since the clean context frames are copies of the truth. Because the min--max map adds a constant offset, NRMSE in the $[0,1]$ space is smaller than NRMSE of the raw vorticity by a fixed factor ($\approx 11.9$ on these trajectories). The factor is the same for every method, so rankings are unaffected.

\subsubsection{Training compute}
\label{app:exp_details:ns:train}
We train the ForcingDAS DiT-B backbone on the NS training set of shape $(950, 50, 128, 128)$ (trajectories $\times$ frames per trajectory $\times$ height $\times$ width) in float32 across 4 NVIDIA A800 GPUs for 500 epochs, with per-GPU batch size 8 (total batch size 32). Total training time is approximately 24 hours.

\subsubsection{Sampling hyperparameters}
\label{app:exp_details:ns:hp}

ForcingDAS sampling on NS uses the DiT-B backbone with 100 DDIM steps, per-frame variance reweighting (Eq.~\ref{eq:reweight}) with parameter $\gamma$, and guidance scale $\zeta$.
Table~\ref{tab:hp_ns} reports the values behind every number in Table~\ref{tab:ns_main} and Table~\ref{tab:ns_cold_start}. ForcingDAS-Pyr uses a 10-frame lag ($u=10$; with $T=100$ DDIM steps the active window is $\min(K,\lceil T/u\rceil)$ frames). For all three schedules, $(\gamma,\zeta)$ were selected on the 4 held-out tuning trajectories, which are disjoint from the 16 evaluation trajectories.

\begin{table}[h]
\centering
\caption{NS sampling hyperparameters (DiT-B backbone, 100 DDIM steps, SO-5\%, $\sigma_y^{\text{data}} = 0.05$). AR = autoregressive (filtering), Pyr = pyramid (fixed-lag smoothing, $u=10$), FS = full-sequence (smoothing). ``Lag'' is the number of frames that are denoised jointly. All $(\gamma,\zeta)$ are selected on the 4 held-out tuning trajectories.}
\label{tab:hp_ns}
\small
\begin{tabular}{@{}lccccc@{}}
\toprule
Protocol & $K$ (frames) & Schedule & Lag & $\gamma$ & $\zeta$ \\
\midrule
With context (10 clean) & 50 & AR  & 1  & 0.01  & 2.0  \\
                        & 50 & Pyr & 10 & 0.01  & 4.0  \\
                        & 50 & FS  & 40 & 0.005 & 12.0 \\
\midrule
Cold start (0 clean)    & 30 & AR  & 1  & 0.05  & 4.0  \\
                        & 30 & Pyr & 10 & 0.05  & 8.0  \\
                        & 30 & FS  & 30 & 0.01  & 12.0 \\
\bottomrule
\end{tabular}
\end{table}

\subsubsection{Baseline configurations}
\label{app:exp_details:ns:baselines}
\textbf{FlowDAS}~\citep{flowdas} uses the original public implementation and its public NS checkpoint, whose training data lie within the 950 training trajectories and do not overlap the evaluation set. It uses a 10-frame conditioning window, 500 Euler--Maruyama steps per assimilated frame, 25 Monte Carlo samples for the guidance estimate, and guidance scale 1.0, the value selected in the original FlowDAS paper. With context, the window is seeded with the 10 clean frames; in the cold start, the released code has no such mode, so we seed the window with standard Gaussian noise. \textbf{SDA}~\citep{rozet2023score} uses a score network trained on the same 950 trajectories with a Markov blanket of window 5, sampled with 256 steps, 1 Langevin correction, $\tau = 0.5$, and $\gamma = 0.01$, selected on the 4 held-out tuning trajectories. SDA has no clean-context mode. Passing the context frames as fully observed frames makes its unnormalized Gaussian likelihood dominated by those frames, and sampling diverged at every likelihood scale we tried. We therefore give SDA the context by replacement conditioning: after every predictor and corrector update, the context frames are overwritten with the ground truth diffused to the current noise level, while the likelihood term is unchanged. The returned context frames are within $1.4\times10^{-4}$ of the truth, so SDA has the same information as ForcingDAS. \textbf{EnKF} and \textbf{EnKS} have no trained parameters; their configuration is in Appendix~\ref{app:enkf_ns}. All four baselines assimilate the shared observations described above.

\subsubsection{Additional results}
\label{app:exp_details:ns:extra}

Table~\ref{tab:ns_main} gives the with-context numbers behind \figNSvis{}; the cold-start counterpart is in Appendix~\ref{app:cold_start:ns}. EnKF and EnKS have access to the \emph{exact} PDE dynamics (with the resolution-induced model error discussed in Appendix~\ref{app:enkf_ns}), whereas ForcingDAS uses only a learned prior and no knowledge of the governing equations. Their results therefore indicate what a physics-based filter or smoother with the true governing equations achieves on this benchmark.

\begin{table}[h]
\centering
\caption{Navier--Stokes vorticity assimilation under SO-5\% with 10 clean context frames ($K=50$, scored on the 40 assimilated frames): NRMSE ($\downarrow$) and the radially averaged kinetic-energy spectrum relative error ($\downarrow$) over mid-$k$ ($k \in [8, 32)$, the inertial range) and all-$k$ ($k \in [0.5, 64)$). Mean {\scriptsize $\pm$ std} over 16 held-out trajectories; every method assimilates identical observations. SDA receives the 10 clean context frames by replacement conditioning (Appendix~\ref{app:exp_details:ns:baselines}). EnKS-FL uses a 10-frame lag, matching ForcingDAS-Pyr. Cold-start results are in Table~\ref{tab:ns_cold_start}.}
\label{tab:ns_main}
\footnotesize
\setlength{\tabcolsep}{1.5pt}
\resizebox{\linewidth}{!}{%
\begin{tabular}{l ccc cc ccc}
\toprule
 & \multicolumn{3}{c}{Filtering} & \multicolumn{2}{c}{Fix-lag Smoothing} & \multicolumn{3}{c}{Full-sequence Smoothing} \\
\cmidrule(lr){2-4} \cmidrule(lr){5-6} \cmidrule(lr){7-9}
Metric & FlowDAS & EnKF & ForcingDAS-AR & EnKS-FL & ForcingDAS-Pyr & EnKS-FS & SDA & ForcingDAS-FS \\
\midrule
NRMSE   & 0.141\pmstd{0.006} & 0.108\pmstd{0.003} & \textbf{0.071}\pmstd{0.004} & 0.113\pmstd{0.003} & \textbf{0.057}\pmstd{0.003} & 0.116\pmstd{0.003} & 0.075\pmstd{0.002} & \textbf{0.067}\pmstd{0.003} \\
mid-$k$ & 0.607\pmstd{0.044} & 0.684\pmstd{0.009} & \textbf{0.160}\pmstd{0.019} & 0.530\pmstd{0.013} & \textbf{0.218}\pmstd{0.015} & 0.430\pmstd{0.016} & 0.193\pmstd{0.017} & \textbf{0.139}\pmstd{0.004} \\
all-$k$ & 0.617\pmstd{0.041} & 0.824\pmstd{0.019} & \textbf{0.215}\pmstd{0.020} & 0.657\pmstd{0.022} & \textbf{0.279}\pmstd{0.016} & 0.550\pmstd{0.023} & 0.251\pmstd{0.027} & \textbf{0.189}\pmstd{0.008} \\
\bottomrule
\end{tabular}}
\end{table}

\subsection{SEVIR Radar Nowcasting}
\label{app:exp_details:sevir}

\subsubsection{Dataset and physical setting}
\label{app:exp_details:sevir:dataset}

SEVIR-LR~\citep{veillette2020sevir} provides spatio-temporal radar imagery sampled from severe-weather events.
We use the Vertically Integrated Liquid (VIL) channel, which encodes precipitation intensity as a single-channel field at $128{\times}128$ resolution and 5-minute temporal cadence.
We use a SEVIR-LR build that keeps all 49 frames of each event. This differs from the standard 25-frame SEVIR-LR build and date-based split that FlowDAS's own pipeline uses. The 19{,}207 VIL events with no missing data are shuffled (seed 42) and split 90/10 into 17{,}287 training and 1{,}920 held-out events. From the held-out events, 30 were sampled (seed 2026) and 6 were discarded for insufficient convective activity (fewer than 5\% of pixels above VIL 16 on the uint8 scale over all 49 frames), leaving 24 curated events. We evaluate on 16 of them (4 inspected during early development plus the 12 most active of the remaining 20), which are not used for model selection. Their activity ranges from 9\% to 81\% of pixels above VIL 16. Four further held-out events, disjoint from the 16 and from all training data, form the tuning set. Trajectories have $K=49$ frames with context (6 clean, 43 assimilated) and $K=30$ frames in the cold start (the first 30 frames of the same events).
SEVIR has no closed-form physical model for VIL evolution, so no PDE-based DA baseline exists; FlowDAS~\citep{flowdas} is the only learned filter we are aware of that has been applied at this resolution and modality.

The raw VIL data are stored as 8-bit integer pixel intensities and are loaded as floating-point values rescaled to the unit interval $[0,1]$ by dividing by $255$.
This unit interval is the data space used by both the diffusion model and the observation operator: in the convention where ``raw data'' means $[0,1]$ VIL, no further rescaling is applied (i.e.\ $b=1$).
For comparability with the standard SEVIR CSI thresholds, which are reported on the $0$--$255$ uint8 scale, we give both conventions below.

\subsubsection{Observation model and noise calibration}
\label{app:exp_details:sevir:obs}

\paragraph{Operators.}
For SEVIR we evaluate four observation operators (cf.\ Setup in \S\ref{sec:exp:sevir}): SO-10\% and SO-20\% (random pixel masks retaining $10\%$ resp.\ $20\%$ of pixels; one seed-42 mask shared across frames and events, observing 1{,}637 and 3{,}295 of the $16{,}384$ pixels) and SR2x, SR4x (bilinear down-sampling by $2$ resp.\ $4$).

\paragraph{Noise.}
Gaussian noise of standard deviation $\sigma_y^{\text{data}} = 0.05$ is added to the clean operator output, identically to NS. The noisy observations are generated once per event (seed $12345+i$ for event $i$) and stored; every method (ForcingDAS, FlowDAS, SDA) assimilates these identical values, and the evaluation script checks this before scoring. Cold-start runs use the first 30 frames of the same observations.

\paragraph{Equivalent noise in physical units.}
Because the dataset's data space coincides with the $[0,1]$ VIL convention ($b = 1$), Eq.~\eqref{eq:noise_scale_conversion} gives $\sigma_y^{\text{raw, [0,1]}} = 0.05$: the data-space and raw-space noise levels coincide.
Translated into the $0$--$255$ uint8 SEVIR convention (multiply by $255$):
\begin{equation}
    \sigma_y^{\text{raw, [0,255]}} \;=\; 0.05 \times 255 \;=\; 12.75.
\end{equation}
For reference, the lowest CSI threshold in the SEVIR evaluation suite is $16$ on the same uint8 scale (Table~\ref{tab:sevir_csi}), so the noise level is $\approx 80\%$ of the smallest nowcasting bin width and $\sim 6\%$ of the largest.

\subsubsection{Training compute}
\label{app:exp_details:sevir:train}
We train the ForcingDAS DiT-B backbone on the SEVIR-VIL training set of shape $(17287, 49, 128, 128)$ (trajectories $\times$ frames $\times$ height $\times$ width) in float32 across 4 NVIDIA A800 GPUs for 25 epochs, with per-GPU batch size 8 (total batch size 32). Total training time is approximately 22 hours.

\subsubsection{Sampling hyperparameters}
\label{app:exp_details:sevir:hp}

Table~\ref{tab:hp_sevir} lists the $(\gamma,\zeta)$ behind every ForcingDAS number. ForcingDAS-Pyr runs with a 20-frame lag ($u=5$). For all three schedules, $(\gamma,\zeta)$ were selected on the 4 held-out tuning events, which are disjoint from the 16 evaluation events, by the lowest mean NRMSE over the assimilated frames.

\begin{table}[h]
\centering
\caption{SEVIR sampling hyperparameters $(\gamma, \zeta)$ of ForcingDAS (DiT-B backbone, 100 DDIM steps, $\sigma_y^{\text{data}} = 0.05$). AR = autoregressive (filtering), Pyr = pyramid (fixed-lag smoothing, 20-frame lag, $u=5$), FS = full-sequence (smoothing).}
\label{tab:hp_sevir}
\small
\begin{tabular}{@{}llcccc@{}}
\toprule
Protocol & Schedule & SO-10\% & SO-20\% & SR2x & SR4x \\
\midrule
With context   & AR  & (0.005, 1.0) & (0.007, 0.7) & (0.015, 0.5) & (0.010, 0.7) \\
($K=49$, 6 clean) & Pyr & (0.010, 8.0) & (0.005, 4.0) & (0.005, 2.0) & (0.010, 2.0) \\
               & FS  & (0.005, 8.0) & (0.015, 8.0) & (0.015, 4.0) & (0.005, 4.0) \\
\midrule
Cold start     & AR  & (0.010, 1.0) & (0.005, 0.7) & (0.005, 0.5) & (0.015, 0.7) \\
($K=30$, 0 clean) & Pyr & (0.010, 8.0) & (0.005, 4.0) & (0.015, 2.0) & (0.015, 4.0) \\
               & FS  & (0.005, 8.0) & (0.015, 8.0) & (0.015, 4.0) & (0.005, 4.0) \\
\bottomrule
\end{tabular}
\end{table}

\subsubsection{Baseline configurations}
\label{app:exp_details:sevir:baselines}
\textbf{FlowDAS}~\citep{flowdas} was retrained on our 49-frame SEVIR build and training split (99{,}250 steps, 23\,h on one A40 GPU) and samples with 500 Euler--Maruyama steps and 25 Monte Carlo draws for the guidance estimate. Its guidance scale was first selected on the 4 tuning events from two values per setting, then re-searched over $\zeta \in \{0.05, 0.1, 0.2, 0.5, 1.0, 1.5, 2.0\}$. The re-search confirmed every original value: $\zeta = 0.1/0.05/0.1/0.1$ with context and $0.2/0.1/0.2/0.2$ in the cold start (SO-10\%/SO-20\%/SR2x/SR4x). Using short 3-step rollouts as a cheap tuning proxy picked larger steps in four settings, and each of these was worse at full length on the tuning events: a large step corrects quickly over a few frames but accumulates error over 30--43 frames. \textbf{SDA}~\citep{rozet2023score} uses a score network trained on our training split (Markov blanket window 5, channels $(96,192,384)$, 3 blocks per level, embedding dimension 64, AdamW with learning rate $2\times10^{-4}$ and weight decay $10^{-3}$, batch size 32; 119 epochs, 10.4\,h on one A40), sampled with 256 steps, 1 Langevin correction, $\tau = 0.5$, and $\gamma = 0.01$, selected on the 4 held-out tuning events. SDA reconstructs the full trajectory jointly and has no clean-context mode. As on NS (Appendix~\ref{app:exp_details:ns:baselines}), passing the context frames as fully observed frames makes sampling diverge, so with context we use replacement conditioning: after every predictor and corrector update, the 6 context frames are overwritten with the ground truth diffused to the current noise level, and the likelihood term is unchanged. Because the last replacement still carries a small amount of diffusion noise, SDA's returned context frames are within $5\times10^{-3}$ of the truth, whereas the other methods copy them exactly; this slightly disadvantages SDA. No method's prediction is clamped; values are clipped to $[0,255]$ only when computing CSI. For reference, ForcingDAS was trained for 25 epochs ($\approx$88 GPU-hours on A800; Appendix~\ref{app:exp_details:sevir:train}).

\subsubsection{Per-threshold CSI}
\label{app:sevir_csi}
\label{app:exp_details:sevir:extra}

Table~\ref{tab:sevir_csi} reports the per-threshold Critical Success Index (CSI) of all methods on SEVIR VIL across the four observation settings with context; the cold-start counterpart is Table~\ref{tab:sevir_csi_cold}. Thresholds are SEVIR VIL pixel intensities on the uint8 scale; higher thresholds indicate heavier precipitation, which is increasingly rare and harder to reconstruct. CSI is pooled: hits, misses, and false alarms are summed over all assimilated frames of all 16 events, following the SEVIR convention, and the {\scriptsize $\pm$} is the std over 2{,}000 bootstrap resamples of the events. Per-event CSI is noisy at high thresholds because several events contain almost no pixels above VIL 133.

\begin{table}[h]
\centering
\caption{SEVIR per-threshold pooled CSI ($\uparrow$; {\scriptsize $\pm$} std over bootstrap resamples of the 16 events) with 6 clean context frames, grouped by DA regime. ForcingDAS-Pyr uses a 20-frame lag; SDA receives the context frames by replacement conditioning. Cold-start results are in Table~\ref{tab:sevir_csi_cold}.}
\label{tab:sevir_csi}
\footnotesize
\setlength{\tabcolsep}{2pt}
\resizebox{\linewidth}{!}{%
\begin{tabular}{l l l ccccccc}
\toprule
Obs & Regime & Method & CSI-16 & CSI-74 & CSI-133 & CSI-160 & CSI-181 & CSI-219 & Mean \\
\midrule
  \multirow{5}{*}{SO-10\%} & \multirow{2}{*}{Filtering} & FlowDAS & 0.668\pmstd{0.049} & 0.682\pmstd{0.041} & 0.502\pmstd{0.027} & 0.305\pmstd{0.013} & 0.196\pmstd{0.029} & 0.066\pmstd{0.040} & 0.403\pmstd{0.020} \\
   &  & ForcingDAS-AR & \textbf{0.820}\pmstd{0.023} & \textbf{0.787}\pmstd{0.032} & \textbf{0.624}\pmstd{0.032} & \textbf{0.457}\pmstd{0.024} & \textbf{0.394}\pmstd{0.035} & \textbf{0.257}\pmstd{0.037} & \textbf{0.557}\pmstd{0.024} \\
   & Fix-lag & ForcingDAS-Pyr & 0.840\pmstd{0.020} & 0.809\pmstd{0.028} & 0.649\pmstd{0.026} & 0.490\pmstd{0.021} & 0.426\pmstd{0.030} & 0.293\pmstd{0.031} & 0.584\pmstd{0.021} \\
   & \multirow{2}{*}{Full-seq.} & SDA & 0.820\pmstd{0.025} & 0.797\pmstd{0.031} & 0.629\pmstd{0.025} & 0.477\pmstd{0.021} & 0.425\pmstd{0.031} & \textbf{0.294}\pmstd{0.037} & 0.573\pmstd{0.021} \\
   &  & ForcingDAS-FS & \textbf{0.853}\pmstd{0.018} & \textbf{0.815}\pmstd{0.028} & \textbf{0.659}\pmstd{0.027} & \textbf{0.498}\pmstd{0.021} & \textbf{0.437}\pmstd{0.032} & 0.291\pmstd{0.033} & \textbf{0.592}\pmstd{0.021} \\
\midrule
  \multirow{5}{*}{SO-20\%} & \multirow{2}{*}{Filtering} & FlowDAS & 0.711\pmstd{0.047} & 0.736\pmstd{0.040} & 0.564\pmstd{0.032} & 0.384\pmstd{0.021} & 0.300\pmstd{0.029} & 0.190\pmstd{0.032} & 0.481\pmstd{0.023} \\
   &  & ForcingDAS-AR & \textbf{0.850}\pmstd{0.020} & \textbf{0.834}\pmstd{0.026} & \textbf{0.692}\pmstd{0.025} & \textbf{0.543}\pmstd{0.021} & \textbf{0.493}\pmstd{0.032} & \textbf{0.340}\pmstd{0.038} & \textbf{0.625}\pmstd{0.020} \\
   & Fix-lag & ForcingDAS-Pyr & 0.884\pmstd{0.015} & 0.861\pmstd{0.022} & 0.730\pmstd{0.020} & 0.595\pmstd{0.017} & 0.546\pmstd{0.027} & 0.411\pmstd{0.034} & 0.671\pmstd{0.017} \\
   & \multirow{2}{*}{Full-seq.} & SDA & 0.849\pmstd{0.022} & 0.837\pmstd{0.027} & 0.691\pmstd{0.020} & 0.557\pmstd{0.018} & 0.516\pmstd{0.027} & 0.384\pmstd{0.033} & 0.639\pmstd{0.017} \\
   &  & ForcingDAS-FS & \textbf{0.879}\pmstd{0.015} & \textbf{0.853}\pmstd{0.023} & \textbf{0.721}\pmstd{0.021} & \textbf{0.585}\pmstd{0.018} & \textbf{0.537}\pmstd{0.027} & \textbf{0.399}\pmstd{0.032} & \textbf{0.662}\pmstd{0.017} \\
\midrule
  \multirow{5}{*}{SR2x} & \multirow{2}{*}{Filtering} & FlowDAS & 0.798\pmstd{0.039} & 0.841\pmstd{0.028} & 0.720\pmstd{0.020} & 0.575\pmstd{0.020} & 0.514\pmstd{0.029} & 0.373\pmstd{0.030} & 0.637\pmstd{0.017} \\
   &  & ForcingDAS-AR & \textbf{0.872}\pmstd{0.017} & \textbf{0.865}\pmstd{0.022} & \textbf{0.745}\pmstd{0.019} & \textbf{0.616}\pmstd{0.022} & \textbf{0.575}\pmstd{0.032} & \textbf{0.431}\pmstd{0.035} & \textbf{0.684}\pmstd{0.017} \\
   & Fix-lag & ForcingDAS-Pyr & 0.898\pmstd{0.014} & 0.886\pmstd{0.019} & 0.774\pmstd{0.015} & 0.657\pmstd{0.017} & 0.618\pmstd{0.025} & 0.488\pmstd{0.031} & 0.720\pmstd{0.014} \\
   & \multirow{2}{*}{Full-seq.} & SDA & 0.853\pmstd{0.022} & 0.850\pmstd{0.026} & 0.715\pmstd{0.019} & 0.592\pmstd{0.019} & 0.559\pmstd{0.026} & 0.433\pmstd{0.031} & 0.667\pmstd{0.015} \\
   &  & ForcingDAS-FS & \textbf{0.893}\pmstd{0.014} & \textbf{0.881}\pmstd{0.020} & \textbf{0.766}\pmstd{0.017} & \textbf{0.645}\pmstd{0.018} & \textbf{0.609}\pmstd{0.027} & \textbf{0.476}\pmstd{0.032} & \textbf{0.712}\pmstd{0.014} \\
\midrule
  \multirow{5}{*}{SR4x} & \multirow{2}{*}{Filtering} & FlowDAS & 0.703\pmstd{0.045} & 0.726\pmstd{0.039} & 0.558\pmstd{0.033} & 0.376\pmstd{0.023} & 0.286\pmstd{0.028} & 0.178\pmstd{0.030} & 0.471\pmstd{0.024} \\
   &  & ForcingDAS-AR & \textbf{0.830}\pmstd{0.022} & \textbf{0.792}\pmstd{0.031} & \textbf{0.623}\pmstd{0.033} & \textbf{0.460}\pmstd{0.027} & \textbf{0.392}\pmstd{0.038} & \textbf{0.252}\pmstd{0.036} & \textbf{0.558}\pmstd{0.026} \\
   & Fix-lag & ForcingDAS-Pyr & 0.833\pmstd{0.018} & 0.811\pmstd{0.028} & 0.645\pmstd{0.032} & 0.458\pmstd{0.028} & 0.370\pmstd{0.042} & 0.237\pmstd{0.044} & 0.559\pmstd{0.026} \\
   & \multirow{2}{*}{Full-seq.} & SDA & 0.802\pmstd{0.028} & 0.782\pmstd{0.033} & 0.608\pmstd{0.026} & 0.454\pmstd{0.020} & 0.398\pmstd{0.031} & 0.261\pmstd{0.034} & 0.551\pmstd{0.021} \\
   &  & ForcingDAS-FS & \textbf{0.847}\pmstd{0.019} & \textbf{0.816}\pmstd{0.028} & \textbf{0.653}\pmstd{0.026} & \textbf{0.491}\pmstd{0.022} & \textbf{0.426}\pmstd{0.035} & \textbf{0.279}\pmstd{0.037} & \textbf{0.585}\pmstd{0.022} \\
\bottomrule
\end{tabular}}
\end{table}

\subsection{ERA5 Global Atmospheric Reanalysis}
\label{app:exp_details:era5}

\subsubsection{Dataset and physical setting}
\label{app:exp_details:era5:dataset}

We use a 4-channel multi-variable subset of WeatherBench2 ERA5~\citep{rasp2020weatherbench, hersbach2020era5} at $\sim$$1.5^\circ$ resolution with 6-hourly temporal resolution.
The native grid is $240{\times}121$ (longitude $\times$ latitude); we drop the North-Pole row, so every model and metric works on $240{\times}120$. The four channels are Z500, T850, U10, V10 (Appendix~\ref{app:era5_vars}). We train on 1979--2015. 2016 is held out from training; the 16 evaluation windows and the 4 tuning windows below are disjoint. We evaluate on 16 trajectories of $K=30$ frames (7.5 days) from 2016, one window every 92 time steps (23 days) starting at 00\,UTC, so the windows cover all seasons. A separate tuning set of 4 windows from 2016 (starting at time steps 136, 500, 864, 1228) is verified not to overlap the 16 evaluation windows, and it is the only data used for the hyperparameter selections described in Appendix~\ref{app:exp_details:era5:hp}.
Appendix~\ref{app:related:da} surveys related learned-DA work on ERA5, and Appendix~\ref{app:classical_da} describes the 3D-Var, FlowDAS, and Tensor-Var baselines.

The dataset stores values that are $z$-score normalized per channel using climatological statistics from the training period:
$\tilde x_{t,c} = (x_{t,c}^{\text{raw}} - \mu_c) / \sigma_c$.
Hence the data space loaded by the model is the $z$-score-normalized space (per-channel mean $0$, std $1$), with no further rescaling inside the algorithm.
The per-channel climatological scales used to convert between the data space and the raw physical space are summarized in Table~\ref{tab:era5_stats}.

\begin{table}[h]
\centering
\caption{Per-channel climatological mean $\mu_c$ and standard deviation $\sigma_c$ (in raw physical units), computed from the ERA5 1979--2015 training period. These are the conversion factors between the data space (where the model and the observation operator act) and the raw physical space, and every method and metric uses them.}
\label{tab:era5_stats}
\small
\begin{tabular}{@{}lccl@{}}
\toprule
Variable & $\mu_c$ & $\sigma_c$ & Units \\
\midrule
Z500 & $54\,085.09$ & $3{,}363.11$ & m$^2$\,s$^{-2}$ \\
T850 & $274.23$     & $15.72$      & K \\
U10  & $-0.053$     & $5.440$      & m\,s$^{-1}$ \\
V10  & $0.188$      & $4.671$      & m\,s$^{-1}$ \\
\bottomrule
\end{tabular}
\end{table}

\subsubsection{Variable descriptions}
\label{app:era5_vars}

Table~\ref{tab:era5_vars} summarizes the four ERA5 variables used in \S\ref{sec:exp:era5}.

\begin{table}[h]
\centering
\caption{ERA5 variables used in the multi-variable experiment. Typical ranges are approximate global values across seasons.}
\label{tab:era5_vars}
\small
\begin{tabular}{@{}llllp{6.5cm}@{}}
\toprule
\textbf{Variable} & \textbf{Level} & \textbf{Units} & \textbf{Typical range} & \textbf{Physical meaning} \\
\midrule
Z500 & 500\,hPa & m$^2$\,s$^{-2}$ & 48{,}000--58{,}000 & Geopotential at 500\,hPa. Encodes the height of the mid-tropospheric pressure surface, capturing synoptic-scale Rossby waves, troughs, ridges, and the jet stream. The single most-reported variable in DA evaluation. \\
\addlinespace
T850 & 850\,hPa & K & 250--300 & Temperature at 850\,hPa ($\approx$1.5\,km altitude). Captures low-level thermal structure including warm/cold air advection and frontal zones. Strongly coupled to surface weather. \\
\addlinespace
U10 & 10\,m & m\,s$^{-1}$ & $-20$ to $+20$ & Zonal (east--west) component of 10-meter wind. Positive values indicate eastward flow. Exhibits sharper spatial gradients than upper-air fields, particularly near coastlines and orography. \\
\addlinespace
V10 & 10\,m & m\,s$^{-1}$ & $-20$ to $+20$ & Meridional (north--south) component of 10-meter wind. Together with U10, completes the surface wind vector. \\
\bottomrule
\end{tabular}
\end{table}

\paragraph{Why these four variables.}
Z500 and T850 are the standard upper-air evaluation pair in the DA literature, testing both dynamical (pressure) and thermodynamic (temperature) fidelity.
U10 and V10 are surface wind components that are directly observable by weather stations, buoys, and satellite scatterometers, making them natural candidates for the observation operator in DA experiments.
The surface wind field is also physically coupled to the upper-air geopotential via \emph{geostrophic balance}: large-scale winds flow approximately parallel to geopotential contours, with speed proportional to the geopotential gradient, so the four channels are not independent.

\paragraph{Normalization.}
The four variables span three orders of magnitude in raw values (Table~\ref{tab:era5_vars}).
We apply per-variable z-score standardization:
\begin{equation}
    \tilde{x}_{t,c} = \frac{x_{t,c} - \mu_c}{\sigma_c}, \quad c \in \{\text{Z500},\, \text{T850},\, \text{U10},\, \text{V10}\},
\end{equation}
where $\mu_c$ and $\sigma_c$ are the climatological mean and standard deviation of variable $c$, computed over the training period (Table~\ref{tab:era5_stats}).
After normalization, all channels are approximately $\cN(0, 1)$-distributed, allowing the diffusion model to treat them on equal footing.
This follows standard practice in the learned weather prediction literature~\citep{pmlr-v235-huang24h, andry2025appa, xu2024fuxida}.

\subsubsection{Observation model and noise calibration}
\label{app:exp_details:era5:obs}

\paragraph{Operator.}
We use a sparse pixel observation (SO) operator: a single random binary mask of shape $H{\times}W$ is shared across all $4$ channels, modeling co-located meteorological measurements. The mask observes $10\%$ of grid points (seed 42), and it is the same mask for every frame, trajectory, and method.

\paragraph{Noise.}
The data-space noise level is $\sigma_y^{\text{data}} = 0.05$. The noise is i.i.d.\ in space and across channels. The observations $\bm{y} = \bm{m}\odot\bm{x} + 0.05\,\bm{\varepsilon}$ are generated once per trajectory and stored, and every method (ForcingDAS, 3D-Var, FlowDAS, Tensor-Var) assimilates these identical values; the evaluation script checks this, and checks that context frames equal the ground truth, before scoring. Per-trajectory differences between methods are therefore exactly paired.

\paragraph{Equivalent per-channel noise in raw physical units.}
For ERA5, Eq.~\eqref{eq:noise_scale_conversion} applies \emph{per channel}, with $b = \sigma_c$ taken from Table~\ref{tab:era5_stats}: $\sigma_y^{\text{raw}, c} = \sigma_c \cdot \sigma_y^{\text{data}}$.
The operator (the same shared mask) is unchanged in raw space; the noise becomes anisotropic, with a different physical std per variable.
Table~\ref{tab:era5_noise_raw} reports the resulting raw-space noise standard deviations.
For example, $\sigma_y^{\text{data}} = 0.05$ corresponds to a noise of $\sim\!168$\,m$^2$\,s$^{-2}$ on Z500 (a small fraction of typical synoptic anomalies, which are $\mathcal{O}(10^3)$\,m$^2$\,s$^{-2}$) and $\sim\!0.79$\,K on T850 (well below typical regional temperature variability).

\begin{table}[h]
\centering
\caption{ERA5 observation noise standard deviation in raw physical units, derived from $\sigma_y^{\text{data}} = 0.05$ via $\sigma_y^{\text{raw}, c} = \sigma_c \cdot \sigma_y^{\text{data}}$ using the climatological standard deviations of Table~\ref{tab:era5_stats}.}
\label{tab:era5_noise_raw}
\small
\begin{tabular}{@{}lcccc@{}}
\toprule
$\sigma_y^{\text{data}}$ & $\sigma_y^{\text{Z500}}$ & $\sigma_y^{\text{T850}}$ & $\sigma_y^{\text{U10}}$ & $\sigma_y^{\text{V10}}$ \\
                         & (m$^2$\,s$^{-2}$)        & (K)                      & (m\,s$^{-1}$)            & (m\,s$^{-1}$)            \\
\midrule
$0.05$  & $168.16$ & $0.786$  & $0.272$   & $0.234$   \\
\bottomrule
\end{tabular}
\end{table}

\subsubsection{Evaluation metrics: NRMSE and ACC}
\label{app:exp_details:era5:metrics}

We report two per-variable evaluation metrics for ERA5 assimilation, both evaluated on the equiangular $240 \times 120$ longitude--latitude grid (after dropping the North-Pole row) and both applying latitude weights that compensate for grid-cell area shrinking toward the poles.
All metrics are computed identically for every method in Table~\ref{tab:era5_main}.

\paragraph{Notation.}
Let $\hat{s}_{k,b,i,j}$ denote the assimilated value at frame $k$ of trajectory $b$, longitude index $j \in \{1,\ldots,J\}$, latitude index $i \in \{1,\ldots,I\}$; let $s_{k,b,i,j}$ denote the ERA5 reference; and let $c_{k,b,i,j}$ denote the per-grid-point climatology at the valid time of frame $k$ (varying with latitude, longitude, day-of-year, and hour-of-day). Metrics are computed over the scored frames $k \in \mathcal{S}$ only: the 24 assimilated frames with context (the 6 context frames are copies of the truth), and frames 1--29 in the cold start (Appendix~\ref{app:cold_start:era5}).
All quantities are in z-score-normalized data space (\S\ref{app:exp_details:era5:dataset}).

\paragraph{Latitude weights.}
We use cosine-of-latitude weights normalized so that their mean over the $I$ latitude rows equals one:
\begin{equation}
    w(i) = \frac{\cos\theta_i}{\frac{1}{I}\sum_{i'=1}^{I}\cos\theta_{i'}},
    \label{eq:lat_weights}
\end{equation}
where $\theta_i$ is the latitude of row $i$.
This is the standard WeatherBench2 weighting; it is numerically equivalent (relative error $<$0.005\%) to the exact cell-area formula $w(i) \propto \sin\theta_i^u - \sin\theta_i^l$ on our $1.5^{\circ}$ equiangular grid.

\paragraph{Latitude-weighted NRMSE.}
Because the data is z-score normalized per channel (per-channel std $\approx 1$), the latitude-weighted RMSE in data space equals the normalized RMSE.
For each variable we compute the per-frame RMSE, average it over the scored frames of trajectory $b$, and report the mean and standard deviation of these per-trajectory values over the $B=16$ trajectories:
\begin{equation}
    \mathrm{NRMSE}_b = \frac{1}{|\mathcal{S}|}\sum_{k\in\mathcal{S}}\sqrt{\frac{1}{IJ}\sum_{i=1}^{I}\sum_{j=1}^{J} w(i)\,\big(\hat{s}_{k,b,i,j} - s_{k,b,i,j}\big)^2}.
    \label{eq:nrmse_def}
\end{equation}

\paragraph{Anomaly Correlation Coefficient (ACC).}
We subtract the per-grid-point, per-time climatology to obtain anomalies $a_{k,b,i,j} = \hat{s}_{k,b,i,j} - c_{k,b,i,j}$ and $r_{k,b,i,j} = s_{k,b,i,j} - c_{k,b,i,j}$, and then compute the latitude-weighted Pearson correlation between predicted and reference anomalies jointly over the scored frames and spatial points of each trajectory $b$, again reporting mean $\pm$ std over trajectories:
\begin{equation}
    \mathrm{ACC}_b = \frac{\sum_{k\in\mathcal{S},i,j} w(i)\, a_{k,b,i,j}\, r_{k,b,i,j}}{\sqrt{\sum_{k\in\mathcal{S},i,j} w(i)\, a_{k,b,i,j}^2}\,\sqrt{\sum_{k\in\mathcal{S},i,j} w(i)\, r_{k,b,i,j}^2}}.
    \label{eq:acc_def}
\end{equation}
ACC takes values in $[-1, 1]$; in operational practice, $\mathrm{ACC} > 0.6$ is considered useful skill and $\mathrm{ACC} > 0.8$ high skill. ACC is dimensionless and invariant to per-variable affine rescaling, so it is unaffected by the choice of normalization. Pooling the sums of Eq.~\eqref{eq:acc_def} over all trajectories instead changes the reported values by at most $0.001$ for the learned methods and by up to $0.004$ for 3D-Var (T850).

\paragraph{Climatology source.}
The climatology $c_{k,b,i,j}$ is the \href{https://console.cloud.google.com/storage/browser/weatherbench2/datasets/era5-hourly-climatology}{WeatherBench2 hourly ERA5 climatology} (1990--2019 base period, $240 \times 121$ equiangular grid, cropped to $240\times120$), indexed by the day-of-year and hour-of-day matching each frame's valid time. It varies with latitude, longitude, and calendar time (the full latitude--longitude climatology, as in WeatherBench2). The same fixed climatology is used for all methods; in z-score data space it is $c^{(\text{z})}_{k,b,i,j} = (c^{(\text{raw})}_{k,b,i,j} - \mu_c) / \sigma_c$ using the per-channel scales of Table~\ref{tab:era5_stats}.

\subsubsection{Training compute}
\label{app:exp_details:era5:train}
We train the ForcingDAS DiT-B backbone on the ERA5 training set of shape $(1351, 40, 4, 240, 120)$ (trajectories $\times$ frames $\times$ channels $\times$ longitude $\times$ latitude) in float32 across 4 NVIDIA A800 GPUs for 300 epochs, with per-GPU batch size 4 (total batch size 16). Total training time is approximately 33 hours.

\subsubsection{Sampling hyperparameters}
\label{app:exp_details:era5:hp}

For ERA5 we use the same per-frame variance-reweighted guidance as on NS and SEVIR. Table~\ref{tab:hp_era5} lists the values behind every ForcingDAS number. ForcingDAS-Pyr runs with a 10-frame lag ($u=10$). For all three schedules, $(\gamma,\zeta)$ were selected on the 4 held-out tuning windows; no value was selected on the 16 evaluation trajectories.

\begin{table}[h]
\centering
\caption{ERA5 sampling hyperparameters (DiT-B backbone, 100 DDIM steps, SO-10\%, $\sigma_y^{\text{data}} = 0.05$). AR = autoregressive (filtering), Pyr = pyramid (fixed-lag smoothing, $u=10$), FS = full-sequence (smoothing). ``Lag'' is the number of frames that are denoised jointly.}
\label{tab:hp_era5}
\small
\begin{tabular}{@{}lccccc@{}}
\toprule
Protocol & $K$ (frames) & Schedule & Lag & $\gamma$ & $\zeta$ \\
\midrule
With context (6 clean) & 30 & AR  & 1  & 0.005 & 1.0 \\
                       & 30 & Pyr & 10 & 0.005 & 2.0 \\
                       & 30 & FS  & 24 & 0.001 & 4.0 \\
\midrule
Cold start (0 clean)   & 30 & AR  & 1  & 0.005 & 1.0 \\
                       & 30 & Pyr & 10 & 0.005 & 2.0 \\
                       & 30 & FS  & 30 & 0.005 & 8.0 \\
\bottomrule
\end{tabular}
\end{table}

\subsection{Effect of Causality-Aware Training on Probabilistic Forecasting}
\label{app:forecasting}

This subsection isolates the effect of \emph{Causality-Aware Training} (CAT, \S\ref{sec:method:cat}) on free-running probabilistic forecasting, conditioning on a clean context window but \emph{without} observation guidance. Standard Diffusion Forcing samples per-frame noise levels i.i.d.\ uniformly at training time, while every DA scheduling matrix imposes a causally monotone pattern at inference; the resulting train--test gap exists already in pure generation, so probabilistic forecasting is the cleanest setting in which to isolate it.

\subsubsection{Setup}
\label{app:forecasting:setup}

\paragraph{Protocol.}
We run free-running probabilistic forecasting with ForcingDAS-AR ($u = T$): the first $C$ frames of each test trajectory are clean context, and the model autoregressively samples the next $H = 3$ frames from the per-frame diffusion prior, with no observation guidance applied. Following the conventions of the main experiments, $C = 10$ for NS, $C = 6$ for SEVIR, and $C = 6$ for ERA5. We evaluate on 4 held-out trajectories per benchmark, a smaller set than the 16 of \S\ref{sec:exp}. For each starting condition we draw an ensemble of $M = 16$ samples by re-running the reverse process with independent random seeds; all samples share the same clean context.

\paragraph{Models compared.}
For each benchmark we train ForcingDAS networks that are identical in architecture, optimizer, learning-rate schedule, and total training steps; the \emph{only} difference is the per-frame noise distribution at training time:
\begin{itemize}[nosep, leftmargin=*]
    \item \textbf{Without CAT} ($\rho = 0$): the per-frame noise vector $\bm{t}$ is drawn i.i.d.\ from $\mathrm{Uniform}\{1,\ldots,T\}$, recovering the original DF training scheme.
    \item \textbf{With CAT} ($\rho > 0$): the CAT mixture of Eq.~\eqref{eq:cat_loss} is used. We sweep $\rho \in \{0.25, 0.75\}$ on NS and SEVIR and report both rows in Tables~\ref{tab:cat_ns}--\ref{tab:cat_sevir}; on ERA5 we report $\rho = 0.25$. The remaining mixture parameters $(\rho_c, C_{\max})$ are shared across the compared models.
\end{itemize}
Any difference in forecasting quality is therefore attributable to the training noise distribution alone.

\paragraph{Metrics.}
We report two metrics per dataset: the Continuous Ranked Probability Score (CRPS, \S\ref{app:forecasting:crps}) for the full forecast distribution, and NRMSE on the ensemble mean for deterministic skill. Both are reported per lead time $h \in \{1, 2, 3\}$ and as a mean over the three lead times. For ERA5 we report both metrics per variable (Z500, T850, U10, V10) with cosine-of-latitude weights, matching the main-text protocol (Appendix~\ref{app:exp_details:era5:metrics}).

\subsubsection{Continuous Ranked Probability Score}
\label{app:forecasting:crps}

CRPS~\citep{gneiting2007strictly} is the standard strictly proper scoring rule for univariate probabilistic forecasts. Given a predictive cumulative distribution function (CDF) $F$ and an observation $y \in \mathbb{R}$, it is defined as
\begin{equation}
    \mathrm{CRPS}(F, y) = \int_{-\infty}^{\infty} \big(F(z) - \mathbf{1}\{y \leq z\}\big)^2 \, \mathrm{d}z,
    \label{eq:crps_cdf}
\end{equation}
the integrated squared distance between the forecast CDF and the step CDF of the truth. Lower values are better; CRPS reduces to the mean absolute error when $F$ is a point mass, so it generalizes deterministic skill in a calibration-aware way.

\paragraph{Ensemble form.}
We have access to $F$ only through an ensemble of $M$ samples $\{x^{(1)}, \ldots, x^{(M)}\}$. The standard sample-form estimator~\citep{gneiting2007strictly, hersbach2000decomposition} is
\begin{equation}
    \widehat{\mathrm{CRPS}}\!\left(\{x^{(i)}\}_{i=1}^M, y\right)
    = \frac{1}{M}\sum_{i=1}^{M} \big|x^{(i)} - y\big|
    - \frac{1}{2 M^{2}} \sum_{i=1}^{M} \sum_{j=1}^{M} \big|x^{(i)} - x^{(j)}\big|,
    \label{eq:crps_ensemble}
\end{equation}
a difference between an accuracy term (mean absolute deviation of samples from the truth) and a sharpness term (mean pairwise spread). For $M = 1$, the second term vanishes and Eq.~\eqref{eq:crps_ensemble} reduces to $|x^{(1)} - y|$, the deterministic absolute error. We use $M = 16$ throughout.

\paragraph{Aggregation across pixels and trajectories.}
For a 2D field, we compute Eq.~\eqref{eq:crps_ensemble} per pixel and per channel and then average over pixels and held-out trajectories at fixed lead time $h$. For ERA5 the spatial average uses the cosine-of-latitude weights of Eq.~\eqref{eq:lat_weights}, so the reported per-variable CRPS is a latitude-weighted mean analogous to the latitude-weighted NRMSE of Eq.~\eqref{eq:nrmse_def}.

\subsubsection{Results}
\label{app:cat}

\begin{table}[h]
\centering
\caption{Effect of CAT on probabilistic forecasting on Navier--Stokes vorticity. ForcingDAS-AR ($u = T$), $C = 10$ clean context frames, $H = 3$-frame forecast horizon, ensemble size $M = 16$, averaged over 4 held-out trajectories. CRPS and ensemble-mean NRMSE are reported per lead time and as a mean over $h \in \{1,2,3\}$; lower is better for both. Two CAT mixing ratios are reported.}
\label{tab:cat_ns}
\footnotesize
\setlength{\tabcolsep}{4pt}
\begin{tabular}{l cccc cccc}
\toprule
 & \multicolumn{4}{c}{CRPS ($\downarrow$)} & \multicolumn{4}{c}{NRMSE on ensemble mean ($\downarrow$)} \\
\cmidrule(lr){2-5} \cmidrule(lr){6-9}
Method & $h{=}1$ & $h{=}2$ & $h{=}3$ & mean & $h{=}1$ & $h{=}2$ & $h{=}3$ & mean \\
\midrule
Without CAT ($\rho = 0$)    & 0.033 & \textbf{0.046} & \textbf{0.049} & \textbf{0.043} & 0.106 & \textbf{0.149} & \textbf{0.158} & \textbf{0.137} \\
With CAT ($\rho = 0.25$)    & 0.035 & 0.048 & 0.050 & 0.044 & 0.111 & 0.155 & 0.160 & 0.142 \\
With CAT ($\rho = 0.75$)    & \textbf{0.032} & 0.048 & 0.050 & \textbf{0.043} & \textbf{0.102} & 0.154 & 0.162 & 0.140 \\
\bottomrule
\end{tabular}
\end{table}

\begin{table}[h]
\centering
\caption{Effect of CAT on probabilistic forecasting on SEVIR-VIL. ForcingDAS-AR ($u = T$), $C = 6$ clean context frames, $H = 3$-frame forecast horizon, ensemble size $M = 16$, averaged over 4 held-out trajectories. Same column convention as Table~\ref{tab:cat_ns}.}
\label{tab:cat_sevir}
\footnotesize
\setlength{\tabcolsep}{4pt}
\begin{tabular}{l cccc cccc}
\toprule
 & \multicolumn{4}{c}{CRPS ($\downarrow$)} & \multicolumn{4}{c}{NRMSE on ensemble mean ($\downarrow$)} \\
\cmidrule(lr){2-5} \cmidrule(lr){6-9}
Method & $h{=}1$ & $h{=}2$ & $h{=}3$ & mean & $h{=}1$ & $h{=}2$ & $h{=}3$ & mean \\
\midrule
Without CAT ($\rho = 0$)    & \textbf{0.013} & 0.019 & 0.025 & 0.019 & \textbf{0.175} & 0.258 & 0.339 & 0.257 \\
With CAT ($\rho = 0.25$)    & 0.014 & 0.020 & 0.026 & 0.020 & \textbf{0.175} & 0.251 & 0.314 & 0.247 \\
With CAT ($\rho = 0.75$)    & \textbf{0.013} & \textbf{0.018} & \textbf{0.022} & \textbf{0.018} & 0.177 & \textbf{0.243} & \textbf{0.294} & \textbf{0.238} \\
\bottomrule
\end{tabular}
\end{table}

\begin{table}[ht]
\centering
\caption{Effect of CAT on probabilistic forecasting on ERA5. ForcingDAS-AR ($u = T$), $C = 6$ clean context frames, $H = 3$-frame forecast horizon, ensemble size $M = 16$, averaged over 4 held-out length-30 trajectories from 2016. Per-variable, latitude-weighted CRPS and ensemble-mean NRMSE in z-score-normalized data space; lower is better for both. Per lead time $h \in \{1,2,3\}$ and as a mean over the three lead times.}
\label{tab:cat_era5}
\footnotesize
\setlength{\tabcolsep}{2.5pt}
\begin{tabular}{ll cccc cccc}
\toprule
 & & \multicolumn{4}{c}{CRPS ($\downarrow$)} & \multicolumn{4}{c}{NRMSE on ensemble mean ($\downarrow$)} \\
\cmidrule(lr){3-6} \cmidrule(lr){7-10}
Method & Variable & $h{=}1$ & $h{=}2$ & $h{=}3$ & mean & $h{=}1$ & $h{=}2$ & $h{=}3$ & mean \\
\midrule
\multirow{4}{*}{Without CAT ($\rho = 0$)}
 & Z500 & 0.030 & 0.052 & 0.075 & 0.053 & 0.050 & 0.086 & 0.122 & 0.086 \\
 & T850 & 0.039 & 0.057 & 0.076 & 0.058 & 0.065 & 0.090 & 0.116 & 0.091 \\
 & U10  & 0.091 & 0.119 & 0.152 & 0.121 & 0.165 & 0.217 & 0.279 & 0.220 \\
 & V10  & 0.120 & 0.154 & 0.199 & 0.158 & 0.215 & 0.278 & 0.358 & 0.284 \\
\midrule
\multirow{4}{*}{With CAT ($\rho = 0.25$)}
 & Z500 & \textbf{0.018} & \textbf{0.027} & \textbf{0.037} & \textbf{0.027} & \textbf{0.033} & \textbf{0.049} & \textbf{0.067} & \textbf{0.050} \\
 & T850 & \textbf{0.031} & \textbf{0.038} & \textbf{0.046} & \textbf{0.038} & \textbf{0.056} & \textbf{0.069} & \textbf{0.084} & \textbf{0.070} \\
 & U10  & \textbf{0.086} & \textbf{0.107} & \textbf{0.135} & \textbf{0.109} & \textbf{0.160} & \textbf{0.203} & \textbf{0.260} & \textbf{0.208} \\
 & V10  & \textbf{0.109} & \textbf{0.137} & \textbf{0.168} & \textbf{0.138} & \textbf{0.201} & \textbf{0.257} & \textbf{0.321} & \textbf{0.260} \\
\bottomrule
\end{tabular}
\end{table}

\paragraph{The train--test gap.}
Standard DF training samples the per-frame diffusion-step vector $\bm{t} = (t_1, \ldots, t_K)$ i.i.d.\ from $\mathrm{Uniform}\{1,\ldots,T\}$. Inference, in contrast, follows a scheduling matrix of the form Eq.~\eqref{eq:scheduling}, in which earlier frames are always at no-higher diffusion steps than later ones. The model is therefore trained on noise configurations that are, with overwhelming probability, very different from the configurations it sees at test time. This gap matters for scientific dynamical systems, where the future is tightly determined by the past: forecasting quality depends on the model's ability to use a clean past as a reliable conditioning signal, and training with i.i.d.\ noise levels spends most of its capacity on configurations with a noisy past.

\paragraph{CAT noise distribution.}
With probability $\rho$ (the \emph{causal ratio}) the i.i.d.\ noise vector is sorted into a non-decreasing staircase, $\bm{t}^{\text{sorted}} = \mathrm{sort}(t_1, \ldots, t_K)$ with $t_1^{\text{sorted}} \leq \cdots \leq t_K^{\text{sorted}}$; with probability $1-\rho$ the i.i.d.\ draw is kept. With probability $\rho_c$ a random number of leading frames are additionally clamped to diffusion step zero, exposing the model to the clean-context conditioning it sees at inference. The mixture preserves robustness to arbitrary noise patterns while biasing capacity toward the causally monotone configurations that dominate inference; setting $\rho = 0$ recovers standard DF training.

\paragraph{Results.}
Tables~\ref{tab:cat_ns}--\ref{tab:cat_era5} report CRPS and ensemble-mean NRMSE per lead time on the three benchmarks. CAT is most effective on ERA5 (Table~\ref{tab:cat_era5}): every per-variable, per-lead-time cell improves over the i.i.d.-trained baseline, with the largest reductions on the synoptic-scale fields Z500 and T850 (mean CRPS on Z500 roughly halves, $0.053 \to 0.027$; T850, $0.058 \to 0.038$). On SEVIR (Table~\ref{tab:cat_sevir}), $\rho = 0.75$ gives consistent gains that grow with the forecast horizon: NRMSE at $h=3$ drops from $0.339$ to $0.294$ (a $13\%$ relative reduction) and mean CRPS from $0.019$ to $0.018$. On NS (Table~\ref{tab:cat_ns}), CAT is essentially flat: the i.i.d.\ baseline and the two CAT settings track each other within a few percent on every metric and lead time.

\paragraph{Why CAT helps where the observed sequence is non-Markovian.}
The pattern across the three benchmarks mirrors the Markovian / non-Markovian distinction of \S\ref{sec:exp}. On NS, the vorticity field is the \emph{full} state of 2D incompressible flow with known dynamics: once a clean current frame is available, future frames are determined by it alone, and additional past frames carry no extra information. The clean-past conditioning that AR inference relies on is therefore no more informative than what an i.i.d.-trained network already learns to use, leaving little headroom for CAT. SEVIR and ERA5 are different: the observed sequences are low-dimensional projections of much higher-dimensional latent states (a vertically integrated radar field; a 4-channel slice of a 37-level atmosphere). Past frames carry information about hidden degrees of freedom that the current frame alone does not, so the inference-time pattern of clean past and noisy future is exactly the regime in which the model needs to be trained, and matching it through CAT translates into clear forecasting gains. That the gain on SEVIR widens with the forecast horizon, while ERA5 already shows large gains at $h=1$, is consistent with this picture: SEVIR's per-step dependence on hidden state is moderate, so the gap compounds; ERA5's dependence is strong from the first step, since most of the atmospheric state is unobserved.

\newpage
\section{Cold-Start (No-Context) Data Assimilation}
\label{app:cold_start}

The main text reports results under the with-context protocol, where the model receives a short clean prefix to seed assimilation. This matches how learned DA is typically deployed in operational pipelines, in which a recent reanalysis or forecast supplies the initial condition. We additionally evaluate ForcingDAS in the more demanding \emph{cold-start} setting: no clean frames are provided, and the entire trajectory is assimilated from observations alone. Many learned DA methods cannot operate in this regime, or degrade severely; we use it as additional evidence of the framework's versatility.

We report cold-start results on the same three benchmarks as the main text. The evaluation protocol (16 held-out trajectories of length $K=30$, observation operators, noise levels, and metrics) is unchanged from \S\ref{sec:exp}; only the context length is set to $0$. Sampling hyperparameters are in Tables~\ref{tab:hp_ns}, \ref{tab:hp_sevir}, and~\ref{tab:hp_era5}.

\subsection{Navier--Stokes Vorticity}
\label{app:cold_start:ns}

Table~\ref{tab:ns_cold_start} reports cold-start NS results at SO-5\% ($K=30$, all 30 frames assimilated and scored), with the same eight methods and three regimes as Table~\ref{tab:ns_main}. Setup, shared observations, and metrics are as in Appendix~\ref{app:exp_details:ns}; Fig.~\ref{fig:ns_visualization} visualizes a representative trajectory.

\begin{table}[h]
\centering
\caption{Navier--Stokes vorticity, \emph{cold-start} at SO-5\% (no clean context, $K=30$): NRMSE ($\downarrow$) and the radially averaged kinetic-energy spectrum relative error ($\downarrow$) over mid-$k$ ($k\in[8,32)$) and all-$k$ ($k\in[0.5,64)$). Mean {\scriptsize $\pm$ std} over 16 held-out trajectories; every method assimilates identical observations.}
\label{tab:ns_cold_start}
\footnotesize
\setlength{\tabcolsep}{1.5pt}
\resizebox{\linewidth}{!}{%
\begin{tabular}{l ccc cc ccc}
\toprule
 & \multicolumn{3}{c}{Filtering} & \multicolumn{2}{c}{Fix-lag Smoothing} & \multicolumn{3}{c}{Full-sequence Smoothing} \\
\cmidrule(lr){2-4} \cmidrule(lr){5-6} \cmidrule(lr){7-9}
Metric & FlowDAS & EnKF & ForcingDAS-AR & EnKS-FL & ForcingDAS-Pyr & EnKS-FS & SDA & ForcingDAS-FS \\
\midrule
NRMSE   & 0.145\pmstd{0.004} & 0.108\pmstd{0.003} & \textbf{0.080}\pmstd{0.005} & 0.112\pmstd{0.003} & \textbf{0.069}\pmstd{0.003} & 0.114\pmstd{0.003} & 0.077\pmstd{0.003} & \textbf{0.073}\pmstd{0.003} \\
mid-$k$ & 0.820\pmstd{0.028} & 0.686\pmstd{0.008} & \textbf{0.319}\pmstd{0.038} & 0.541\pmstd{0.013} & \textbf{0.254}\pmstd{0.021} & 0.479\pmstd{0.016} & 0.213\pmstd{0.019} & \textbf{0.161}\pmstd{0.012} \\
all-$k$ & 0.869\pmstd{0.031} & 0.822\pmstd{0.023} & \textbf{0.323}\pmstd{0.033} & 0.665\pmstd{0.023} & \textbf{0.307}\pmstd{0.040} & 0.598\pmstd{0.025} & 0.280\pmstd{0.025} & \textbf{0.208}\pmstd{0.015} \\
\bottomrule
\end{tabular}}
\end{table}

\begin{figure}[t]
    \centering
    \includegraphics[width=\linewidth]{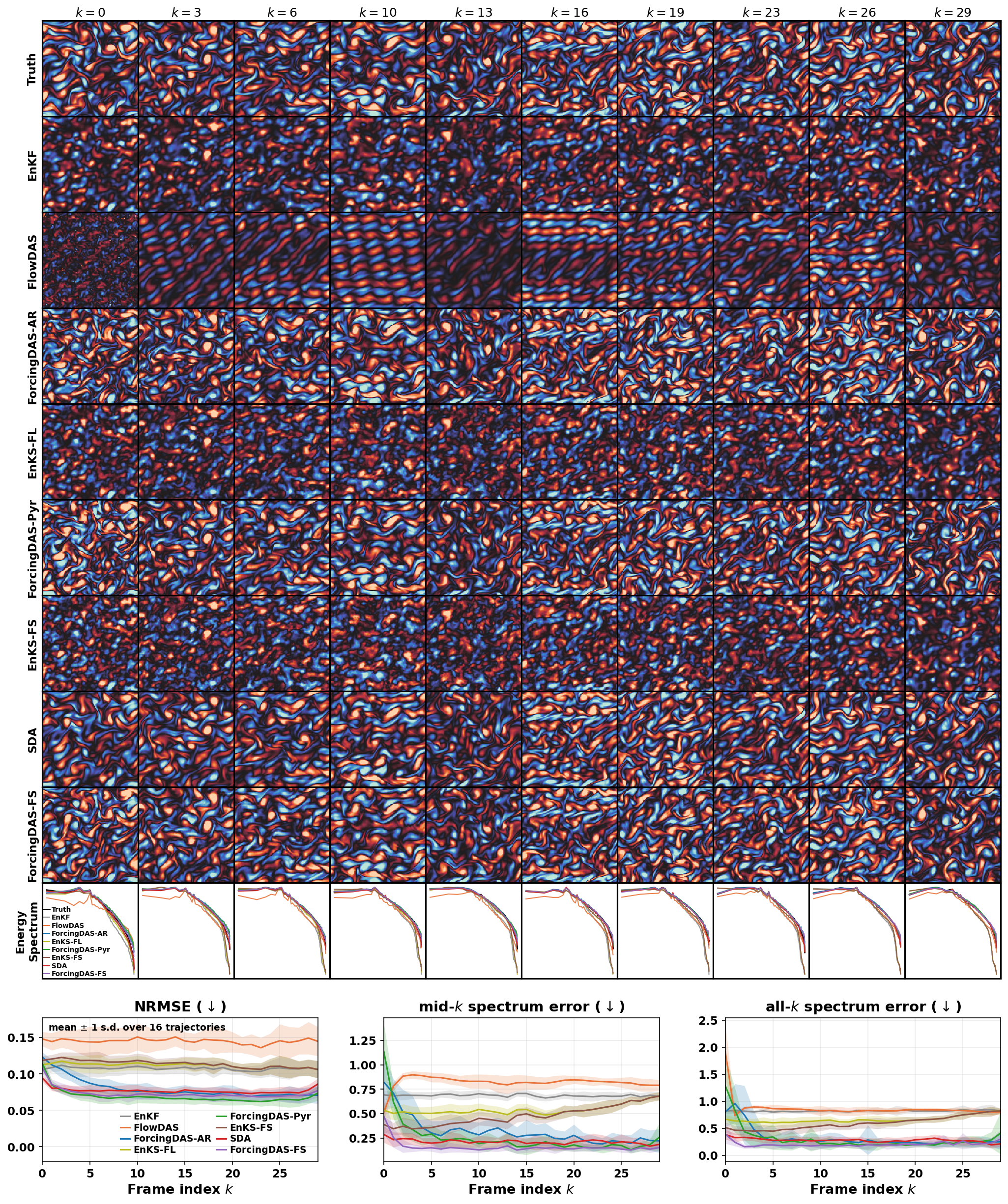}
    \caption{NS assimilation in the \emph{cold-start} protocol (SO-5\%, no clean context, $K=30$). \textbf{Top nine rows:} ground-truth vorticity and the estimates of every method, grouped by regime: the filters EnKF, FlowDAS, and ForcingDAS-AR; the fixed-lag smoothers EnKS-FL and ForcingDAS-Pyr; and the full-sequence smoothers EnKS-FS, SDA~\citep{rozet2023score}, and ForcingDAS-FS, on a representative held-out trajectory, at $k \in \{0, 3, 6, 10, 13, 16, 19, 23, 26, 29\}$. \textbf{Tenth row:} per-frame radially averaged kinetic-energy spectrum on log--log axes. \textbf{Bottom:} per-frame NRMSE, mid-$k$, and all-$k$ spectrum relative error ($\downarrow$), mean $\pm$ 1 s.d.\ over all 16 held-out trajectories. Numbers are in Table~\ref{tab:ns_cold_start}.}
    \label{fig:ns_visualization}
\end{figure}

Among the full-sequence smoothers, ForcingDAS-FS is best on all three metrics. Its NRMSE margin over SDA is small in absolute terms ($0.073$ vs.\ $0.077$) but systematic: paired over the 16 trajectories, ForcingDAS-FS is better on every trajectory for NRMSE, mid-$k$, and all-$k$ (Wilcoxon $p = 3.1\times10^{-5}$; mean NRMSE difference $-0.004$, 95\% paired-bootstrap CI $[-0.005, -0.003]$), and its spectrum errors are clearly lower ($0.161$ vs.\ $0.213$ mid-$k$; $0.208$ vs.\ $0.280$ all-$k$). EnKS-FS, which runs the exact PDE, is well behind ($0.114$), consistent with the severely rank-deficient sample covariance of its 100-member ensemble over a 30-frame augmented state (Appendix~\ref{app:enkf_ns}). In filtering, ForcingDAS-AR again roughly halves FlowDAS's NRMSE and improves on the EnKF ($0.080$ vs.\ $0.108$). In fixed-lag smoothing, ForcingDAS-Pyr improves on EnKS-FL over the same 10-frame lag ($0.069$ vs.\ $0.112$). Removing the clean context costs ForcingDAS little (AR $0.071 \to 0.080$, FS $0.067 \to 0.073$ against Table~\ref{tab:ns_main}).

Fig.~\ref{fig:ns_visualization} shows the cold-start behavior qualitatively. Without a clean prefix, FlowDAS never locks onto the flow: its fields show smooth periodic patterns unrelated to the truth, and its NRMSE stays near $0.15$ over the whole window. EnKS-FS produces grainy fields, and its spectrum decays too fast at high wavenumbers. SDA and ForcingDAS-FS both track the truth. The ForcingDAS regimes need a few frames to settle when no context is given. This is most visible for AR, whose first few frames are noticeably worse; after the first 5--8 frames all three regimes hold a flat, low error.

\subsection{Precipitation Nowcasting (SEVIR-VIL)}
\label{app:cold_start:sevir}

\begin{figure}[t]
    \centering
    \includegraphics[width=\linewidth]{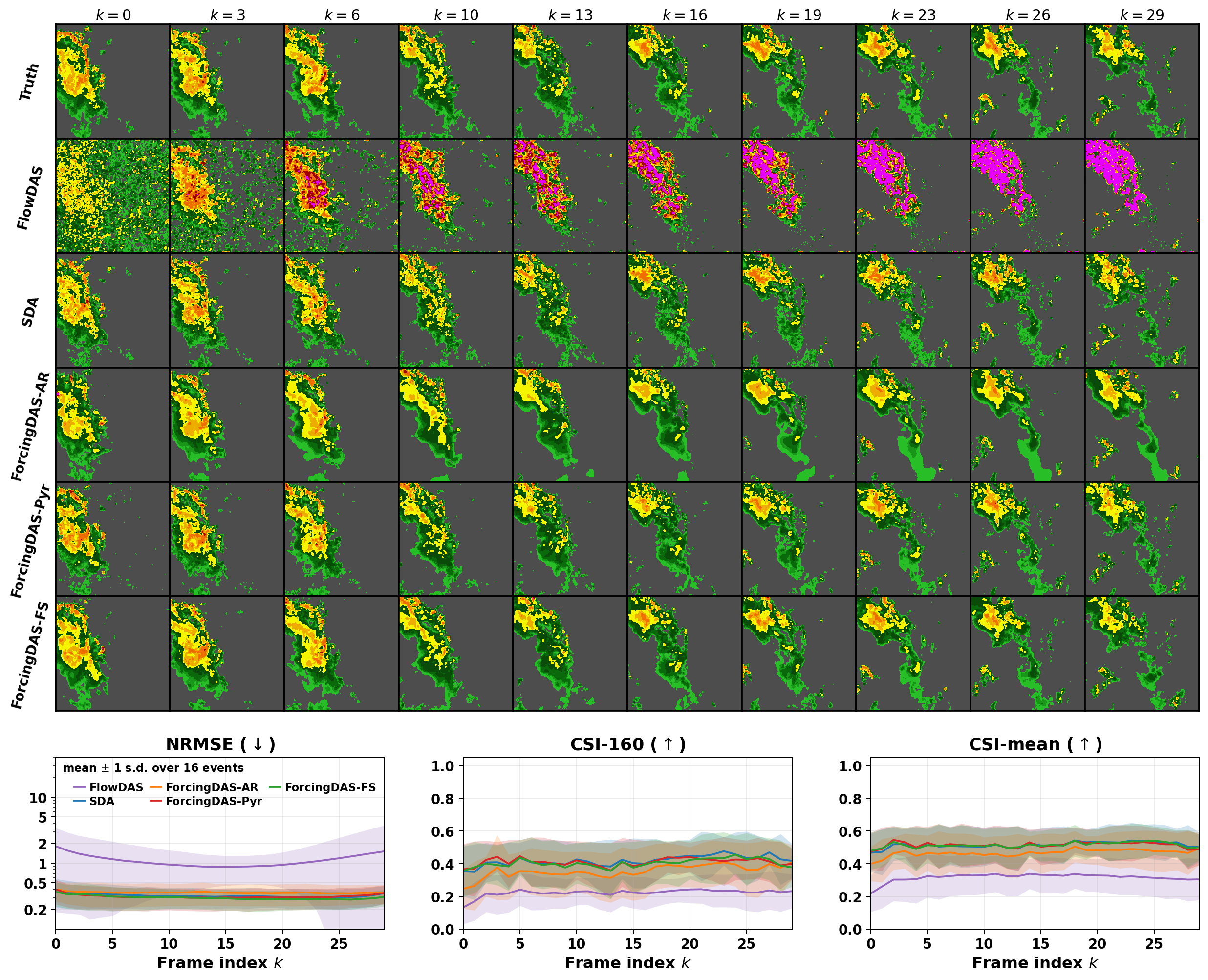}
    \caption{SEVIR VIL assimilation in the \emph{cold-start} protocol (SO-10\%, no clean context, $K=30$). \textbf{Top six rows:} ground truth and the estimates of the learned filter FlowDAS, the learned smoother SDA~\citep{rozet2023score}, and ForcingDAS in its three regimes (AR, Pyr with a 20-frame lag, FS) on the same event as \protect\figSEVIRvis{}, at $k \in \{0, 3, 6, 10, 13, 16, 19, 23, 26, 29\}$. \textbf{Bottom:} per-frame NRMSE ($\downarrow$, log scale), CSI-160 ($\uparrow$), and CSI-mean ($\uparrow$), mean $\pm$ 1 s.d.\ over all 16 events. Numbers are in Table~\ref{tab:sevir_cold_start}.}
    \label{fig:sevir_cold_start}
\end{figure}

Table~\ref{tab:sevir_cold_start} reports cold-start SEVIR results on the first 30 frames of the same 16 events, with the same shared observations; Table~\ref{tab:sevir_csi_cold} gives the per-threshold CSI and Fig.~\ref{fig:sevir_cold_start} visualizes a representative event. Here SDA runs in its native mode, without any clean context.

\begin{table}[h]
  \centering
  \caption{SEVIR VIL nowcasting, \emph{cold-start} (no clean context, $K=30$): NRMSE ($\downarrow$; mean {\scriptsize $\pm$ std} over 16 events) and CSI-mean ($\uparrow$) over thresholds $\{16, 74, 133, 160, 181, 219\}$ (pooled over all events and frames; {\scriptsize $\pm$} std over bootstrap resamples of the events). ForcingDAS-Pyr uses a 20-frame lag. Best in \emph{Filtering} and best in \emph{Full-sequence Smoothing} per column in \textbf{bold}. Per-threshold CSI is in Table~\ref{tab:sevir_csi_cold}.}
  \label{tab:sevir_cold_start}
  \footnotesize
  \setlength{\tabcolsep}{1.5pt}
  \resizebox{\linewidth}{!}{%
\begin{tabular}{l cc cc cc cc}
\toprule
 & \multicolumn{2}{c}{SO-10\%} & \multicolumn{2}{c}{SO-20\%} & \multicolumn{2}{c}{SR2x} & \multicolumn{2}{c}{SR4x} \\
\cmidrule(lr){2-3} \cmidrule(lr){4-5} \cmidrule(lr){6-7} \cmidrule(lr){8-9}
Method & NRMSE & CSI-m & NRMSE & CSI-m & NRMSE & CSI-m & NRMSE & CSI-m \\
\midrule
\multicolumn{9}{l}{\textit{Filtering}} \\
\quad FlowDAS & 1.107\pmstd{0.721} & 0.334\pmstd{0.025} & 0.709\pmstd{0.492} & 0.451\pmstd{0.029} & 0.379\pmstd{0.262} & 0.652\pmstd{0.020} & 0.736\pmstd{0.463} & 0.420\pmstd{0.027} \\
\quad ForcingDAS-AR & \textbf{0.357}\pmstd{0.132} & \textbf{0.540}\pmstd{0.032} & \textbf{0.268}\pmstd{0.096} & \textbf{0.626}\pmstd{0.025} & \textbf{0.200}\pmstd{0.062} & \textbf{0.692}\pmstd{0.020} & \textbf{0.356}\pmstd{0.137} & \textbf{0.535}\pmstd{0.035} \\
\multicolumn{9}{l}{\textit{Fix-lag Smoothing}} \\
\quad ForcingDAS-Pyr & 0.313\pmstd{0.116} & 0.580\pmstd{0.026} & 0.225\pmstd{0.077} & 0.672\pmstd{0.022} & 0.167\pmstd{0.051} & 0.721\pmstd{0.018} & 0.318\pmstd{0.130} & 0.572\pmstd{0.028} \\
\multicolumn{9}{l}{\textit{Full-sequence Smoothing}} \\
\quad SDA & 0.322\pmstd{0.115} & 0.575\pmstd{0.024} & 0.252\pmstd{0.089} & 0.643\pmstd{0.020} & 0.218\pmstd{0.083} & 0.672\pmstd{0.017} & 0.343\pmstd{0.126} & 0.553\pmstd{0.024} \\
\quad ForcingDAS-FS & \textbf{0.300}\pmstd{0.105} & \textbf{0.582}\pmstd{0.026} & \textbf{0.233}\pmstd{0.077} & \textbf{0.661}\pmstd{0.021} & \textbf{0.173}\pmstd{0.051} & \textbf{0.716}\pmstd{0.017} & \textbf{0.301}\pmstd{0.101} & \textbf{0.569}\pmstd{0.029} \\
\bottomrule
\end{tabular}}
\end{table}

Under matched context, ForcingDAS-FS improves on SDA on every operator, in NRMSE and in pooled CSI-mean and CSI-160 (Table~\ref{tab:sevir_csi_cold}). Because the spread across events ($\pm0.08$--$0.13$ NRMSE) is much larger than the gap between methods, the paired per-event comparison is the informative one: ForcingDAS-FS has the lower NRMSE on 11, 13, 16, and 15 of the 16 events (SO-10\%, SO-20\%, SR2x, SR4x; Wilcoxon $p = 0.05$, $0.02$, $3\times10^{-5}$, $9\times10^{-5}$). The gain is therefore consistent on SO-20\%, SR2x, and SR4x and only a trend on SO-10\%. Without a clean prefix, FlowDAS degrades sharply (NRMSE $0.616 \to 1.107$ on SO-10\%, with a large spread across events, because it diverges on several of them), while ForcingDAS changes little (Pyr $0.293 \to 0.313$). Fig.~\ref{fig:sevir_cold_start} shows this on the event of \figSEVIRvis{}: FlowDAS starts from noise and then saturates the storm core, while SDA and the three ForcingDAS regimes all recover the storm from the first frame. On SEVIR-VIL, VIL is a vertically integrated radar product, a low-dimensional projection of a much higher-dimensional atmospheric state, which favors a joint trajectory prior over a frame-to-frame transition model.

\begin{table}[h]
\centering
\caption{SEVIR per-threshold pooled CSI ($\uparrow$; {\scriptsize $\pm$} std over bootstrap resamples of the 16 events) in the \emph{cold-start} protocol (no clean context, $K=30$), grouped by DA regime. ForcingDAS-Pyr uses a 20-frame lag. With-context results are in Table~\ref{tab:sevir_csi}.}
\label{tab:sevir_csi_cold}
\footnotesize
\setlength{\tabcolsep}{2pt}
\resizebox{\linewidth}{!}{%
\begin{tabular}{l l l ccccccc}
\toprule
Obs & Regime & Method & CSI-16 & CSI-74 & CSI-133 & CSI-160 & CSI-181 & CSI-219 & Mean \\
\midrule
  \multirow{5}{*}{SO-10\%} & \multirow{2}{*}{Filtering} & FlowDAS & 0.576\pmstd{0.047} & 0.559\pmstd{0.053} & 0.436\pmstd{0.031} & 0.253\pmstd{0.018} & 0.143\pmstd{0.032} & 0.039\pmstd{0.029} & 0.334\pmstd{0.025} \\
   &  & ForcingDAS-AR & \textbf{0.814}\pmstd{0.028} & \textbf{0.784}\pmstd{0.035} & \textbf{0.602}\pmstd{0.043} & \textbf{0.424}\pmstd{0.035} & \textbf{0.368}\pmstd{0.046} & \textbf{0.247}\pmstd{0.047} & \textbf{0.540}\pmstd{0.032} \\
   & Fix-lag & ForcingDAS-Pyr & 0.837\pmstd{0.026} & 0.814\pmstd{0.030} & 0.642\pmstd{0.035} & 0.478\pmstd{0.028} & 0.418\pmstd{0.038} & 0.290\pmstd{0.037} & 0.580\pmstd{0.026} \\
   & \multirow{2}{*}{Full-seq.} & SDA & 0.821\pmstd{0.029} & 0.805\pmstd{0.031} & 0.628\pmstd{0.031} & 0.473\pmstd{0.024} & \textbf{0.426}\pmstd{0.035} & \textbf{0.299}\pmstd{0.040} & 0.575\pmstd{0.024} \\
   &  & ForcingDAS-FS & \textbf{0.849}\pmstd{0.023} & \textbf{0.815}\pmstd{0.030} & \textbf{0.640}\pmstd{0.035} & \textbf{0.476}\pmstd{0.029} & 0.420\pmstd{0.040} & 0.293\pmstd{0.039} & \textbf{0.582}\pmstd{0.026} \\
\midrule
  \multirow{5}{*}{SO-20\%} & \multirow{2}{*}{Filtering} & FlowDAS & 0.642\pmstd{0.052} & 0.661\pmstd{0.051} & 0.560\pmstd{0.038} & 0.386\pmstd{0.023} & 0.288\pmstd{0.045} & 0.170\pmstd{0.064} & 0.451\pmstd{0.029} \\
   &  & ForcingDAS-AR & \textbf{0.856}\pmstd{0.022} & \textbf{0.841}\pmstd{0.026} & \textbf{0.686}\pmstd{0.032} & \textbf{0.528}\pmstd{0.029} & \textbf{0.486}\pmstd{0.042} & \textbf{0.359}\pmstd{0.045} & \textbf{0.626}\pmstd{0.025} \\
   & Fix-lag & ForcingDAS-Pyr & 0.884\pmstd{0.018} & 0.866\pmstd{0.022} & 0.726\pmstd{0.026} & 0.589\pmstd{0.024} & 0.546\pmstd{0.036} & 0.420\pmstd{0.039} & 0.672\pmstd{0.022} \\
   & \multirow{2}{*}{Full-seq.} & SDA & 0.851\pmstd{0.025} & 0.845\pmstd{0.026} & 0.690\pmstd{0.025} & 0.554\pmstd{0.022} & 0.523\pmstd{0.033} & 0.393\pmstd{0.037} & 0.643\pmstd{0.020} \\
   &  & ForcingDAS-FS & \textbf{0.877}\pmstd{0.019} & \textbf{0.857}\pmstd{0.023} & \textbf{0.712}\pmstd{0.027} & \textbf{0.574}\pmstd{0.024} & \textbf{0.533}\pmstd{0.036} & \textbf{0.413}\pmstd{0.037} & \textbf{0.661}\pmstd{0.021} \\
\midrule
  \multirow{5}{*}{SR2x} & \multirow{2}{*}{Filtering} & FlowDAS & 0.795\pmstd{0.038} & 0.849\pmstd{0.025} & 0.722\pmstd{0.022} & 0.588\pmstd{0.023} & 0.543\pmstd{0.037} & 0.411\pmstd{0.040} & 0.652\pmstd{0.020} \\
   &  & ForcingDAS-AR & \textbf{0.880}\pmstd{0.018} & \textbf{0.875}\pmstd{0.022} & \textbf{0.745}\pmstd{0.024} & \textbf{0.617}\pmstd{0.027} & \textbf{0.581}\pmstd{0.039} & \textbf{0.453}\pmstd{0.038} & \textbf{0.692}\pmstd{0.020} \\
   & Fix-lag & ForcingDAS-Pyr & 0.899\pmstd{0.016} & 0.893\pmstd{0.018} & 0.773\pmstd{0.021} & 0.651\pmstd{0.023} & 0.616\pmstd{0.033} & 0.492\pmstd{0.037} & 0.721\pmstd{0.018} \\
   & \multirow{2}{*}{Full-seq.} & SDA & 0.856\pmstd{0.025} & 0.860\pmstd{0.024} & 0.715\pmstd{0.023} & 0.593\pmstd{0.022} & 0.567\pmstd{0.032} & 0.443\pmstd{0.032} & 0.672\pmstd{0.017} \\
   &  & ForcingDAS-FS & \textbf{0.894}\pmstd{0.016} & \textbf{0.887}\pmstd{0.019} & \textbf{0.763}\pmstd{0.021} & \textbf{0.645}\pmstd{0.022} & \textbf{0.614}\pmstd{0.033} & \textbf{0.496}\pmstd{0.033} & \textbf{0.716}\pmstd{0.017} \\
\midrule
  \multirow{5}{*}{SR4x} & \multirow{2}{*}{Filtering} & FlowDAS & 0.634\pmstd{0.048} & 0.667\pmstd{0.047} & 0.536\pmstd{0.036} & 0.345\pmstd{0.026} & 0.232\pmstd{0.035} & 0.103\pmstd{0.037} & 0.420\pmstd{0.027} \\
   &  & ForcingDAS-AR & \textbf{0.823}\pmstd{0.028} & \textbf{0.776}\pmstd{0.037} & \textbf{0.585}\pmstd{0.047} & \textbf{0.420}\pmstd{0.038} & \textbf{0.360}\pmstd{0.048} & \textbf{0.246}\pmstd{0.048} & \textbf{0.535}\pmstd{0.035} \\
   & Fix-lag & ForcingDAS-Pyr & 0.806\pmstd{0.030} & 0.815\pmstd{0.029} & 0.637\pmstd{0.034} & 0.475\pmstd{0.029} & 0.409\pmstd{0.042} & 0.288\pmstd{0.047} & 0.572\pmstd{0.028} \\
   & \multirow{2}{*}{Full-seq.} & SDA & 0.803\pmstd{0.032} & 0.791\pmstd{0.032} & 0.607\pmstd{0.033} & 0.450\pmstd{0.024} & 0.394\pmstd{0.035} & 0.270\pmstd{0.040} & 0.553\pmstd{0.024} \\
   &  & ForcingDAS-FS & \textbf{0.833}\pmstd{0.024} & \textbf{0.811}\pmstd{0.031} & \textbf{0.629}\pmstd{0.036} & \textbf{0.467}\pmstd{0.031} & \textbf{0.403}\pmstd{0.044} & \textbf{0.274}\pmstd{0.048} & \textbf{0.569}\pmstd{0.029} \\
\bottomrule
\end{tabular}}
\end{table}

Per-event CSI is noisy at high thresholds. Under a per-event mean instead of pooling, SDA is level with or slightly ahead of ForcingDAS-FS on CSI-160 for cold-start SO-10\% and SR4x (by at most $0.013$), while ForcingDAS-FS remains ahead on the other two operators; pooled, ForcingDAS-FS is ahead on all four.

\subsection{Global Atmospheric State Estimation (ERA5)}
\label{app:cold_start:era5}

\begin{figure}[t]
    \centering
    \includegraphics[width=\linewidth]{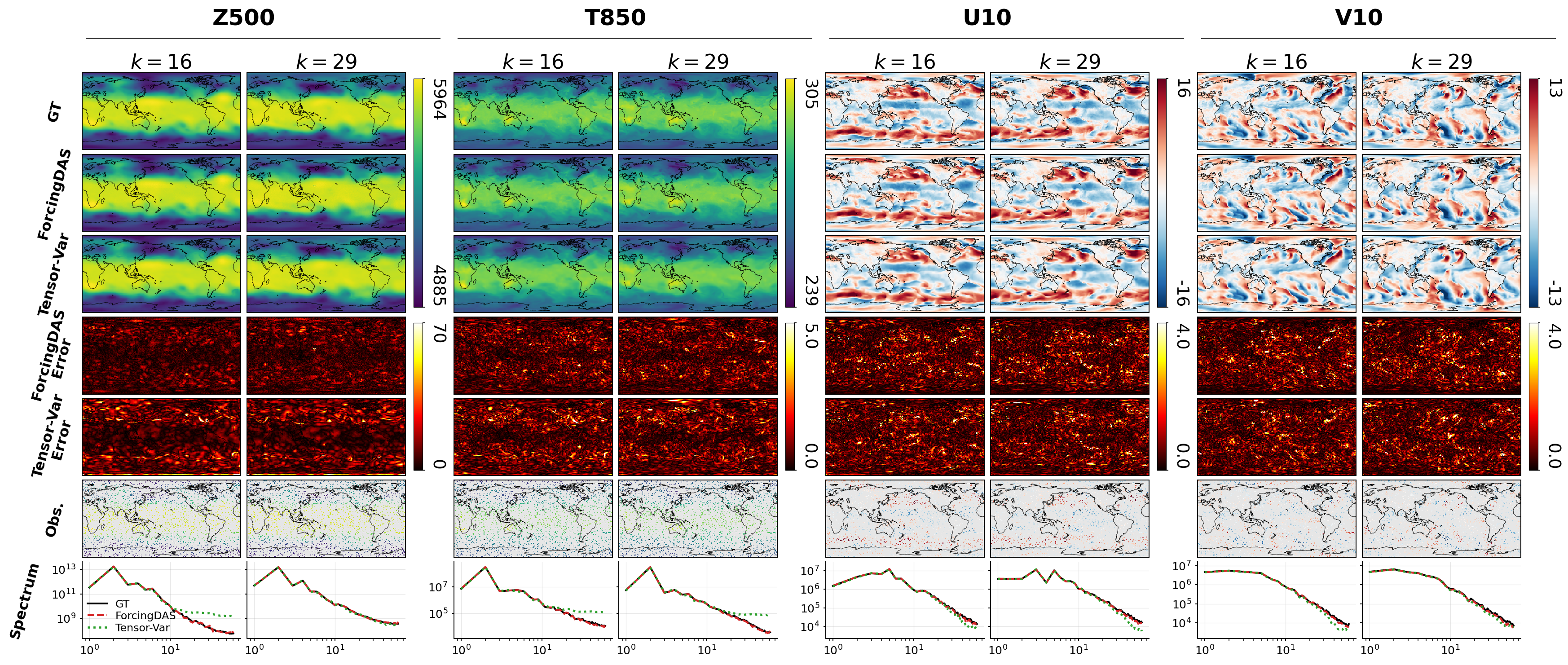}
    \caption{ERA5 SO-10\% assimilation in the \emph{cold-start} protocol (no clean context, $K=30$), on the same trajectory and in the same layout as Fig.~\ref{fig:era5_z500_comparison}; ForcingDAS runs as a full-sequence smoother (FS). Numbers are in Table~\ref{tab:era5_cold_start}.}
    \label{fig:era5_cold_start}
\end{figure}

Table~\ref{tab:era5_cold_start} reports cold-start ERA5 results (no clean context, $K=30$) on the same 16 trajectories and shared observations as the main text. A cold-start run begins with no clean state, so we treat the first assimilated frame as a spin-up period and exclude it from every metric, identically for all methods; scores cover frames 1--29.

\begin{table}[h]
\centering
\caption{ERA5 SO-10\% assimilation, \emph{cold-start} (no clean context, $K=30$, frame 0 excluded as spin-up): latitude-weighted NRMSE ($\downarrow$, z-score units) and ACC ($\uparrow$) per variable. Mean {\scriptsize $\pm$ std} over 16 held-out trajectories from 2016. Best in \emph{Filtering} and best in \emph{Full-sequence Smoothing} per column in \textbf{bold}.}
\label{tab:era5_cold_start}
\footnotesize
\setlength{\tabcolsep}{2pt}
\resizebox{\linewidth}{!}{%
\begin{tabular}{l cc cc cc cc}
\toprule
 & \multicolumn{2}{c}{Z500} & \multicolumn{2}{c}{T850} & \multicolumn{2}{c}{U10} & \multicolumn{2}{c}{V10} \\
\cmidrule(lr){2-3} \cmidrule(lr){4-5} \cmidrule(lr){6-7} \cmidrule(lr){8-9}
Method & NRMSE & ACC & NRMSE & ACC & NRMSE & ACC & NRMSE & ACC \\
\midrule
\multicolumn{9}{l}{\textit{Filtering}} \\
\quad 3D-Var           & 0.058\pmstd{0.007} & 0.971\pmstd{0.009} & 0.098\pmstd{0.007} & 0.896\pmstd{0.030} & 0.380\pmstd{0.008} & 0.853\pmstd{0.010} & 0.417\pmstd{0.010} & 0.875\pmstd{0.007} \\
\quad FlowDAS          & 0.084\pmstd{0.004} & 0.860\pmstd{0.026} & 0.143\pmstd{0.004} & 0.699\pmstd{0.046} & 0.273\pmstd{0.006} & 0.915\pmstd{0.005} & 0.320\pmstd{0.006} & 0.921\pmstd{0.007} \\
\quad ForcingDAS-AR    & \textbf{0.027}\pmstd{0.001} & \textbf{0.994}\pmstd{0.001} & \textbf{0.056}\pmstd{0.001} & \textbf{0.968}\pmstd{0.004} & \textbf{0.165}\pmstd{0.003} & \textbf{0.973}\pmstd{0.002} & \textbf{0.199}\pmstd{0.003} & \textbf{0.972}\pmstd{0.002} \\
\multicolumn{9}{l}{\textit{Fix-lag Smoothing}} \\
\quad ForcingDAS-Pyr   & 0.025\pmstd{0.003} & 0.995\pmstd{0.002} & 0.050\pmstd{0.002} & 0.974\pmstd{0.005} & 0.153\pmstd{0.004} & 0.977\pmstd{0.001} & 0.187\pmstd{0.004} & 0.975\pmstd{0.002} \\
\multicolumn{9}{l}{\textit{Full-sequence Smoothing}} \\
\quad Tensor-Var       & 0.035\pmstd{0.001} & 0.990\pmstd{0.001} & 0.064\pmstd{0.002} & 0.956\pmstd{0.006} & 0.156\pmstd{0.003} & 0.976\pmstd{0.001} & \textbf{0.186}\pmstd{0.003} & \textbf{0.976}\pmstd{0.002} \\
\quad ForcingDAS-FS    & \textbf{0.022}\pmstd{0.000} & \textbf{0.996}\pmstd{0.000} & \textbf{0.050}\pmstd{0.001} & \textbf{0.974}\pmstd{0.004} & \textbf{0.154}\pmstd{0.003} & \textbf{0.977}\pmstd{0.001} & 0.187\pmstd{0.003} & 0.975\pmstd{0.002} \\
\bottomrule
\end{tabular}}
\end{table}

Without a clean context, ForcingDAS-AR again improves on 3D-Var on all four variables and both metrics, on every trajectory. The other learned filter, FlowDAS, is the method that suffers most from losing the clean prefix: its NRMSE rises from $0.035/0.070/0.206/0.249$ with context (Table~\ref{tab:era5_main}) to $0.084/0.143/0.273/0.320$, a factor of $2.4$ on Z500 and $2.0$ on T850, against $1.03$ on every variable for ForcingDAS-AR. That leaves it behind even the classical 3D-Var on Z500 ($0.084$ vs.\ $0.058$) and T850 ($0.143$ vs.\ $0.098$), although it stays well ahead on the surface winds ($0.273$ vs.\ $0.380$, $0.320$ vs.\ $0.417$); a per-step transition model has no clean state to propagate at the start of a cold-start run, whereas a joint trajectory prior conditions on the whole window. Among the smoothers, ForcingDAS-FS improves on Tensor-Var on Z500 ($0.022$ vs.\ $0.035$), T850 ($0.050$ vs.\ $0.064$), and U10 ($0.154$ vs.\ $0.156$), in each case on all 16 trajectories. V10 is the exception: Tensor-Var is slightly better ($0.186$ vs.\ $0.187$ NRMSE, better on 14 of 16 trajectories). The Z500 spread of ForcingDAS-Pyr comes from a few trajectories whose first frames after the spin-up frame are still unsettled. Removing the clean context costs ForcingDAS little: FS changes from $0.022/0.049/0.149/0.182$ with context (Table~\ref{tab:era5_main}) to $0.022/0.050/0.154/0.187$. The two protocols score different frames (24 vs.\ 29), so this comparison is indicative only. As on the other benchmarks, the four ERA5 channels are a small slice of a much higher-dimensional atmospheric state, and a joint trajectory prior can use past frames to infer what the current frame does not show.

\newpage
\section{Baseline Implementation Details}
\label{app:classical_da}

This appendix describes the ERA5 baselines (3D-Var, FlowDAS, Tensor-Var; \S\ref{sec:exp:era5}) and the Navier--Stokes ensemble baselines (EnKF, EnKS; \S\ref{sec:exp:ns}). The learned NS and SEVIR baselines are described in Appendices~\ref{app:exp_details:ns:baselines} and~\ref{app:exp_details:sevir:baselines}. Every baseline assimilates the same stored observations as ForcingDAS and is scored by the same evaluation code.

\subsection{ERA5: Shared Setup}

\paragraph{Background state.}
In the cold-start setting the background state is the climatological mean, which is $\bm{x}_b = \mathbf{0}$ in normalized space. With context, the 6 clean frames are their own background, and the last context frame is the first-pass background for every assimilated frame.

\paragraph{Background error covariance.}
We model the background error covariance $\bm{B}$ using the \emph{control variable transform} (CVT)~\citep{courtier1998ecmwf}, the standard operational approach for introducing spatial correlations into $\bm{B}$ without forming the full $\mathbb{R}^{d \times d}$ matrix.
A control variable $\bm{v}$ is optimized instead of the state increment directly; the state is recovered via $\bm{x} = \bm{x}_b + \sigma_b\, K \ast \bm{v}$, where $K$ is a Gaussian smoothing kernel applied per channel in the Fourier domain and $\sigma_b = 1$.
This implicitly specifies $\bm{B} = \sigma_b^2\, \bm{K}\bm{K}^\top$, a spatially correlated covariance whose off-diagonal structure allows observations to inform nearby unobserved grid points through the correlation length scale of $K$.
We use per-variable isotropic Gaussian kernels with length scales $\ell = (8, 6, 5, 5)$ grid points for (Z500, T850, U10, V10) respectively, reflecting the decreasing spatial coherence from synoptic-scale geopotential to sub-synoptic surface winds.

\paragraph{Observation model.}
Observations follow the shared binary spatial mask $\bm{m} \in \{0,1\}^{H \times W}$ (identical across all $C$ channels):
\begin{equation}
    \bm{y}_{k,c} = \bm{x}_{k,c} \odot \bm{m} + \bm{\epsilon}, \quad \bm{\epsilon} \sim \cN(\mathbf{0}, \sigma_y^2\, \I),
\end{equation}
where $\sigma_y = 0.05$ in normalized space. The baselines read the stored observations rather than drawing their own noise, so they assimilate exactly the values ForcingDAS assimilates.

\paragraph{Spatial grid.}
All data are cropped from $240\times121$ to $240\times120$ by dropping the North-Pole row, exactly as for ForcingDAS.

\subsection{3D-Var (Filtering)}

3D-Var performs an independent analysis at each time step.
Using the CVT parameterization from the shared setup above, we optimize the control variable $\bm{v}_k$ at each time step:
\begin{equation}
    J_{\text{3D}}(\bm{v}_k) = \frac{1}{2} \|\bm{v}_k\|^2 + \frac{1}{2\sigma_y^2} \sum_{c=1}^{C} \|\bm{y}_{k,c}[\bm{m}] - \bm{x}_{k,c}[\bm{m}]\|^2,
    \quad \bm{x}_k = \bm{x}_{b,k} + \sigma_b\, K_c \ast \bm{v}_{k,c},
    \label{eq:3dvar}
\end{equation}
where $K_c$ is the per-channel Gaussian kernel and $[\bm{m}]$ denotes restriction to observed grid points.
The background penalty $\frac{1}{2}\|\bm{v}_k\|^2$ in control space is equivalent to $\frac{1}{2}(\bm{x}_k - \bm{x}_{b,k})^\top \bm{B}^{-1} (\bm{x}_k - \bm{x}_{b,k})$ in state space.

\paragraph{Cycling.}
Although each frame is analyzed independently (no temporal coupling in the cost function), we cycle the background sequentially in two passes: the first pass uses the background of the shared setup above; the second uses persistence of the first-pass analysis, $\bm{x}_{b,k} = \bm{x}_{a,k-1}$.
This propagates information across time via the background field, allowing observations at time $k{-}1$ to inform the prior at time $k$.

\paragraph{Optimization.}
We use L-BFGS with 80 outer iterations (each with 50 inner line-search steps, history size 50, strong Wolfe conditions).
The Gaussian kernel $K_c$ is applied in the Fourier domain via \texttt{rfft2}/\texttt{irfft2}, making each L-BFGS step $O(CHW \log(HW))$.
For efficiency, all frames are optimized jointly in a single batch; this is equivalent to per-frame optimization since the cost decomposes across time.

\paragraph{Interpretation.}
3D-Var with a Gaussian CVT produces a spatially smoothed analysis: the correlation kernel fills unobserved locations by interpolating from nearby observed pixels. This works reasonably for the smooth synoptic-scale Z500 field (ACC $0.964$ with context), but the Gaussian smoothing removes structure at scales shorter than the correlation length. For the surface winds, whose spatial scales approach the kernel width, its ACC drops to $0.79$--$0.81$ and its NRMSE is about $3\times$ that of ForcingDAS-AR. Since 3D-Var has no dynamical model, the clean context helps it only through the persistence background, and its with-context errors are in fact higher than its cold-start errors (the two protocols score different frames; Tables~\ref{tab:era5_main} and~\ref{tab:era5_cold_start}).

\subsection{FlowDAS (Filtering)}
FlowDAS~\citep{flowdas} learns a one-step stochastic-interpolant forecast model and assimilates each frame by guided sampling from it. On ERA5 it conditions on the 6 clean context frames and rolls out 24 autoregressive steps, assimilating the shared observation of each frame. In the cold-start protocol the conditioning window is replaced by Gaussian noise and all 30 frames are assimilated, using the same checkpoint and sampler. The drift model is a U-Net with 128 base channels and multipliers $1{-}2{-}2{-}2$, 30.2M parameters, taking 28 input channels (6 conditioning frames $\times$ 4 variables, plus the noisy frame) and predicting 4; it was trained for $300{,}000$ steps on the same \texttt{era5\_full} 1979--2015 training set and normalization as ForcingDAS. Sampling uses 500 Euler--Maruyama steps per frame with 25 Monte-Carlo guidance samples and guidance scale $1.0$, selected on the 4 held-out tuning windows.

\subsection{Tensor-Var (Smoothing)}
Tensor-Var~\citep{yang2025tensorvar} has four components: a CNN encoder $K_S$ that maps a state to a 512-dimensional feature, and a matching decoder $K_S^{+}$ with skip connections; a linear feature dynamics matrix $C \in \mathbb{R}^{512\times512}$, estimated by ridge regression on encoded consecutive states; a Restormer-style inverse-observation network $K_O$ that maps a stack of 10 observation frames to a state estimate; and a 4D-Var in feature space with a soft dynamics penalty, solved in closed form over non-overlapping 3-frame windows and decoded with $K_S^{+}$. We re-implemented it from the public code and retrained it on our setting: the four variables on the $240\times120$ grid (replicate-padded to $240\times128$ internally), the 1979--2015 training years with the same normalization, and the SO-10\% observations with $\sigma_y = 0.05$. Training takes 100 epochs for $K_S$, $K_S^{+}$, and $C$, then 100 epochs for $K_O$ with the others frozen, on 4 A40 GPUs ($\approx$113 GPU-hours in total). The model has 33.6M parameters (32.3M for $K_S$ and $K_S^{+}$, 1.2M for $K_O$). We use the published feature-space 4D-Var weights ($B=0.1$, $R=1$, $Q=0.1$). Inference takes about 1\,s per 30-frame trajectory on one GPU. The observation window of $K_O$ for frame $k$ is the ten most recent observations $\bm{y}_{k-9},\ldots,\bm{y}_{k}$, including the current one; with context, the context slots of that window hold the masked clean states. One detail favors Tensor-Var: $K_O$ is trained on the same fixed observation mask used at evaluation, so the model is specific to that mask, whereas ForcingDAS never sees the mask during training.

\paragraph{Fairness considerations.}
The comparisons differ in what each method is given. 3D-Var has a fixed Gaussian spatial prior and no dynamics. ForcingDAS learns its trajectory prior from the 1979--2015 training years and never sees the observation mask during training; FlowDAS likewise learns its forecast model from ERA5 training data and applies the observations only at inference. Tensor-Var learns its dynamics and its inverse-observation network from the same training years, but its inverse network is trained on the evaluation mask. All methods assimilate identical observations and are scored by the same code on the same frames.

\subsection{Ensemble Kalman Filter and Smoother Baselines for Navier--Stokes}
\label{app:enkf_ns}

We implement classical ensemble data assimilation baselines for the 2D Navier--Stokes experiment (\S\ref{sec:exp:ns}):
a stochastic Ensemble Kalman Filter (EnKF)~\citep{evensen2003ensemble} and two variants of the Ensemble Kalman Smoother (EnKS): a fixed-lag variant (EnKS-FL) and a full-sequence variant (EnKS-FS).
All three share the same forward model, observation operator, and core analysis update; they differ only in whether and how future observations are used to refine past state estimates.

\subsubsection{Forward model}

The EnKF/EnKS forecast step integrates the same stochastic vorticity equation~\eqref{eq:ns_vorticity} with the same forcing structure~\eqref{eq:ns_forcing} used to generate the truth data (Appendix~\ref{app:exp_details:ns:dataset}).
The forecast model uses a pseudo-spectral discretization on a $128 \times 128$ grid with a semi-implicit Euler--Maruyama time integrator at $\Delta t_{\mathrm{sim}} = 10^{-3}$:
linear terms (viscosity and drag) are treated implicitly, eliminating CFL restrictions from diffusion;
the nonlinear advection term is treated explicitly with $\tfrac{2}{3}$-rule dealiasing;
stochastic forcing is discretized via the standard Euler--Maruyama scheme (the appropriate choice for SDEs, unlike higher-order deterministic integrators such as RK4).
We use $\Delta t_{\mathrm{sim}} = 10^{-3}$ because the scheme is not stable at $2\times10^{-3}$ over a full observation interval: integrated freely for $\Delta t_{\mathrm{obs}} = 1.0$, the explicit advection term overflows within about 7 cycles. The smaller step is also closer to the $\Delta t = 10^{-4}$ at which the truth was generated.

\paragraph{Model error.}
As described in Appendix~\ref{app:exp_details:ns:dataset}, the truth data were generated at $256 \times 256$ resolution ($\Delta t = 10^{-4}$) and bilinearly downsampled to $128 \times 128$~\citep{pmlr-v235-chen24n}.
The EnKF forecast runs natively at $128 \times 128$, introducing an unavoidable resolution-dependent model error: the downsampled dynamics differ from dynamics simulated natively at the coarser resolution due to missing sub-grid energy transfer.
The exact PDE parameters (viscosity, drag, forcing structure) are known and matched.

\subsubsection{EnKF configuration}

\paragraph{Ensemble.}
We use $N_e = 100$ ensemble members, each initialized as a random snapshot drawn from the training trajectories only, so that no member is taken from the trajectory being assimilated. The EnKF does not use the clean context frames: it starts from this ensemble and assimilates from frame 10 onward.

\paragraph{Forecast.}
Each ensemble member is integrated forward by $\Delta t_{\mathrm{obs}} = 1.0$ time units (1000 steps at $\Delta t_{\mathrm{sim}} = 10^{-3}$), the spacing between consecutive stored frames, using the stochastic NS solver with independent noise realizations per member.
This ensures the ensemble spread captures the inherent forecast uncertainty from the unknown stochastic forcing.

\paragraph{Analysis.}
We use the batch stochastic EnKF formulation.
Let $\bm{X}^f \in \mathbb{R}^{N_e \times n}$ denote the forecast ensemble ($n = 128^2 = 16{,}384$), $\bm{H}$ the observation operator that selects observed grid points, and $\bm{R} = \sigma_y^2 \bm{I}_m$ the observation error covariance.
The Kalman gain is computed as
\begin{equation}
    \bm{K} = \bm{P}^f \bm{H}^\top \left(\bm{H}\bm{P}^f\bm{H}^\top + \bm{R}\right)^{-1},
\end{equation}
where $\bm{P}^f$ is the sample covariance from the ensemble.
Each member is updated with perturbed observations~\citep{burgers1998analysis}: $\bm{\omega}_i^a = \bm{\omega}_i^f + \bm{K}(\tilde{\bm{y}}_i - \bm{H}\bm{\omega}_i^f)$, where $\tilde{\bm{y}}_i = \bm{y} + \bm{\epsilon}_i$, $\bm{\epsilon}_i \sim \mathcal{N}(\mathbf{0},\,\bm{R})$, with the perturbations mean-centered across the ensemble.
The analysis ensemble mean is the state estimate at each observation time.
Before the forecast--analysis cycling begins, one analysis step is performed at $t_0$ using sparse observations of the initial frame, so that the output frame count matches the data-driven methods.

\paragraph{Localization.}
Covariance localization via the Gaspari--Cohn function~\citep{gaspari1999construction} is applied element-wise to both $\bm{P}^f\bm{H}^\top$ and $\bm{H}\bm{P}^f\bm{H}^\top$, with a cutoff radius of 15~grid points.
Periodic boundary conditions are respected in the distance computation.

\paragraph{Inflation.}
Multiplicative prior inflation is applied before each analysis step: $\bm{\omega}_i \leftarrow \bar{\bm{\omega}} + \lambda(\bm{\omega}_i - \bar{\bm{\omega}})$ with $\lambda = 1.10$, to counteract ensemble collapse from undersampling.

\subsubsection{EnKS algorithm}

Both EnKS variants extend the EnKF by updating past state estimates using future observations, via an augmented-state approach.

\paragraph{Mechanism.}
At each analysis step, the current ensemble is concatenated with ensembles stored from past observation times into an augmented state vector:
\begin{equation}
    \bm{x}_{\mathrm{aug}} = \begin{bmatrix} \bm{x}_{t-L} \\ \bm{x}_{t-L+1} \\ \vdots \\ \bm{x}_t \end{bmatrix} \in \mathbb{R}^{(L+1) \cdot n},
\end{equation}
where $L$ is the lag and $n = 128^2 = 16{,}384$ is the state dimension.
The observation operator $\bm{H}$ acts only on the current-time block (last $n$ components).
The standard EnKF analysis is applied to the full augmented state; because the ensemble sample covariance naturally contains cross-time correlations, the Kalman update propagates observation information backward to the earlier blocks.
After analysis, the augmented state is split back into individual time-step ensembles.
Localization is applied by tiling the spatial Gaspari--Cohn matrix across all time blocks in the augmented state.

\paragraph{EnKS-FL (fixed-lag, $L=10$).}
The buffer holds the most recent $L+1$ ensembles (current + $L$ past). Each state estimate is refined by up to $L$ future observations before being finalized and removed from the buffer. We use $L=10$, the same window as ForcingDAS-Pyr, so that the two fixed-lag smoothers are compared over the same lag. It is evaluated in both protocols.

\paragraph{EnKS-FS (full-sequence).}
The buffer holds the ensembles of the entire assimilation window, and each state is refined by all subsequent observations with no early finalization.
With $N_e = 100$ ensemble members and an augmented state of up to $30 \times 16{,}384 = 491{,}520$ dimensions in the cold start, the sample covariance is severely rank-deficient, which limits the smoother's effectiveness and can degrade results compared to smaller windows.

\subsubsection{Observation setup}

EnKF, EnKS-FL, and EnKS-FS assimilate the shared observations of Appendix~\ref{app:exp_details:ns:obs}: the same 795 observed grid points (5\% of the $128\times128$ grid, fixed across time steps) and the same noise realization, converted to raw vorticity, so that $\bm{R} = \sigma_y^2 \I$ with $\sigma_y = 1.829$. Consecutive observations are $\Delta t_{\mathrm{obs}} = 1.0$ time units apart.

\subsubsection{Protocol}

The 16 evaluation trajectories are assimilated independently, and metrics are averaged across trajectories. In the with-context protocol ($K=50$), the first 10 frames (indices 0--9) are the context frames of the other methods and are not scored; assimilation runs for 39 cycles over frames 10--49 (40 output frames including the initial analysis). In the cold-start protocol ($K=30$), assimilation covers all 30 frames (29 cycles, no frames skipped). EnKS-FS uses a lag spanning the whole assimilation window ($L=39$ with context, $L=29$ in the cold start) with no temporal taper, so every state is updated by all later observations.

\newpage
\section{Limitations and Future Work}
\label{app:limitations}

This appendix expands the brief discussion in the main-text Limitations and Conclusion (\S\ref{sec:conclusion}), and closes with directions for future work.

\subsection{Limitations}

\paragraph{Causal-only smoothing.}
Because the denoising network uses causal temporal attention, future observations can only refine past states through the backward gradient of the observation loss; the forward pass carries no future-to-past signal (Appendix~\ref{app:asymmetry}). A model with bidirectional attention could also pass this information through the forward pass, but would no longer support causal filtering from the same weights. Removing this restriction, e.g., through a hybrid causal/bidirectional backbone, while preserving the single-model filtering--smoothing interpolation is an open direction.

\paragraph{Compute at long horizons.}
Pyramid and full-sequence scheduling denoise multiple frames jointly at every DDIM step, with memory and compute that scale with the active window, $\min(K, \lceil T/u \rceil)$ frames for the pyramid schedule and the whole horizon for the full-sequence schedule. For very long trajectories, this cost can become prohibitive relative to the purely autoregressive regime, and blockwise or hierarchical schedules may be needed.

\paragraph{Simulated observations.}
All observations in our experiments are synthetic: a fixed random pixel mask or a down-sampling operator applied to the reference field, plus i.i.d.\ Gaussian noise. Real observing systems have irregular and time-varying geometry, instrument-specific and often nonlinear operators (e.g., satellite radiances), and correlated, non-Gaussian errors. On ERA5 in particular, the reference is itself a reanalysis, so we assimilate synthetic observations of a model-consistent state rather than raw measurements. Assimilating real instrument observations is left to future work.

\paragraph{Resolution and variables on ERA5.}
Our ERA5 study operates at $1.5^{\circ}$ with four variables, sufficient to compare against the 3D-Var, FlowDAS, and Tensor-Var baselines on the same footing but coarser than operational weather systems ($0.25^{\circ}$ with 60+ variables). Scaling the framework to native WeatherBench2-style resolutions will likely require a stronger diffusion backbone, possibly a latent-diffusion variant, and improved memory handling.

\paragraph{Probabilistic calibration.}
We evaluate a single estimated trajectory per run; while the per-frame noise levels provide an \emph{internal} uncertainty representation, we have not systematically evaluated the calibration of ensemble statistics obtained by running multiple seeds of the reverse process. A careful probabilistic evaluation with spread--error diagnostics is left to future work.

\medskip
\subsection{Future Work}
Beyond the limitations above, we see three concrete directions for extending ForcingDAS.

\paragraph{Physics constraints and reward-driven fine-tuning.}
The current trajectory prior is purely data-driven and may produce trajectories that are statistically plausible yet inconsistent with the conservation laws of the underlying PDE. Adding soft physics-residual penalties during training~\citep{bastek2025physics}, using diffusion-style multi-step refinement to stabilize long PDE rollouts~\citep{lippe2023pde}, and projecting sampling updates onto a constraint manifold at inference~\citep{he2024manifold} are all natural ways to close this gap. A complementary direction is reward-driven fine-tuning of the trajectory prior on assimilation-specific objectives such as long-horizon NRMSE or spectral fidelity, via policy-gradient reinforcement learning~\citep{black2024training} or preference-based alignment~\citep{wallace2024diffusion}.

\paragraph{Distillation, latent diffusion, and representation learning.}
Pyramid and full-sequence sampling are expensive at long horizons. Distilling the reverse process into a few-step student, and lifting the trajectory prior into a latent space, would both reduce cost and provide a clean setting in which to study the geometry of the learned representation~\citep{NEURIPS2025_9a3b1949,zhang2025generalization,li2026mclr}.

\paragraph{Spatio-temporal medical imaging.}
ForcingDAS is, at its core, a joint denoising framework for sequences of partial observations of an evolving state, which matches the structure of many dynamic medical imaging problems. Natural targets include accelerated cardiac cine MRI~\citep{qiu2024spatiotemporal}, motion-compensated 4D cone-beam CT for radiotherapy~\citep{motion_4d_CBCT_2023}, SPECT~\citep{li2025shorter,jia202390y,singh2023score}, and dynamic PET imaging~\citep{dynamic_pet_2022}. The trajectory-prior and scheduling-matrix machinery transfer directly; only the observation operator (k-space, projection geometries, Poisson noise) is modality-specific.

\end{document}